\documentclass[prb,preprint]{revtex4}

\pdfoutput=1

\usepackage{graphicx}

\begin{document}

\title{Road to room-temperature superconductivity: A universal model\medskip }

\date{February 14, 2013} \bigskip

\author{Manfred Bucher \\}
\affiliation{\text{\textnormal{Physics Department, California State University,}} \textnormal{Fresno,}
\textnormal{Fresno, California 93740-8031} \\}

\begin{abstract}
In a semiclassical view superconductivity is attributed exclusively to the advance of atoms' outer $s$ electrons through the nuclei of neighbor atoms in a solid.  The necessary progression of holes in the opposite direction has the electric and magnetic effect as if $two$ electrons were advancing instead of each actual one.  Superconductivity ceases when the associated lateral oscillation of the outer $s$ electrons extends between neighbor atoms.  If such overswing occurs already at $T = 0$, then the material is a normal conductor.  Otherwise, lateral overswing can be caused by lattice vibrations at a critical temperature $T_{c}$ or by a critical magnetic field $B_{c}$.  Lateral electron oscillations are reduced---and $T_{c}$ is increased---when the atoms of the outer $s$ electrons are \emph{squeezed}, be it in the bulk crystal, in a thin film, or under external pressure on the sample.  The model is applied to alkali metals and alkali-doped fullerenes.  Aluminum serves as an example of a simple metal with superconductivity.  Application of the model to transition metals, intertransitional alloys and compounds of transition metals with other elements sheds light on the pattern of their critical temperature.  More examples of the squeeze effect are provided by the superconductivity of $PdH$, $MgB_{2}$, borocarbides, ferropnictides, and organic charge-transfer salts.  The model also provides the superconduction mechanism in the oxide superconductors, exemplified by $YBa_{2}Cu_{3}O_{7}$.  Finally the model suggests which steps to take in order to reach superconductivity at room temperature and above.
\end{abstract}

\maketitle

\section{INTRODUCTION}
  The semiclassical Bohr-Sommerfeld model of the atom is extended to electron conductivity in solids. As the model takes into account only the \emph{particle} character of electrons, it lends itself to easy visualization and conceptualization.  It is quantum theoretically incomplete, however, due to the neglect of the electrons' wave character.  Historically, the Bohr-Sommerfeld model has been successful for the energy levels, orbit size and (to some extent) angular momentum of the hydrogen atom.  The latter was plagued by a small but systematic disagreement with experimental observation.  That disagreement has been removed recently by permitting an electron of angular quantum number $\ell  = 0$ ($s$ electron) to swing through the atomic nucleus, called ``Coulomb oscillation''(see Fig. 1).\cite{1}

\emph{\underline {Hydrogen-molecule ion.}}  Applied to the hydrogen-molecule ion, $H_{2}^{+}$, where the original Bohr-Sommerfeld model failed, the concept of the Coulomb oscillator has successfully yielded the molecule's stability, binding energy and nuclear spacing.\cite{1}  Since the semiclassical treatment of $H_{2}^{+}$ serves as a paradigm for the present model of superconductivity, its essential features shall be summarized:  The single electron of $H_{2}^{+}$ has two modes of motion in the electrostatic field of the two nuclei:  In \emph{axial} Coulomb oscillation the electron swings back and forth between the far turning points and through both nuclei and the (empty) midpoint (see Fig. 2). Conversely, in a \emph{lateral} oscillation it swings perpendicularly between lateral turning points and through the midpoint. 

The motion is quantized such that the sum, 
\begin{equation}
{\tilde A} = {{\tilde A}_x} + {{\tilde A}_y},
\end{equation}
\noindent of axial and lateral action integrals, 
\begin{equation}
{\tilde A_x} = m\oint {{v_x}dx,} 
\end{equation}                                                         
\noindent and 
\begin{equation}
{{\tilde A}_y} = m\oint {{v_y}dy,} 
\end{equation}               
\noindent equals an integer multiple of Planck's constant, 
\begin{equation}
{\tilde A} = nh,
\end{equation}
\noindent with $m$ being the electron's mass, $v_{x}$ and $v_{y}$ its axial and lateral speed, respectively, and $n$ its principal quantum number.  This is known as the Einstein quantization condition.\cite{1}  

\includegraphics[width=6in]{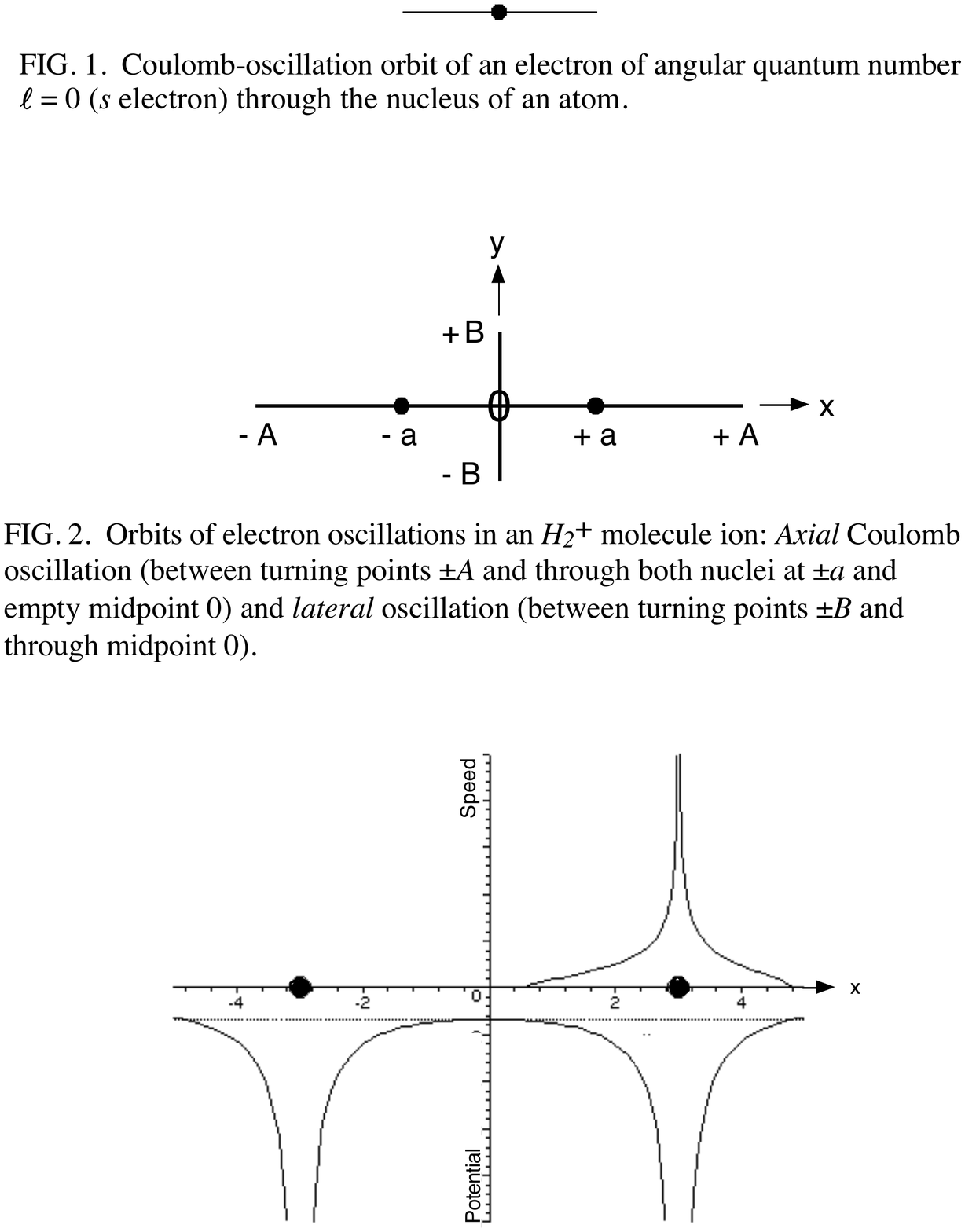}

\noindent FIG. 3. Axial electron speed $v_{x}$ (top curve), axial potential $V_{x}$ (bottom curve) and total energy level $E$ (bottom line) of $H_{2}^{+}$ when nuclei are \emph{far} apart ($R$ = 6 a. u.).

\pagebreak

\includegraphics[width=6in]{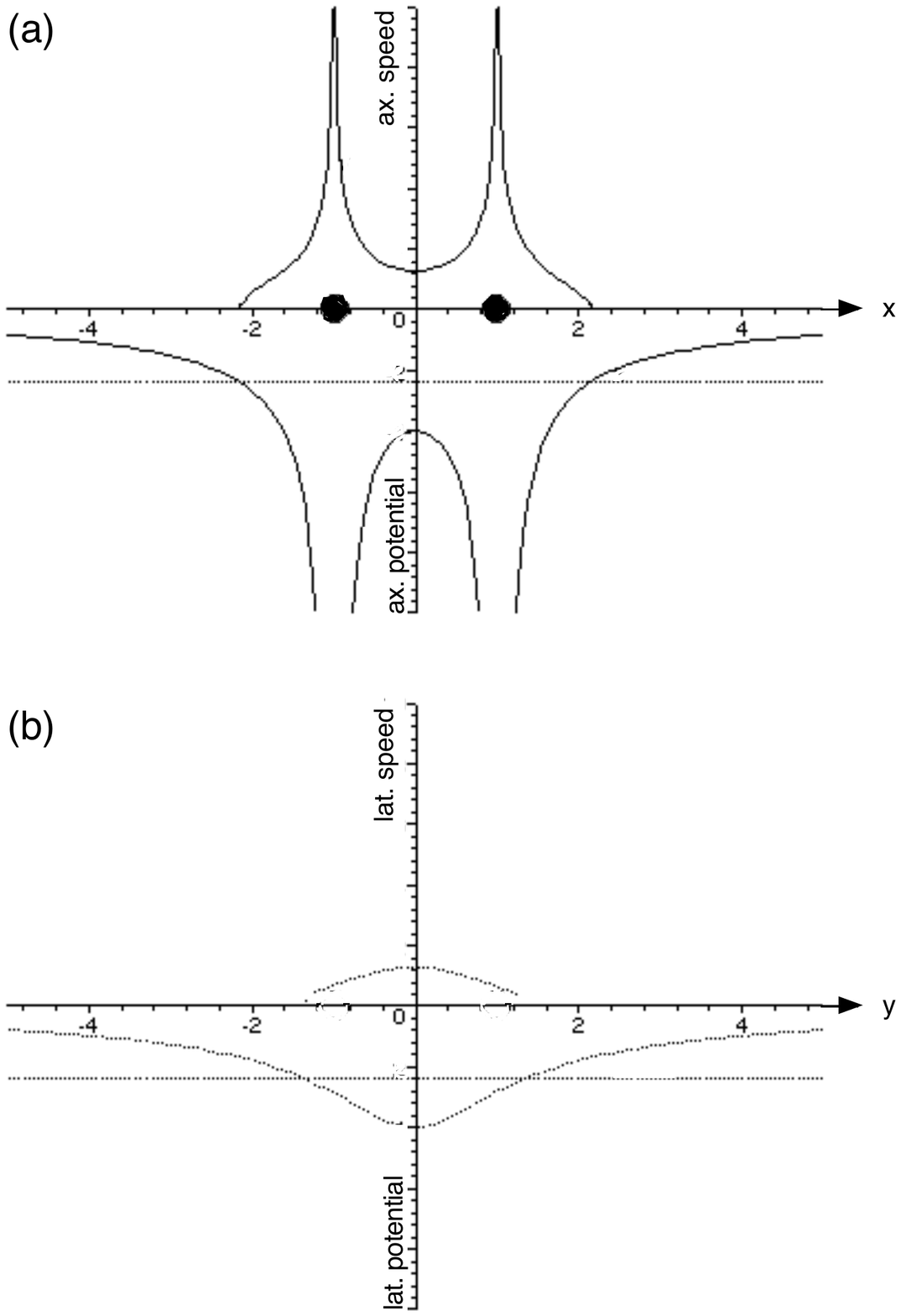}

\noindent FIG. 4. (a) Axial electron speed $v_{x}$ (top curve), axial potential $V_{x}$ (bottom curve) and total energy level $E$ (bottom line) of $H_{2}^{+}$ when nuclei are \emph{close} ($R$ = 2 a. u.). (b) Associated lateral electron speed $v_{y}$ (top curve), lateral potential $V_{y}$ (bottom curve) and same total energy level $E$ (bottom line).

\pagebreak
By ignoring the electron's wave character it is not possible to view both oscillation modes classically as a continuous process.  However, one may imagine that on average the molecule-ion's electron spends half the time in either axial or lateral oscillation.  Both action integrals, Eqs. (2) and (3), are conveniently visualized by the area under the respective axial and lateral speed curve $v_{x}(x)$ and $v_{y}(y)$ (see Figs. 3 and 4). When the nuclei are far apart, as in Fig. 3, then the electron is trapped in the potential funnel about one nucleus through which it oscillates.  On the other hand, when the nuclei are sufficiently close, then axial Coulomb oscillation occurs through both nuclei as well as lateral oscillation through the midpoint, shown in Fig. 4.

While the ($\ell  = 0$ inclusive) Bohr-Sommerfeld model gives the same energies for the hydrogen-molecule ion $H_{2}^{+}$ as quantum mechanics, these theories differ in the \emph{spatial} description of atomic entities.  To phrase it in a casual but memorable analogy, the Bohr-Sommerfeld orbits furnish the ``skeleton,''  and the Schr\"{o}dinger wave function $\psi$ (more precisely, $\psi \psi^{*}$)  provides the ``flesh.''   For $H_{2}^{+}$ the skeleton consists of the axial and lateral oscillation widths A0A and B0B, respectively, shown in Fig. 2.  These quantities provide proportional measures for the size of the quantum mechanical $H_{2}^{+}$ molecular orbital in terms of major and minor axes of an ellipsoid of revolution, 
\begin{equation}
\frac{2}{3}A0A = {\langle |x|\rangle _t} = \int {\psi |x| {\psi ^*}} {d^3}r
\end{equation}
\noindent and 
\begin{equation}
\frac{2}{3}B0B = {\langle |y|\rangle _t} = \int {\psi |y| {\psi ^*}} {d^3}r.  
\end{equation}
\noindent Here the ${\langle |...|\rangle _t}$ are time-averages of the magnitude of displacement of semiclassical oscillations and the $\int {\psi |...|{\psi ^*}} {d^3}r$ are moments of the $H_{2}^{+}$ molecular orbital.  The relation of oscillation widths in the Bohr-Sommerfeld treatment with quantum mechanical molecular-orbital size will be useful for the ``squeeze effect'' introduced below.

\emph{\underline {Hydrogen molecule.}}  The hydrogen molecule, $H_{2}$, has two electrons, each subject to Coulomb attraction to two nuclei at rest (in fixed-nuclei approximation) and Coulomb repulsion from the other electron in motion.  A semiclassical exploration with one electron confined to axial Coulomb oscillation and the other one to lateral oscillation with phases that provide maximum electron-electron distance yields stability and reasonable values of binding energy and nuclear spacing. A more accurate treatment of the problem is in progress.
\pagebreak

\includegraphics[width=5.75in]{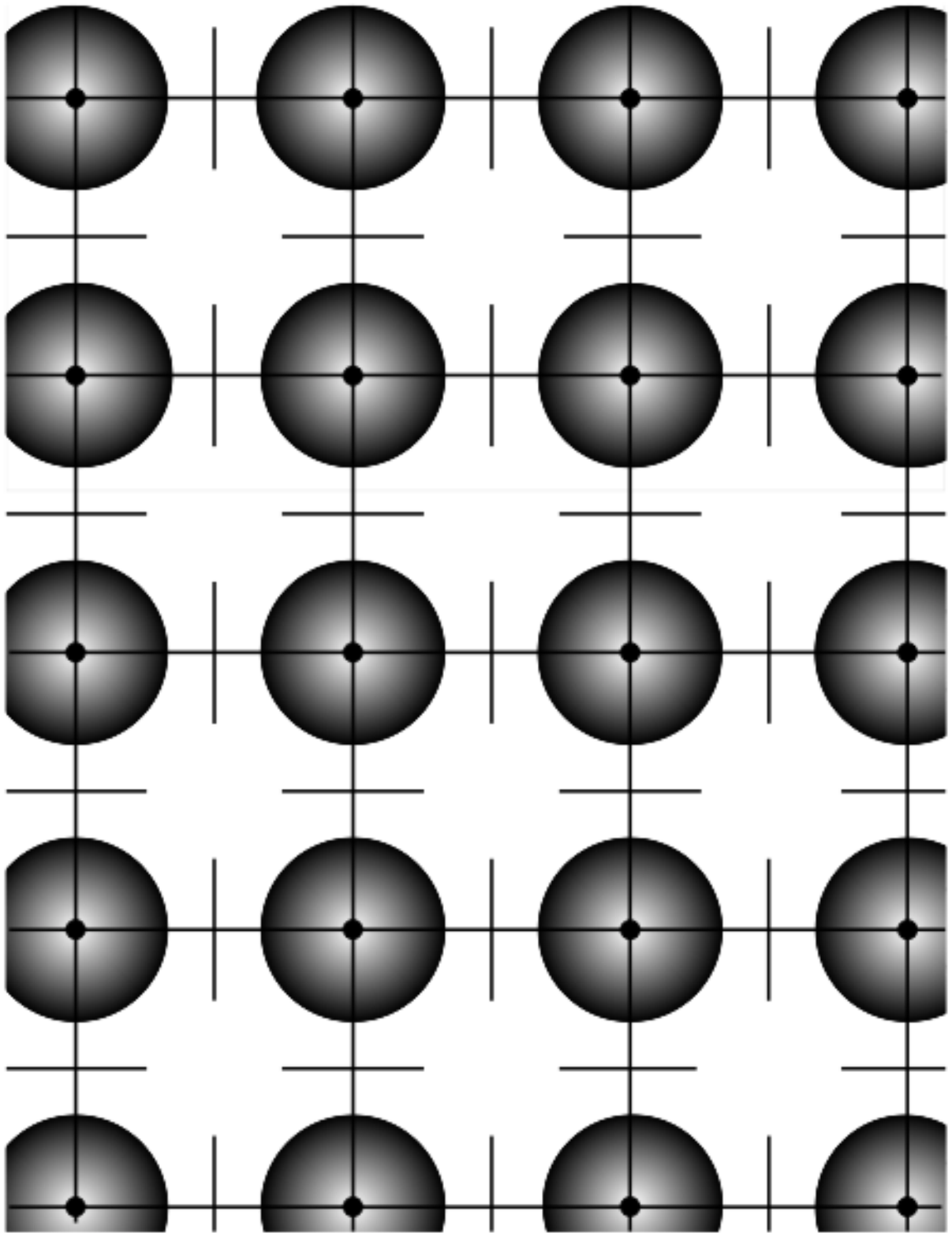}

\noindent FIG. 5.  Nuclei (dots) and inner electron shells of atoms (shaded) in a simple cubic lattice. The lines show possible orbits of the single outer electron of each atom, both axially through neighbor nuclei, and laterally about the midpoints. In the depicted case the solid would be superconducting. Conversely, if the lateral oscillations overswing, then the solid would be a normal conductor.

\pagebreak

\section{COULOMB OSCILLATIONS IN SOLIDS}  
We now apply insights from the Coulomb-oscillator treatment of the $H_{2}^{+}$ molecule ion to a solid.  First under the simplest conditions conceivable: (i) all atoms are of the same kind (elemental crystal), (ii) each atom has only one outer electron ($s$ electron), (iii) the atoms are arranged in a simple cubic lattice, at (iv) absolute zero temperature, $T=0$ (see Fig. 5).  The assumptions are overly artificial in order to highlight the basic concept of the superconductivity mechanism---more realistic examples will follow.

Three cases are possible:  (1) If the Coulomb oscillation of each atom's outer electron is only through its \emph{parent nucleus}, then the solid is an an electrical isolator.  In contrast, if each outer electron also swings through the nucleus of a neighbor atom, then the solid  is an electrical conductor.  Which \emph{kind} of conductor, depends on the concomitant lateral oscillation:  (2) If the width of the lateral oscillation is less than the nearest-neighbor distance (more generally, if the lateral oscillation does not lead over the crest of the lateral potential profile), then the solid is a superconductor (see Fig. 5).  In this case electron motion under an external electric field can proceed through the crystal only along the network of internuclear paths, delayed but undeterred by the intermediate lateral oscillations.  (3) Conversely, if the width of the lateral oscillation is more than the nearest-neighbor distance (leading over the crest of the lateral potential profile), then the solid is a normal conductor, obeying Ohm's law.  In that case the directional impetus of an electron under an external electric field gets dissipated by sideways scattering.  Without an external field, the outer electrons of all atoms then behave as a ``free electron gas'' in the crystal.

After these general preliminaries, let us approach actual materials with the Coulomb-oscillator model of electric conductivity.  A $free$ alkali atom (denoted by $A$) has, like the hydrogen atom, a single outer electron subject to the electric field of its nucleus, but now also of inner closed electron shells.  The combined electrostatic effect of nucleus and inner shells---together called the ``ion core''---can approximately be expressed by a fictitious nucleus of effective-charge number $Z$* instead of the atomic number $Z$ of the actual nucleus.  One way of fixing $Z$* is by the atom's ionization energy,
\begin{equation}
\hat E = \frac{{{Z^{*2}}}}{{{n^2}}}{R_y},
\end{equation}                                      
\noindent where $n$ is the principal quantum number of the outer electron and $R_{y} = (m e^{4}\hbar)/(2 n^{2})$ = 13.6 eV is the Rydberg energy (see Table I).
\pagebreak

\includegraphics[width=6.75 in]{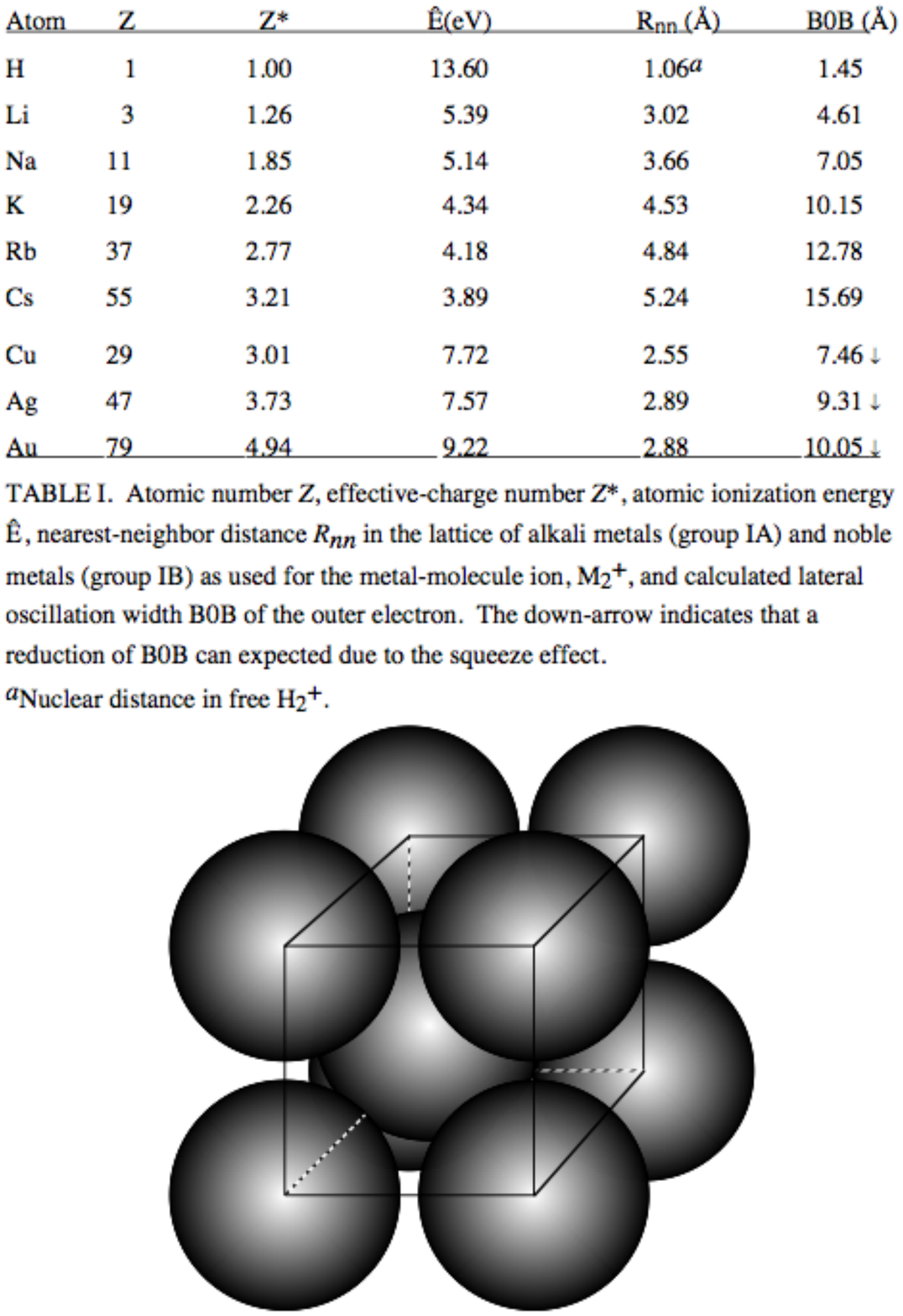}

\noindent FIG. 6. Hard-sphere display of atoms in the $bcc$ crystal lattice of an alkali metal.

\pagebreak

\includegraphics[width=5.75in]{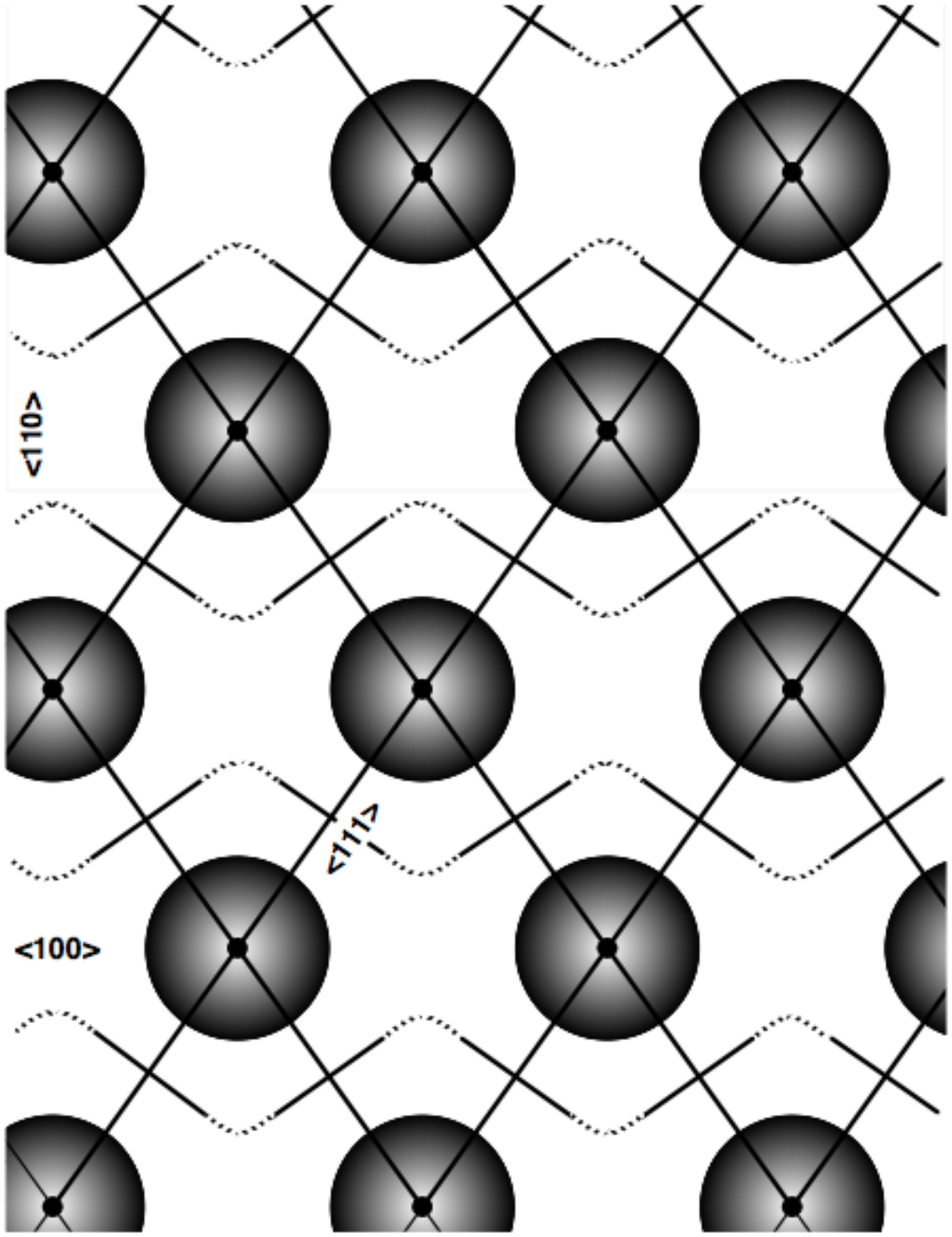}

\noindent FIG. 7. Nuclei (dots) and inner electron shells of alkali atoms (shaded) in the (110) plane of a $bcc$  lattice. Prominent crystal directions are indicated by $<$...$>$. Solid lines show possible orbits of the single outer electron of each atom, both axially through neighbor nuclei, and laterally about the midpoints. In the depicted case the solid would be superconducting. Conversely, if the lateral oscillations would overswing, as indicated by dotted curved paths, then the solid would be a normal conductor.

\pagebreak

All alkali metals crystallize in the body-centered cubic ($bcc$) lattice (see Fig. 6).  In order to view Coulomb oscillations in a $bcc$ lattice, we regard a ($110$) plane which holds nearest-neighbor atoms, as displayed in Fig. 7.  The solid lines show internuclear electron orbits of axial Coulomb oscillations between nearest-neighbor nuclei and sideways orbits of lateral oscillations about the midpoints.  In the situation depicted in Fig. 7, the solid would be a superconductor as the lateral oscillations are too short to pass the lateral potential barrier (here, in curved paths).  

In order to  get an estimate of the width of lateral oscillation, the calculation of Coulomb oscillations in $H_{2}^{+}$ was extended to alkali-molecule ions, $A_{2}^{+}$, using effective-charge numbers $Z$* obtained from ionization energies $\hat{E}$, and nearest-neighbor distances $R_{nn}$ from alkali metal lattices.  As Table I shows, the calculated widths of lateral oscillation B0B (double amplitude by the notation in Fig. 2) far exceed the nearest-neighbor distance $R_{nn}$.  It can be expected that refined calculations for isolated alkali \emph{molecules}, $A_{2}$, or lattice arrays of $A$ atoms, do not much alter this finding. This would mean that the lateral oscillations do \emph{overswing} in cystalline alkali metals and thus render these materials Ohmian conductors, in agreement with observation.  

The same calculation was also done for molecule ions $M_{2}^{+}$ of noble metals (element group IB) whose atoms, like the alkali atoms (element group IA) possess one outer $s$ electron.  Again, the calculated lateral oscillation width B0B in Table I far exceeds the nearest-neighbor distance $R_{nn}$ in the crystals' face-centered cubic ($fcc$) lattice, in agreement with the lack of superconductivity in noble metals.  As will be argued below, the B0B values of the noble metals can be expected to be reduced by the squeeze effect.  Although insufficient to make the noble metals superconducting, this trend would qualitatively explain the different kind of conductivity of alkaline-earth metals (group IIA, normal conductors) and of the group-IIB metals ($Zn, Cd, Hg$) which are superconducting.

\section{ALKALI-DOPED FULLERENES}

\emph{\underline {Larger separation of doped alkali atoms in fullerenes.}}  Whereas in alkali metals lateral oscillations of the atoms' outer electron overswing due to small nearest-neighbor distance, a new situation arises when alkali atoms are further separated. This is the case in alkali-doped fullerenes, denoted here as $\underbar{A}${\textbf{\small A}}$_2$$C_{60}$ and illustrated in Fig. 8 where the dopant atoms $A$ are 
\pagebreak

\includegraphics[width=5.65in]{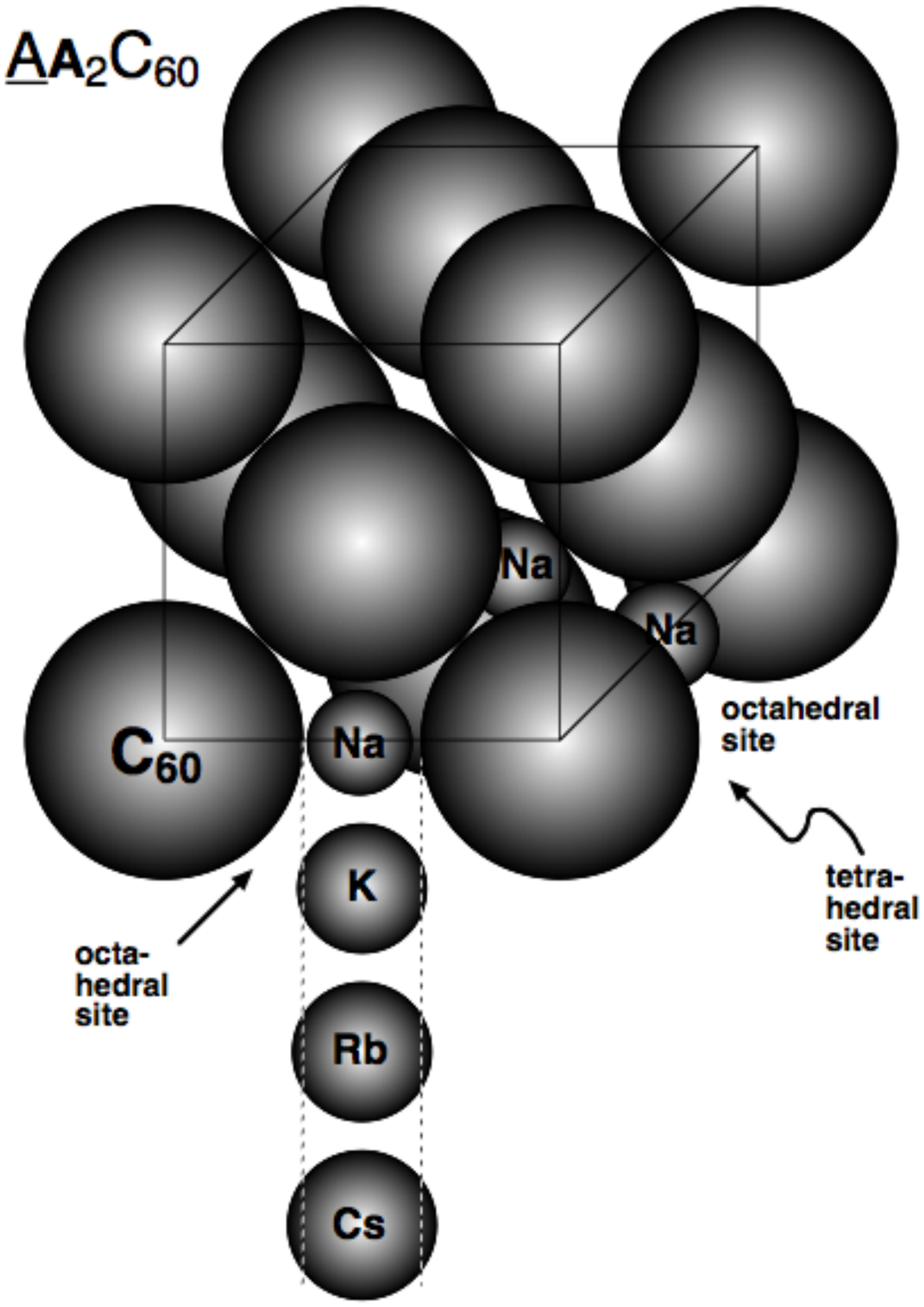}

\noindent FIG. 8.  Fullerene host crystal of $C_{60}$ molecules (big balls) at $fcc$ lattice sites, together with some alkali atoms (here, sodium) at interstitial sites.  In the chemical formula $\underbar{A}${\textbf{\small A}}$_2$$C_{60}$ alkali atoms at octahedral sites are denoted by underline, $\underbar{A}$, and at tetrahedral sites by small bold print, \textbf{\small A}.  The shown interstitial atoms are relevant in the formation of Coulomb oscillators $\underbar{A}$-\textbf{\small A} or $\underbar{A}$-$\underbar{A}$.  The lower part of the figure gives a size comparison (in hard-sphere display) of alkali atoms with respect to the octahedral vacancy.
\pagebreak

\noindent depicted as hard spheres, just as was done in Fig. 6 (with appropriate scale) for the $bcc$ alkali metals.  A comparison of Figs. 6 and 8 shows immediately a considerably larger separation of alkali atoms $A$ in doped fullerenes caused by the intervening $C_{60}$ host molecules. Quantitatively, the nearest-neighbor distances are in the range from $R_{nn}$($Li$) = 3.02 {\AA} to $R_{nn}$($Cs$) = 5.24 {\AA} in alkali metals, but from $\underbar{Li}$-\textbf{\small Li} = 6.01 {\AA} to $\underbar{Cs}$-\textbf{\small Cs} = 6.32 {\AA} in fullerene hosts, $\underbar{A}${\textbf{\small A}}$_2$$C_{60}$.  This opens the possibility, that lateral oscillations may $not$ overswing which in turn renders these materials superconducting.

\emph{\underline {Fullerene geometry.}}  A $C_{60}$ molecule, called fullerene, has 60 carbon nuclei positioned at the vertices of a truncated icosahedron.  The shape of the $C_{60}$ molecule resembles a soccer ball (when focused on internuclear lines) or, more realistically, a fat blackberry (accounting for the spheres of the $C$ atoms).  In Fig. 8 the $C_{60}$ molecules are displayed, for simplicity, as hard spheres. The $C_{60}$ molecules crystalize in an $fcc$ lattice, shown in Fig. 8. Although pure fullerene, $C_{60}$, is no superconductor, when doped with alkali metal, the doped fullerene crystal, $\underbar{A}${\textbf{\small A}}$_2$$C_{60}$, becomes superconducting.\cite{2}

The notational distinction of $\underbar{A}$ (underline) and \textbf{\small A} (small bold print) concerns the residence of dopant alkali atoms $A$ at different interstitial sites in the $C_{60}$ host crystal, being of octahedral (\underbar{A}) or tetrahedral (\textbf{\small A}) neighbor symmetry. The notation is chosen for mnemonic purposes:  The underline in the notation of octahedral interstitial atoms, \underbar{A}, indicates that these atoms reside halfway at the cubic edges of the $C_{60}$ host crystal (see Fig. 8), but also at the center of each cubic cell.  Tetrahedral sites \textbf{\small A} are at $( \pm \frac{1}{4}, \pm \frac{1}{4}, \pm \frac{1}{4})$ positions along the cell's space diagonals.  The notation with small bold print, \textbf{\small A}, indicates that those dopant atoms are compressed due to the smaller size of the tetrahedral vacancies.  There are twice as many tetrahedral than octahedral sites at which alkali atoms of $A$ and $A'$ species can settle; hence the general molecular formula $\underbar{A}${\textbf{\small A}}'$_2$$C_{60}$.  The species $A$ and $A'$ represent any alkali metal of $Li$, $Na$, $K$, $Rb$ or $Cs$.\cite{3}    When different species of alkali dopants are present, $A$ and $A'$, then the atom species of the larger size preferentially occupies the octahedral site due to that site's larger vacancy, as for example in $\underbar{Rb}${\textbf{\small Na}}$_2$$C_{60}$.  This shows up clearly in the triangular display of the compounds in Fig. 10.  The octahedral and tetrahedral vacancies (empty interstitial sites) are symbolized in pictorial notation as $|$\underline{ }\underline{ }$|$ and \underline{/}\underline{\textbackslash}.

\emph{\underline {Atom pairs with Coulomb oscillators.}}  We first consider fullerenes with only \emph{one} dopant species, $\underbar{A}${\textbf{\small A}}$_2$$C_{60}$ = $A_{3}C_{60}$. Concerning the Coulomb oscillation of the outer $s$ electron of each neighboring alkali atom, the smallest separation of intersitial sites is, in increasing order, $|\underbar{A}$-$\textbf{\small A}|$  $<$ $|\textbf{\small A}$-$\textbf{\small A}|$ $<$ $|\underbar{A}$-$\underbar{A}|$.\cite{4}  Table II lists the calculated oscillation width B0B of alkali-molecule ions $A_{2}^{+}$, obtained with the effective-charge numbers $Z^{*}$ from Table I and experimental data of nearest-neighbor distances $R_{nn}$.  The values show that diatomic Coulomb oscillators are formed in all $A_{3}C_{60}$ crystals.  It can be expected that likewise $A_{2}$ \emph{molecules} are formed between these sites.  The attractive force from the bonding of such embedded $A_{2}$ molecules tends to contract the host crystal.  However, an opposing, lattice-widening force arises when the doped $A$ atoms are larger than the interstitial voids.  The tetragonal void, being the smaller one, $r($\underline{/}\underline{\textbackslash}$)$ $<$ $r(|$\underline{ }\underline{ }$|)$, is decisive whether an $A_{3}C_{60}$ crystal is contracted to a lattice constant $a(A_{3}C_{60})$ $<$  $a(C_{60})$, as for the small doping atoms $A = Li, Na$ , or expanded to $a(A_{3}C_{60})$ $>$  $a(C_{60})$ for the larger atoms $A = K$, $Rb$, $Cs$.  The dashed line in Fig. 10 shows that division. In cases of mixed alkali fullerenes, $\underbar{A}${\textbf{\small A}}'{\textbf{\small A}}''$C_{60}$, the size of the octahedral void, $r(|$\underline{ }\underline{ }$|)$, acts as a corrective.  A comparison of the sizes of atoms and voids in Figs. 8 and 10 makes this division plausible.

Not much is known about lithium-doped fullerenes \cite{3} but the lateral oscillation gap, $\Delta y = R_{nn} -$ B0B, in Table II suggests superconductivity of $Li_{3}C_{60}$. For all other alkali-doped fullerenes $A_{3}C_{60}$ ($A = Na, K, Rb, Cs$) the $(\underbar{A}$-$\textbf{\small A})^{+}$ pair shows lateral overswing, B0B $>$ $R_{nn}$, which would render these materials normal conductors. The first case seems unobservable, the other cases are not observed.  The conclusion is premature, though, as another aspect must be considered---the squeeze effect.

In Fig. 9 the difference of the lattice constant, $a$, of alkali-doped and pure fullerene,
\begin{equation}
\Delta{a} = a(A_{3}C_{60}) - a(C_{60}),
\end{equation}
\noindent is plotted \emph{vs} the nearest-neighbor distance $R_{nn}$ in alkali metals.  All $A_{3}C_{60}$ compounds have a negative contribution to $\Delta{a}$ from the bonding of the embedded $A_{2}$ molecules, as mentioned.  Positive contributions to $\Delta{a}$ arise from the size of interstitial alkali atoms $A$ in excess of the interstitial voids, predominantly of the smaller tetragonal void \underline{/}\underline{\textbackslash}.  The trend of the experimental values in Figs. 9 and 10 suggests that the $Li$ atom fits easily into \underline{/}\underline{\textbackslash} but that the $Na$ atom more than fills that void, resulting in a slight squeeze of $Na$.  If the dopant atoms $A$ were hard spheres, then their size in excess of the void, $r(A) -$ $r($\underline{/}\underline{\textbackslash}$)$, would contribute by a geometric projection factor of $1/\sqrt 3  = 58\%$ to the lattice constant $a$, shown by the slope of the solid line in Fig. 9.  Thus if the doped $A = K$, $Rb$ or $Cs$ atoms would retain their size as in the respective metals, the $A_{3}C_{60}$ lattice would widen as shown by the 
\pagebreak

\includegraphics[width=5.65 in]{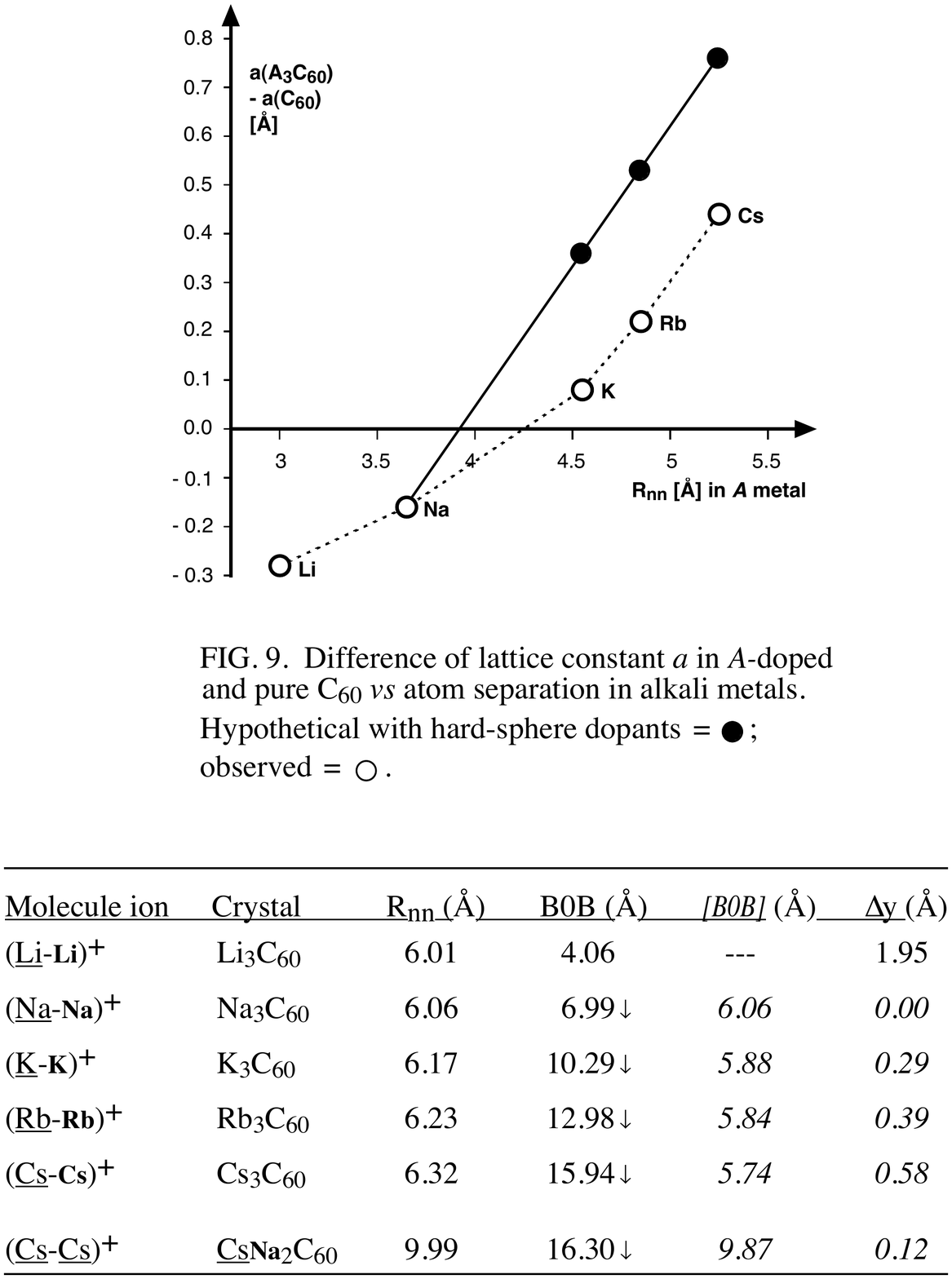}

\noindent TABLE II. Calculated lateral oscillation width B0B of alkali-molecule ions $A_{2}^{+}$ with effective-charge number $Z^{*}$ from Table I and observed nearest-neighbor distance $R_{nn}$ between interstitial sites in alkali-doped fullerenes. The down-arrow ($ \downarrow $) indicates that B0B can be expected to be reduced by the squeeze effect. The \emph{squeezed} lateral oscillation width \emph{[B0B]} was obtained in a model calculation. The lateral oscillation gap $\Delta{y}$ is to be compared with the critical temperature $T_{c}$ in Fig. 10.
\pagebreak

\noindent full dots in Fig. 9.  However, the observed lattice widening (hollow dots) shows that those $A$ atoms are squeezed in the $C_{60}$ host.

\emph{\underline {Atomic squeeze.}}  Consider two free atoms, $A$ and $A'$.  A molecular orbital ${\Psi _{AA'}}$ can be constructed, in good approximation, from one-electron quantum states $\psi _{A}$ and $\psi _{A'}$ of both atoms,
\begin{equation}
{\Psi _{AA'}} = \frac{{{\psi _A} + {\psi _{A'}}}}{{\sqrt {2 + 2{S_{AA'}}} }},
\end{equation}                                               
\noindent with overlap ${S_{AA'}} = \int {{\psi _A}\psi _{A'}^*} {d^3}r$.  Now confine one atom laterally, say $A$, by potential walls so that its free-atom wave function $\psi _{A}$ contracts to a confined wave function ${\psi _{[A]}}$.  Such lateral confinement of atom $A$ then causes a lateral shrinkage of the previous molecular orbital ${\Psi _{AA'}}$ to
\begin{equation}
{\Psi _{[A]A'}} = \frac{{{\psi _{[A]}} + {\psi _{A'}}}}{{\sqrt {2 + 2{S_{[A]A'}}} }}.
\end{equation}  
\noindent To put it simply: \emph{Lateral squeeze of bonded atoms shrinks the width of their molecular orbital.}  

The same mechanism persists when \emph{crystal} atoms $A$ and $A'$ are compressed (``squeezed'') by hard neighbor atoms or molecules (here, $C_{60}$).  The repulsive potential from
those neighbors reduces the lateral extension of the $AA'$ molecular orbital and, by Eq. (6), the width of lateral oscillation, B0B.  This reduces accordingly the lateral action $\tilde{A}_{y}$, Eq. (3).  The length of the axial Coulomb oscillation, A0A, and the axial action $\tilde{A}_{x}$, Eq. (2), increase correspondingly in order to maintain a constant quantum sum, Eqs. (1) and (4).  The lesser lateral oscillation width causes a wider lateral oscillation gap, $\Delta{y}$, (or a higher lateral potential barrier $\Delta{\epsilon}$ to overcome) with increased squeeze of the oscillator's constituent atoms. Such atomic squeeze can result from ``chemical pressure'' (compression of the atom under consideration by hard neighbor atoms in the crystal matrix), from external pressure on the sample, or from asymmetric forces on atoms near the surface of a sample (thin film).

\emph{\underline {Alkali fullerenes with one dopant species.}} In principle the reduction of the lateral oscillation width B0B of free alkali molecule ions $A_{2}^{+}$ is determined by the overlap of quantum mechanical wave functions of dopant alkali atoms $A$ with those of the host molecules $C_{60}$.  It is affected by the size of the dopant atoms, $r(A)$, and of the voids at tetrahedral and octahedral interstitial sites, $r($\underline{/}\underline{\textbackslash}$)$ and $r(|$\underline{ }\underline{ }$|)$, and can be expected to have, like the tails of the overlapping wave functions, an exponential dependence.  The \emph{squeezed} lateral oscillation width may be expressed, in the fashion of Born-Mayer short-range repulsion, as 

\noindent .
\pagebreak
\begin{equation}
\emph{[B0B]} = \text{B0B}\exp ( - \frac{{r(A) - {r(void)}}}{\rho }),             
\end{equation} 

\noindent with atom sizes $r(Li)$ = 1.51 \AA, $r(Na)$ = 1.83 \AA, $r(K) = 2.26$ {\AA}, $r(Rb) = 2.42$ {\AA}, $r(Cs) = 2.62$ {\AA}.  The tetragonal void radius was chosen as $r($\underline{/}\underline{\textbackslash}$) = 1.70$ {\AA} to reproduce the emergence of superconductivity, $T_{c} \approx 0$, in $Na_{3}C_{60}$ in terms of a vanishing oscillation gap, $\Delta{y} = R_{nn} - \emph{[B0B]} = 0$. The hardness parameter $\rho$ is dominated by properties of the $C_{60}$ molecule.  It was chosen here as $\rho = 0.90$ {\AA} to reproduce a gap ratio of $\Delta y(Na_{3}C_{60}):\Delta y(K_{3}C_{60}):\Delta y(Rb_{3}C_{60}):\Delta y(Cs_{3}C_{60}) \approx 0:2:3:4$, comparable to the ratio of their critical temperatures $T_{c}$ shown (by shaded blocks) in Fig. 10.   Achievement of this gap ratio also required the use of a smaller size of the potassium atom, $r(K) = 2.205$ {\AA}, instead of the hard-sphere value from alkali metals, $r(A) \equiv \frac{1}{2}{R_{nn}}$, given above.\cite{5}  

The lateral oscillation gaps thus obtained show, in a crude fashion, the same ordering as the corresponding critical temperatures, $\Delta y \sim {T_c}$ (compare Table II with Fig. 10).  It shall be understood that this simple numerical reduction treatment from unsqueezed to squeezed lateral oscillation width, $\text{B0B}  \to  \emph{[B0B]}$, is merely for the purpose of \emph{illustration}, not for accuracy. It tries to show that the reduction of the lateral oscillation is related, in a systematic manner, to the size and force situation in the cystal.  However, an accurate treatment of lateral-oscillation reduction requires a derivation of the corresponding parameters by quantum mechanical means.

\emph{\underline {Critical temperature.}}  At temperatures above absolute zero, thermal agitation of the atoms arises. This activates overswing of electrons across the lateral oscillation gap $\Delta{y}$ (over the lateral potential barrier $\Delta{\epsilon}$) if the phonon energy is larger than $kT_{c}$.  Here $k$ is Boltzmann's constant and $T{_c}$ is the transition temperature, conventionally called ``critical'' temperature.  By the model's criterion of lateral oscillation, such phonon-assisted lateral overswing destroys superconductivity, which renders the solid a normal (Ohmian) conductor.

If we assume, for simplicity, that the lateral potential barrier is proportional to the lateral gap, $\Delta{\epsilon}  \propto \Delta{y}$, then, together with $\Delta{\epsilon} = kT_{c}$, the lateral gap and the  critical temperature are proportionally related, $\Delta{y} \propto T{_c}$.  Conversely, the critical temperature $T_{c}$ gives a rough measure of the lateral oscillation gap $\Delta{y}$.

\emph{\underline {Critical magnetic field.}} When a superconducting sample at $T=0$ is exposed to a magnetic field $B$, then superconduction ceases if the field strength exceeds
a critical value $B_{c}$.  

\pagebreak

\includegraphics[width=7.5in]{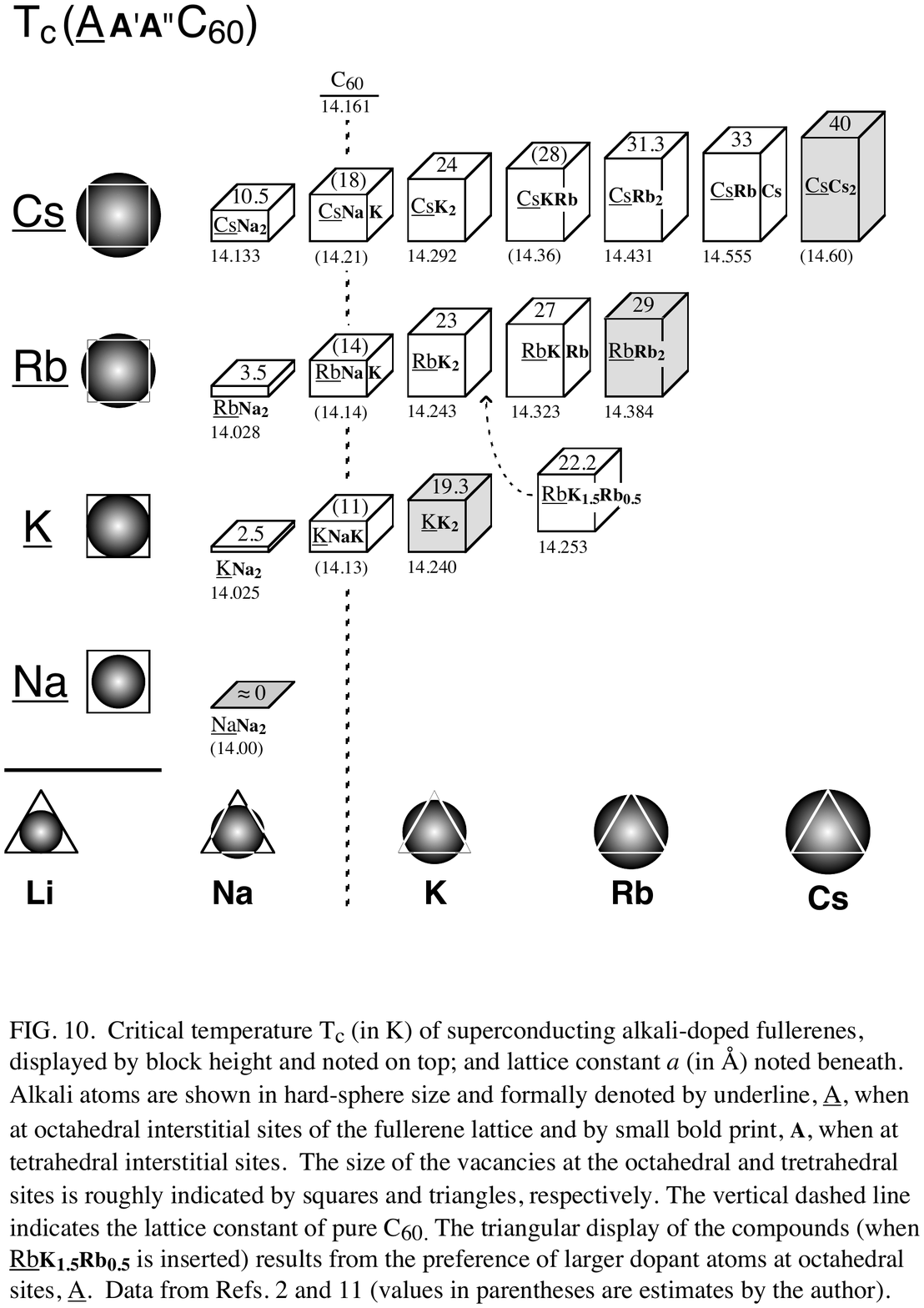}

\pagebreak 
\noindent The explanation is essentially the opposite of the squeeze effect:  The Lorentz force transfers electron motion form the predominant axial direction to the lateral direction, leading to lateral overswing at $B = B_{c}$.

\emph{\underline {Mixed alkali fullerene cases.}}  Figure 10 gives an overview of superconductivity in alkali-doped fullerenes.  The critical temperature $T_{c}$ is shown by block height and noted on top.  The lattice constant, $a$, is noted beneath.  Alkali atoms are depicted in their hard-sphere size and the extension of the interstitial voids roughly by triangles and squares.  The cases with only one dopant species, $A_{3}C_{60}$, already addressed above, are highlighted in gray.  The accuracy of the lattice constants, $a$, is high ($\pm 0.01$ {\AA} or less) but no error margin is found in the literature for $T_{c}$.  The author assumes that the $T_{c}$ data have an uncertainty of $\pm 1$ K which would smoothen the trend of $T_{c}$ in Fig. 10 in both horizontal and vertical progression.

The experimental finding of $T_{c}($\underbar{Na}${\textbf{\small Na}}_{2}$$C_{60}$) $\approx 0$, depicted at the lower left edge of the histogram, is interpreted as reduced lateral $\underbar{Na}$-{\textbf{\small Na}} Coulomb oscillations with vanishing gap, $\Delta{y} = 0$. 
So far, semiclassical calculations have not been extended to \emph{mixed} Coulomb oscillators, $A$-$A'$.  The discussion must therefore continue at a qualitative level.  The vertical progression in Fig. 10, with variable $\underbar{A}$ and constant {\textbf{\small A}}'$_2$, shows the minor influence of atomic squeeze of the $\underbar{A}$ atom in the octahedral void $|$\underline{ }\underline{ }$|$.  Thus the critical temperature changes by only $\Delta T_{c} \approx 2$ K in the upward progressions
$\underbar{Na}${\textbf{\small Na}}$_2$ $\to$ $\underbar{K}${\textbf{\small Na}}$_2$ $\to$ $\underbar{Rb}${\textbf{\small Na}}$_2$, and
$\underbar{K}${\textbf{\small K}}$_2$ $\to$ $\underbar{Rb}${\textbf{\small K}}$_2$ $\to$ $\underbar{Cs}${\textbf{\small K}}$_2$, and $\underbar{Rb}${\textbf{\small Rb}}$_2$ $\to$ $\underbar{Cs}${\textbf{\small Rb}}$_2$. However, the large increase by $\Delta T_{c} = 7$ K for $\underbar{Rb}${\textbf{\small Na}}$_2$ $\to$ $\underbar{Cs}${\textbf{\small Na}}$_2$ at the top left corner of the histogram deviates markedly from that trend and will be discussed shortly.

The horizontal progression, with constant $\underbar{A}$ and variable {\textbf{\small A}}'$_2$,
shows the major influence of atomic squeeze of the {\textbf{\small A}} atom in the tetragonal void \underline{/}\underline{\textbackslash}.  
For $\underbar{Rb}${\textbf{\small K}}$_2$ $\to$ $\underbar{Rb}${\textbf{\small Rb}}$_2$, and
$\underbar{Cs}${\textbf{\small K}}$_2$ $\to$ $\underbar{Cs}${\textbf{\small Rb}}$_2$ $\to$ $\underbar{Cs}${\textbf{\small Cs}}$_2$, the critical temperature changes by $\Delta T_{c} \approx 7 K$.  The change in critical temperature is much larger, $\Delta T_{c} \approx 18 K$,
for $\underbar{K}${\textbf{\small Na}}$_2$ $\to$ $\underbar{K}${\textbf{\small K}}$_2$, and
$\underbar{Rb}${\textbf{\small Na}}$_2$ $\to$ $\underbar{Rb}${\textbf{\small K}}$_2$, due to very little squeeze of {\textbf{\small Na}} but substantial squeeze of {\textbf{\small K}} in the
\underline{/}\underline{\textbackslash} void.  The cases with mixed occupancy at tetragonal sites, {\textbf{\small A}}'{\textbf{\small A}}'', interpolate correspondingly.

Larger atom size $r(A)$ in a given void increases both the lattice constant, $a$, and the critical temperature $T_{c}$.  By the present model the cause for the increase of $T_{c}$ rests dominantly on the atomic squeeze, not on the minor effect from lattice widening as was assumed by early researchers in the field.

The unusually high $T_{c}$ of $\underbar{Cs}${\textbf{\small Na}}$_2$$C_{60}$ at the top left corner of the histogram suggests the presence of a different mechanism than in all other cases where the Coulomb oscillators are provided by nearest $\underbar{A}$-{\textbf{\small A}}' atom pairs.  It is conceivable that in $\underbar{Cs}${\textbf{\small Na}}$_2$$C_{60}$ the Coulomb oscillators are furnished by closest $\underbar{Cs}$-$\underbar{Cs}$ pairs instead of $\underbar{Cs}$-{\textbf{\small Na}} pairs.  The bottom row in Table II shows that its free oscillation width B0B exceeds the $\underbar{Cs}$-$\underbar{Cs}$ separation $R_{nn}$.  A reduction by Eq. (11), with the previous hardness parameter $\rho  = 0.90$ {\AA} and an assumed octahedral void size $r(|$\underline{ }\underline{ }$|) = 2.18$ {\AA}, gives a lateral oscillation gap $\Delta{y}$ in roughly the same proportionality to $T_{c}$ as in the other cases.

In summary, the experimental values of critical temperature $T_{c}$ of alkali-doped fullerenes are qualitatively explained with the Coulomb-oscillator model of superconductivity.  The tenets of that model are (1) destruction of superconductivity by overwhelming the lateral oscillation gap $\Delta{y}$ through thermal agitation with phonon energy larger than $kT_{c}$, and (2) widening of the lateral gap in accordance to the squeeze of the oscillators' constituent atoms---the ''squeeze effect.''  This reduces the problem largely to geometric considerations, namely the size of dopant alkali atoms and the size of interstitial voids in the $C_{60}$ host.  

\section{ALUMINUM}

New aspects of the Coulomb-oscillator model of superconductivity arise for elements where the atoms' outer electrons have non-vanishing angular momentum, $\ell  \ne 0$.  Such electrons have, in Bohr-Sommerfeld treatment of free atoms, genuine elliptical orbits.  Aluminum metal, although of no practical importance as a superconductor, $T_{c} = 1$  K, is a good case to illustrate the new features.

The aluminum atom, $Al$, has an electron configuration $[1{s^2}2{s^2}2{p^6}]3{s^2}3{p^1}$ where $[...]$ denotes the inner electron shells.  Of the three valence electrons, the $3p$ electron is less bound to the ion core than the two $3s$ electrons, $\epsilon(3s)$ $<$  $\epsilon(3p)$ $<$ 0.  When the $Al$ atoms crystallize to form a metallic solid, those $3p$ electrons delocalize and behave as a free electron gas in the crystal.  In contrast, the $3{s^2}$ electrons of each $Al$  atom perform axial Coulomb oscillations through neighbor nuclei. At $T = 0$ the lateral oscillations are shy of overswing, shown schematically in Fig. 11.  This renders the $3{s^2}$ electrons superconducting.   By the two-fluid model of superconductivity, \cite{7} the flow of super electrons short-circuits the flow of normal electrons (here, $3s$ and $3p$, respectively) in a mixture of both types.  This then leaves aluminum metal at $T = 0$ a superconductor.

\pagebreak

\includegraphics[width=5.9in]{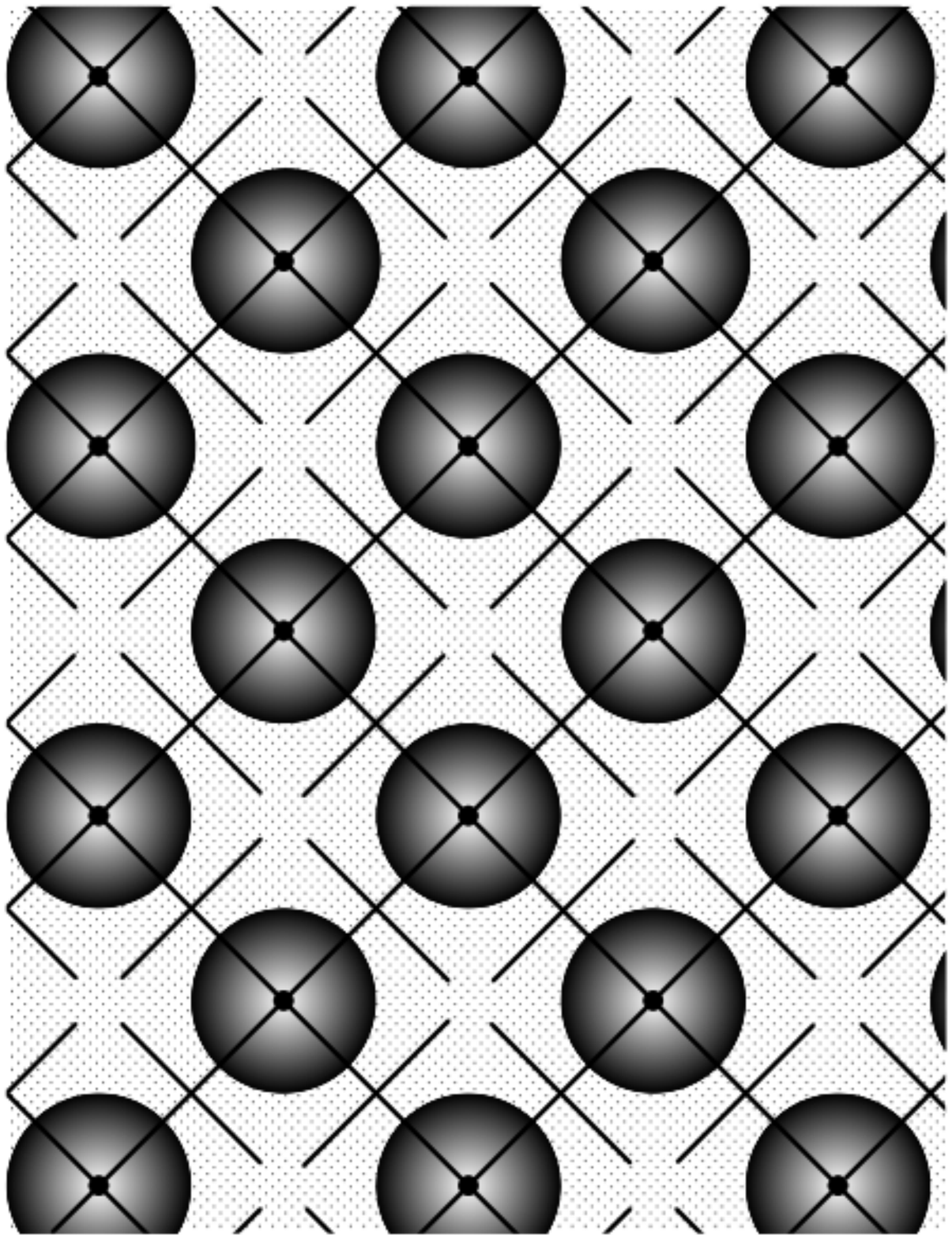}

\noindent FIG. 11. Nuclei (dots), inner electrons (dark shade), outer $s$-electron orbits (lines) and delocalized outer $p$ electrons (dotted shading) in the $fcc$ lattice of metallic $Al$.  In the situation shown the metal is superconducting due to lateral oscillation gaps.  Conversely, sufficient thermal activation at $T$ $>$ $T_{c}$ will cause lateral overswing, rendering the metal a normal conductor.

\pagebreak

The lateral oscillation gap $\Delta{y}$ (lateral energy barrier $\Delta{\epsilon}$) is overcome by thermal agitation at elevated temperature, $T  >  T_{c}$.  Phonon assistance with energy $\Delta{\epsilon}$ $>$ $kT_{c}$, enables lateral overswing of the $3s$ electrons and thereby destroys superconductivity.  In this case $both$ the $3s$ and $3p$ electrons constitute the free electron gas of normal aluminum metal.

\emph{\underline {Isotope effect.}} In order to relate energy gap $\Delta{\epsilon}$ and critical temperature  $T_{c}$ quantitatively, we invoke Einstein's model of specific heat which quantizes the atoms' thermal oscillation, $\tilde{\epsilon} = \hbar \omega$,  assumed to have the same (angular) frequency for all atoms, $\omega  = \sqrt {\kappa /M} $.  Here $M$ is is the mass of the oscillating atom and $\kappa$ is the spring constant of the interatomic force.  Equating $\Delta{\epsilon} = \tilde{\epsilon} = \hbar \omega = kT_{c}$ gives a dependence of the critical temperature on the inverse square root of the atomic mass,  $T_{c} \propto 1/\sqrt M $.  This mechanism accounts for the experimentally observed ``isotope effect'' of superconductivity, $T_{c} \propto {M^{ - \alpha }}$, with $\alpha  \approx \frac{1}{2}$ in elemental metals.

%$T_{c} \propto \[{M^{ - \alpha }}$, did the format change here?

\emph{\underline {Charge transport.}} So far we have considered only the \emph{possibility} of superconductivity by Coulomb oscillations.  Now we turn to the more relevant question of charge transport through a superconducting sample between the electrodes of an external power source (battery).  Suppose a $3s$ electron of the $Al$ atom labeled $A$ (at the left of Fig. 12) is attracted and absorbed by the $\oplus$ electrode (anode) of the battery.  In the electric field between the electrodes, the missing  $3s$ electron of atom $A$ will be  be replaced preferentially by a $3s$ electron from a neighbor atom, say $B$.  The cascade of $3s$ electron replacement from neighbor atoms will extend across the sample, all the way to the (-) electrode (cathode) of the battery (highlighted with hatched lines in Fig. 12).  This amounts to an electron flow in the direction to the left.  (An equivalent charge-transport process will occur if, instead of superconducting aluminum, superconducting alkali fullerene was between the electrodes.)  However, a non-classical and rather counter-intuitive constraint holds for superconductors.  It is easiest explained when superconductivity is carried by electrons from a closed (doubly occupied) $s$ shell (such as $3{s^2}$ in $Al$) through axial Coulomb oscillation and concomitant lateral oscillation.  But it holds equally when ${s^1}$ electrons from neighboring atoms (as in alkali fullerenes) form doubly occupied molecular orbitals.  A manifestation of the Pauli exclusion principle, the \emph{availability} of a quantum state is of equal importance as the occupation of a quantum state by an electron.  The above scenario of $3s$ electron-hopping from $Al$ atom to neighbor $Al$ atom is possible only because of the availability of an unoccupied $3s$ quantum state, called a ``hole'', in the neighbor atom.  Thus the flow of replacement electrons to the
\pagebreak

\includegraphics[width=5.6in]{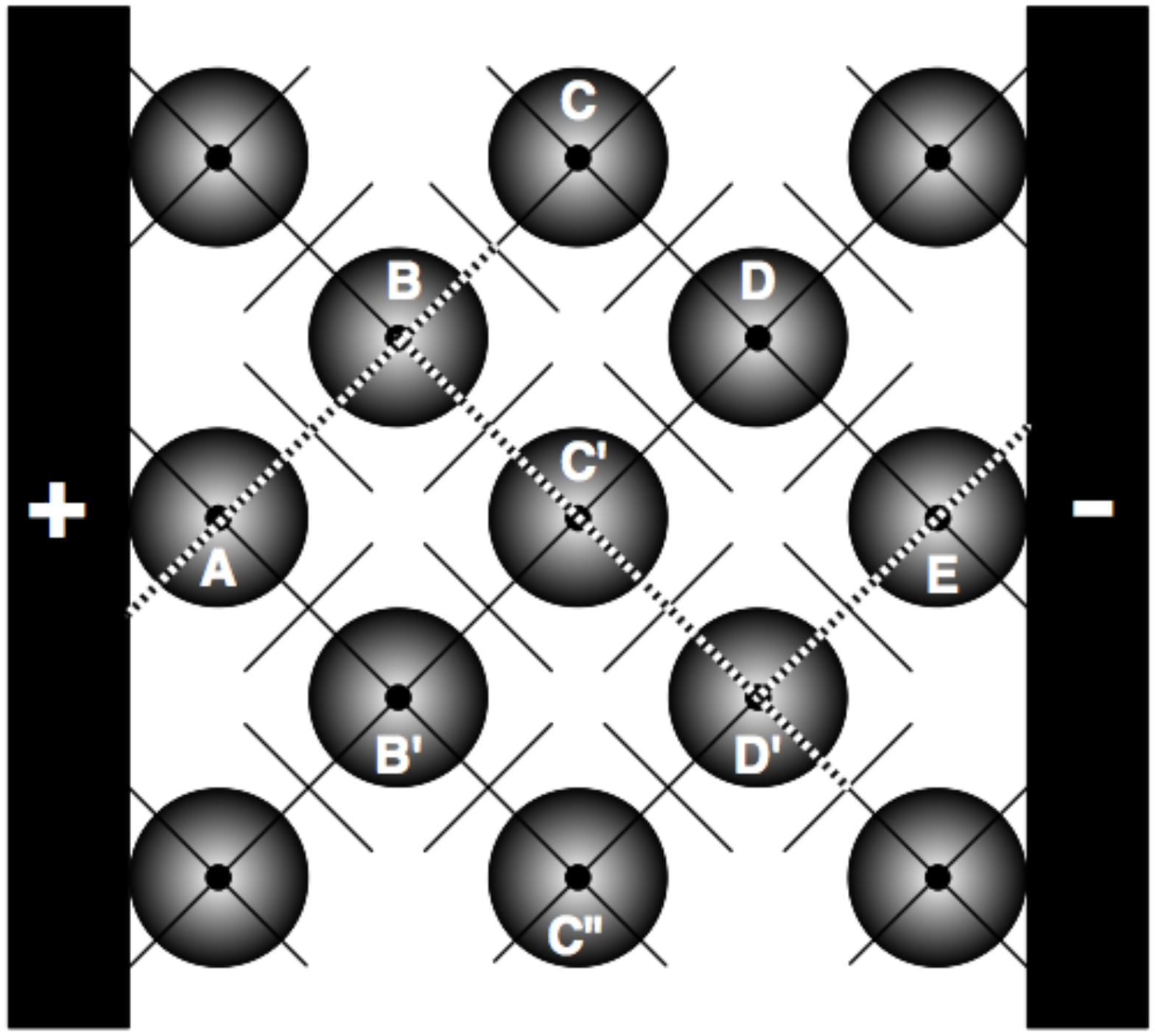}

\noindent FIG. 12. Charge transport in a Coulomb-oscillator superconductor composed of atoms with closed outer $s$ shells, here $3{s^2}$ in metallic $Al$.  Gray spheres indicate the cores of inner electron shells and black dots represent atomic nuclei. Thin lines show orbits of axial Coulomb oscillations and lateral oscillations.  Hatched lines show a possible path of progressive electron replacement between neighbor atoms.  (For clarity, the free-electron gas background from delocalized $3p$ electrons is omitted in the display.)

\pagebreak
\noindent left is accompanied by a flow of holes to the right by the same amount (but of positive charge).  As a consequence of the holes' opposite charge and flow, the electric and magnetic effect of the hole current is exactly the same as that of the electron current.  The result of both flows together is such as if the current was carried by electron \emph{pairs}.  The electron-pair effect of superconductivity shows up in various experiments, most directly in flux quantization by units of ${\Phi _0} = \frac{{hc}}{{2e}}$ (where $c$ is the speed of light and $e$ the elementary electric charge) as well as in supercurrent tunneling (Josephson effect).  Electron pairs (Cooper pairs) are also the backbone of the $BCS$ theory of superconductivity, although with quite a different meaning.\cite{8}

\section{ELEMENTAL TRANSITION METALS }
Of all elemental superconductors, most are transition metals, abbreviated here by $Tr$. Characteristic of their free-atom electron configuration is an open outer $n{d^\delta }$ shell, with principal quantum number $n = 3, 4, 5$, and electron occupancy $\delta  = 1,2, ..., 8.$  (We exclude the elements with completely filled outer $d$ shell---the IB group, $Cu, Ag, Au$, and the IIB group, $Zn, Cd, Hg$---from the discussion of transition metals. Furthermore, $La$ and $Lu$ both have a $6{s^2}5{d^1}$ configuration but with an underlying closed $4f$ shell in the latter atom.  This makes $Lu$ rather than $La$ the natural first member of the third transition series.)  The $n{d^\delta }$ shell of $Tr$ atoms must be regarded in concert with an $s$ shell of lower energy, $(n+1) s^{\sigma}$, closed in the majority of cases ($\sigma = 2$) or half filled otherwise ($\sigma = 1$)---besides $Pd$ with $\sigma = 0$---as noted in Fig. 13.  That figure shows, by block height, a comparison of the critical temperature $T_{c}$.  At first view the distribution of the metals' $T_{c}$ seems complicated.  But soon two features emerge:  (1) No superconductivity is found for the elements in the triangle cornered by $Cr$, $Ni$, and $Pt$ (apart from the insignificantly low superconductivity of $Mn$, $T_{c}=0.04$ K).  (2) Several drastic changes of $T_{c}$ occur from neighbor to neighbor element in the same row.  Such abrupt neighbor variation is in stark contrast to the gradual change of many other properties of $Tr$'s, such as the nearest-neighbor distance $R_{nn}$ (see Fig. 14) or the ionization energy of their free atoms.

The goal of this discussion is to provide a qualitative understanding of the pattern of the critical temperature $T_{c}$  of elemental $Tr$'s in the framework of the Coulomb-oscillator model, specifically the lateral oscillation gap $\Delta{y}$.  This quantity is the difference of lateral oscillation width, now denoted as $2Y$ (instead of previously B0B), of the corresponding $s$
\pagebreak

\includegraphics[width=7.5in]{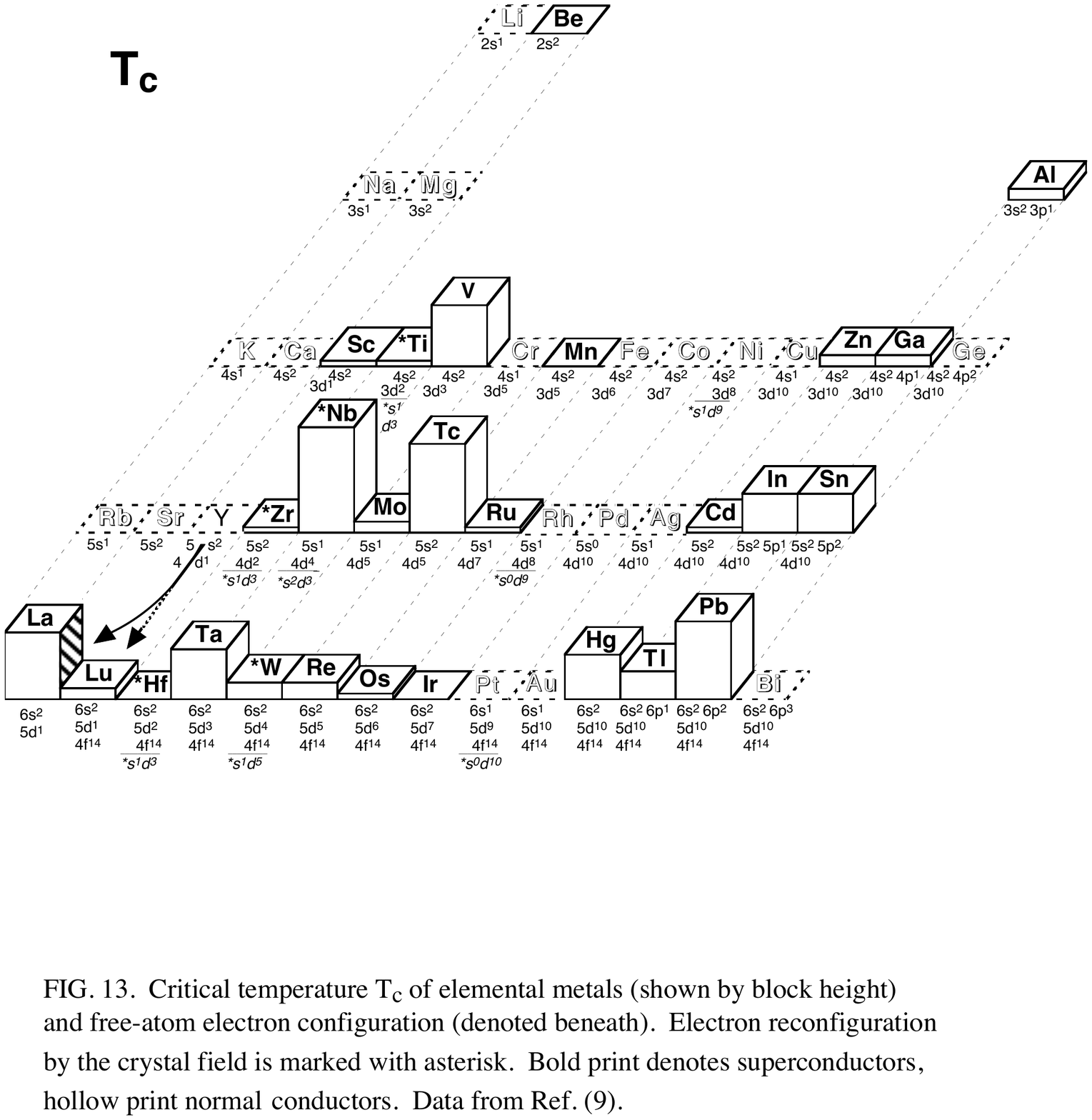}

\pagebreak
\noindent electron(s) and the nearest-neighbor distance of crystal atoms, $\Delta{y} \equiv R_{nn} - 2Y$.  The nearest-neighbor distances $R_{nn}$ are well known from experiment.  In contrast, the lateral oscillation widths $2Y$ are guessed in their qualitative trend, based on properties of a Coulomb oscillator under systematic changes of atomic number $Z$ and screening effects by $d$ electrons. Such screening necessitates a consideration of the $d$ electrons' actual electron charge density distribution, $\rho(\textbf{r})  = e \psi(\textbf{r})\psi^{*}(\textbf{r})$, based on the corresponding wave functions $\psi$.  Thus, in a hybrid fashion, quantum mechanical information becomes necessary in the semiclassical treatment of \emph{screened} Coulomb oscillators.  The predicted qualitative trends need to be confirmed quantitatively by first-principles calculations.

In an elemental cystal the nearest-neighbor distance $R_{nn}$ is closely related to the ``size'' of a crystal atom (identical under the assumption of hard-sphere atoms).  A comparison of the observed $R_{nn}$'s in the various lattice types of the $Tr$'s is provided by the histogram of Fig 14.  A common trend in all three transition series, the dependence of $R_{nn}$ on $d$-shell occupancy $\delta = 1, ..., 10$, is seen in concave curves, with a minimum near half $d$-shell filling, $\delta = 5$.  This trend is contrary to a steady shrinking of atomic size in the periods of the elements without $d$ electrons---from $Li$ to $Ne$ and from $Na$ to $Ar$ (not shown in Fig. 14).
 
The concave trend is easiest explained for the third $Tr$ series which comprises elements with successively filling $5d$ shell and underlying closed $6s^{2}$ and $4f^{14}$ shells, that is, the transition metals $Lu-Ir$, shown in Fig. 15.  The solid curve, representing the nearest-neighbor distance $R_{nn}$, looks like a wide, flat bowl, already familiar from Fig. 14.  Also bowl-shaped but more pronounced---steeper walls, more depth---is the dashed curve for the lateral oscillation width $2Y$ accompanying the outer $s$ electrons' Coulomb oscillation.  It is the \emph{difference} between these different bowl-shaped curves that gives rise to the pattern of lateral oscillation gaps, $\Delta{y} = R_{nn} - 2Y$, and thereby to the pattern of critical temperatures, $T_{c} \propto \Delta{y}$.

The key to an understanding of the concave curves lies with the Coulomb oscillations of the outer $s$ electrons of \emph{free} $Tr$ atoms.  This affects directly the lateral oscillation width $2Y$ of those $s$ electrons when such atoms reside in a crystal.  It also affects, indirectly and more moderately, the neighbor distance $R_{nn}$ through the contribution of those $s$ electrons to the atomic size.  The influence of atomic number $Z$ and $d$-shell occupancy $\delta $ on outer $s$ electrons in free $Tr$ atoms gives rise to two opposing effects:  One is a contraction of their Coulomb-oscillator amplitude with increasing atomic number $Z$ (within a given period of 
\pagebreak

\includegraphics[width=7in]{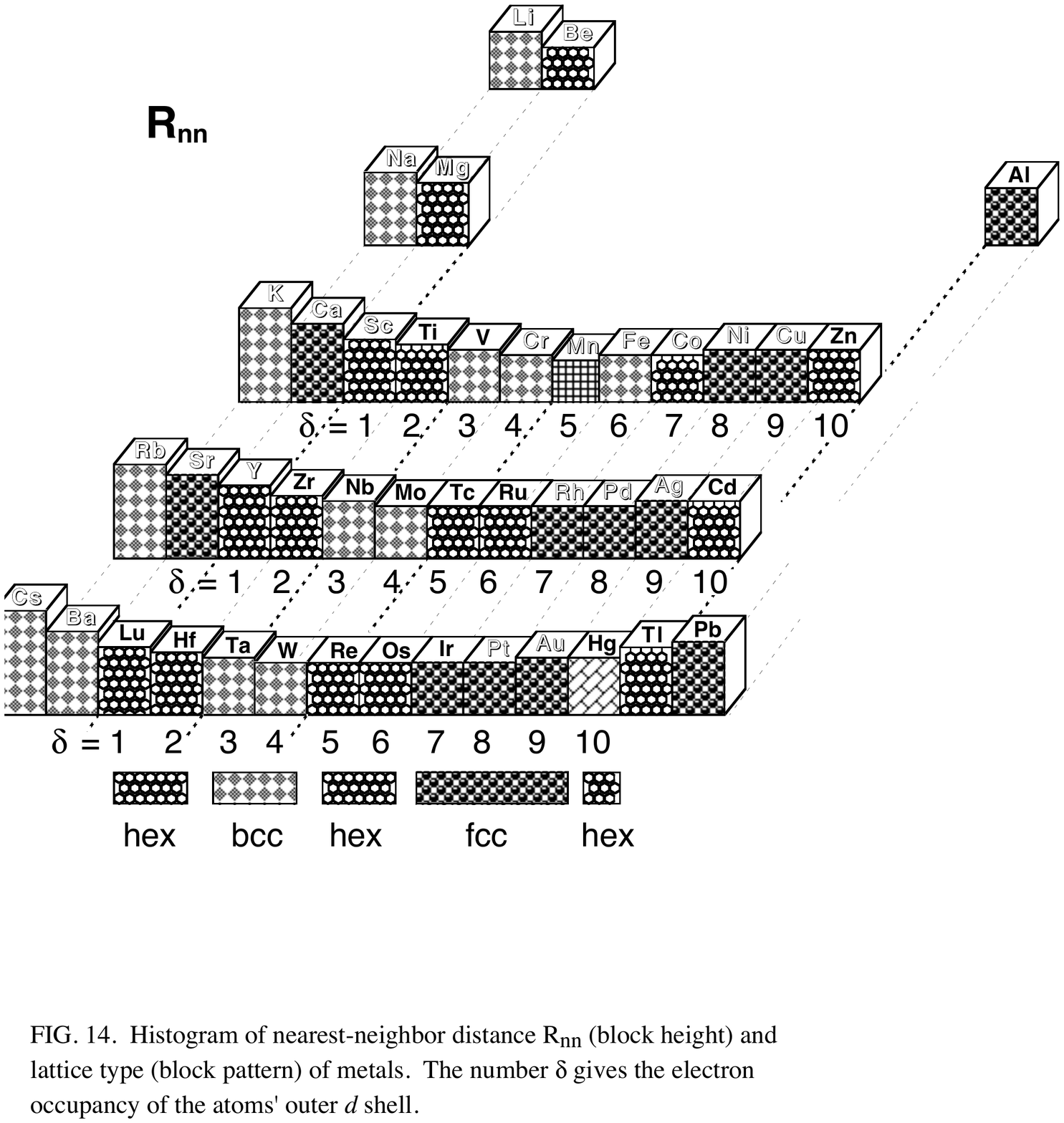}

\includegraphics[width=6in]{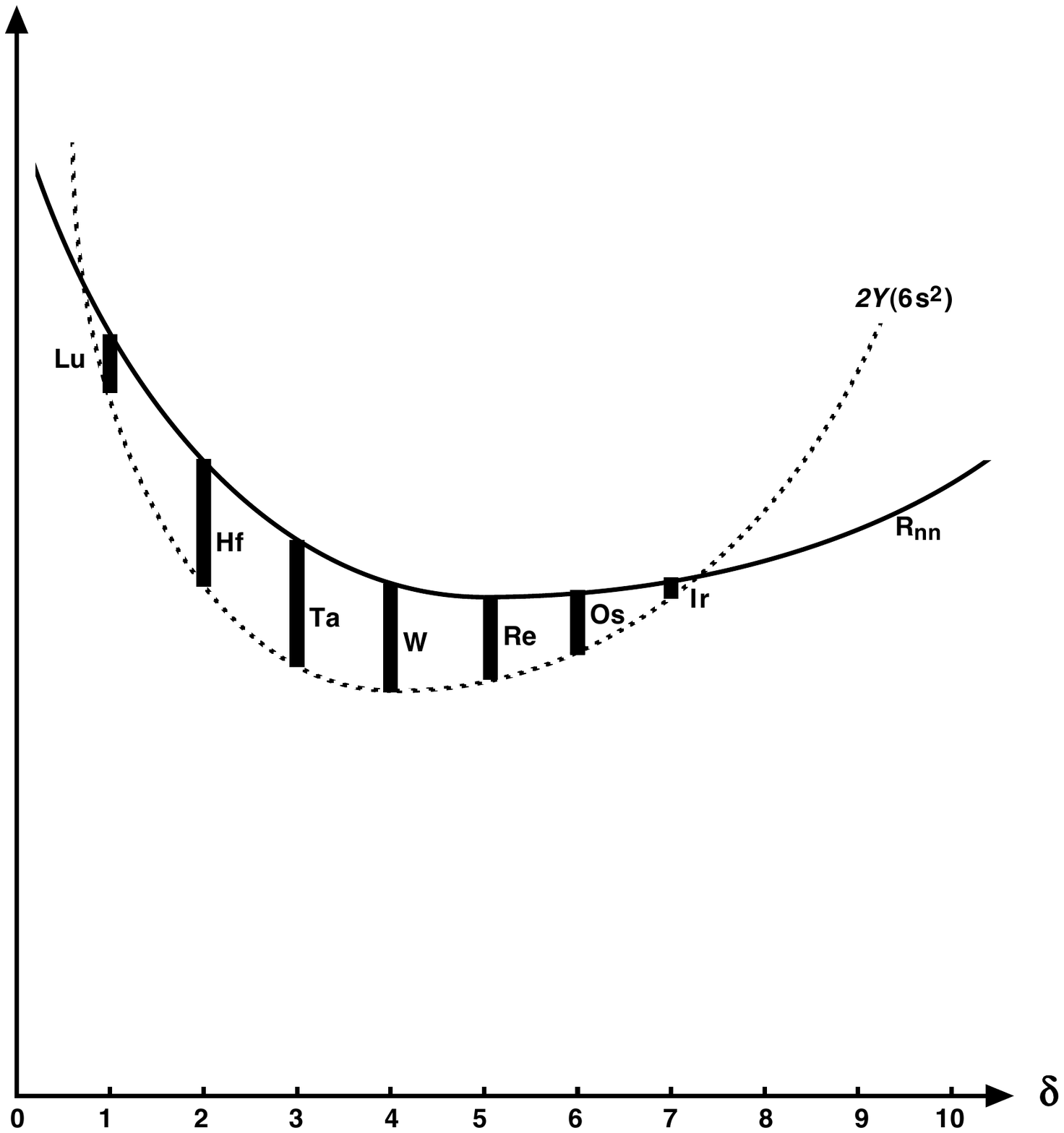}

\noindent FIG. 15.  Lattice spacing and Coulomb oscillations in the $third$ transition-metal series $vs$. $d$-shell occupation $\delta $.  Curve $R_{nn}$ shows the observed nearest-neighbor distance. Curve $2Y(6s^{2})$ schematically shows the lateral oscillation width for atom species with closed $s$ shell. Black bars represent the hypothetical lateral oscillation gap, $\Delta{y} \propto T_{c}$.

\pagebreak
\noindent elements) due to incomplete screening by inner electrons.  Such contraction is most distinct for $s$ electrons as they experience the full (unscreened) nuclear charge $Ze$ when very close to the nucleus in their nucleus-traversing swing.  Other electrons ($p, d, ...$) don't get that close to the nucleus and are therefore always partially screened by inner electrons.  The other effect is an expansion of the Coulomb oscillations of the outer $s$ electron(s) with increasing $d$-shell occupancy due to increased screening of the nuclear charge $Ze$ by the $d$ shell.  Note that the two effects act on different parts of a Coulomb oscillator:  The nuclear-based contraction acts inside the $d$ shell whereas the $d$ shell-based expansion occurs outside the shielding $d$ shell.  

How do the contracting and expanding effects in free $Tr$ atoms depend on atomic number $Z$ and $d$-shell occupancy $\delta $?  The  nuclear-based contraction increases steadily with increasing $Z$ but  the expansion increases more drastically with $d$-shell filling $\delta $.  Therefore contraction is dominant when $d$-shell filling is low and expansion is dominant when $d$-shell filling is high (see slopes in the left and right regions of Fig. 15, respectively).  Both effects cancel each other near the middle of a $Tr$ series where $d$ shells are half filled, $\delta = 5$.  

Large or small Coulomb-oscillator amplitudes in free atoms give rise to correspondingly large or small amplitudes of both axial Coulomb oscillations and accompanying lateral oscillations of $s$ electrons in a \emph{crystal}, indicated by the deeper-bowl curve of $2Y(6s^{2})$ in Fig. 15.  On the other hand, since the nuclear-based contraction is less pronounced with other electron orbitals ($p, d, ...$), the contribution of those orbitals to the atomic size gives rise to a more moderate dependence on $d$-shell occupancy and consequently to the wider-bowl curve of the $R_{nn}$ curve in Fig. 15.

The size of the lateral oscillation gap, $\Delta{y} = R_{nn} - 2Y$, is represented by the vertical bars in Fig. 15.  A comparison with critical temperatures $T_{c}$ in Fig. 13 shows, except for $Hf$ and $W$, a rough proportionality, $T_{c} \propto \Delta{y}$, with a maximum at $Ta$.  The stark discrepancy for $Hf$ and $W$ would be removed, however, if the atoms' electron configuration, $6s^{2}5d^{2}$ and $6s^{2}5d^{4}$, were to change to $6s^{1}5d^{3}$ and $6s^{1}5d^{5}$, respectively, when free $Hf$ and $W$ atoms become crystal atoms (see Fig. 18 below). A justification for such an electron reconfiguration upon crystallization will follow.

Generally, the electron configuration $(n+1)s^{\sigma } nd^{\delta }$ of $Tr$ atoms with outer $d$ shell of principal quantum number $n = 3, 4, 5$, and $d$-shell occupancy $\delta = 1, ..., 8$, results from a fine balance of counteracting quantum-mechanical effects.  One can see in Fig. 13 that the $s$ shell is fully occupied (closed), $\sigma = 2$, in about two thirds of the free $Tr$ atoms under consideration and half full otherwise, $\sigma = 1$---mostly in the second transition series, except $\sigma = 0$ for $Pd$. When free atoms become crystal atoms, the $s/d$-shell occupancy may change due to crystal-field effects.  In this study the free-atom electron configuration, noted in Fig. 13, was adopted for an assessment of lateral oscillation width $2Y$ of the $(n+1)s$ electrons.  In cases where this leads to a gross discrepancy in the proportionality of $2Y$ with the critical temperature, $\Delta{y} \propto T_{c}$, as for $Hf$ and $W$ above, then the alternate possibility of $s$-shell occupancy $\sigma$ is used (denoted by an asterisk).

We now turn to Fig. 16 for the \emph{first} transition series. Here the free atoms have a closed $4s$ shell except for $Cr$, $\sigma = 1$, caused by preemptive filling of the $d$ half-shell.  The asterisk at $Ti^{*}$ denotes that  an alteration of electron configuration from free to crystal atom was assumed, $4s^{2}3d^{2} \to 4s^{1}3d^{3}$, for the reason stated above.  The dotted curve $2Y(4s^{1})$  indicates the lateral oscillation width of electrons from a half-filled $4s$ shell, which applies to $Ti^{*}$ and $Cr$.  The size of the corresponding lateral oscillation gap, $\Delta{y} = R_{nn} - 2Y$ (black bar), is in reasonable proportionality with $T_{c}$ of $Ti^{*}$ (see Fig. 13).  Also, the negative gap (hollow bar), indicating lateral overswing, agrees with the lack of superconductivity in $Cr$.  The dashed curve  $2Y(4s^{2}$) in Fig 16 indicates the lateral oscillation width of closed 4$s$-shell crystal electrons.  The size of the black bars, representing lateral oscillation gaps $\Delta{y}$, is in reasonable agreement with very low $T_{c}$ of $Sc$ and $Mn$ as well as with the considerable $T_{c}$ of $V$ (see Fig. 13).  Negative oscillation gaps (hollow bars) in Fig. 16 for $Fe$, $Co$ and (off scale) $Ni$, are in agreement with the lack of superconductivity in those metals.

Notice that the lateral-oscillation curve for closed $s$ shells,  $2Y(s^{2})$, is more pronounced---steeper and deeper---than for half-filled $s$ shells, $2Y(s^{1})$.  The cause originates with Coulomb oscillators in \emph{free} $Tr$ atoms.  The counteracting effects that act on a single Coulomb oscillator ($\sigma = 1$)---nuclear-based contraction and $d$ shell-based screening---are enhanced when the two electrons of a closed $s$ shell ($\sigma = 2$) are in simultaneous Coulomb oscillations.  Considering their actual charge-density distribution, one can think that one of the two $s$ electrons incompletely screens the nuclear charge for the other $s$ electron, causing additional contraction.  However, outside the outer $d$ shell, the repulsion between those $s$ electrons causes additional expansion.

Figure 17 concerns the \emph{second} transition series.  Most free atoms of those elements have 
a singly occupied $s$ shell, $5s^{1}4d^{\delta}$, as mentioned.  The corresponding lateral oscillation width 
\pagebreak

\includegraphics[width=5.7in]{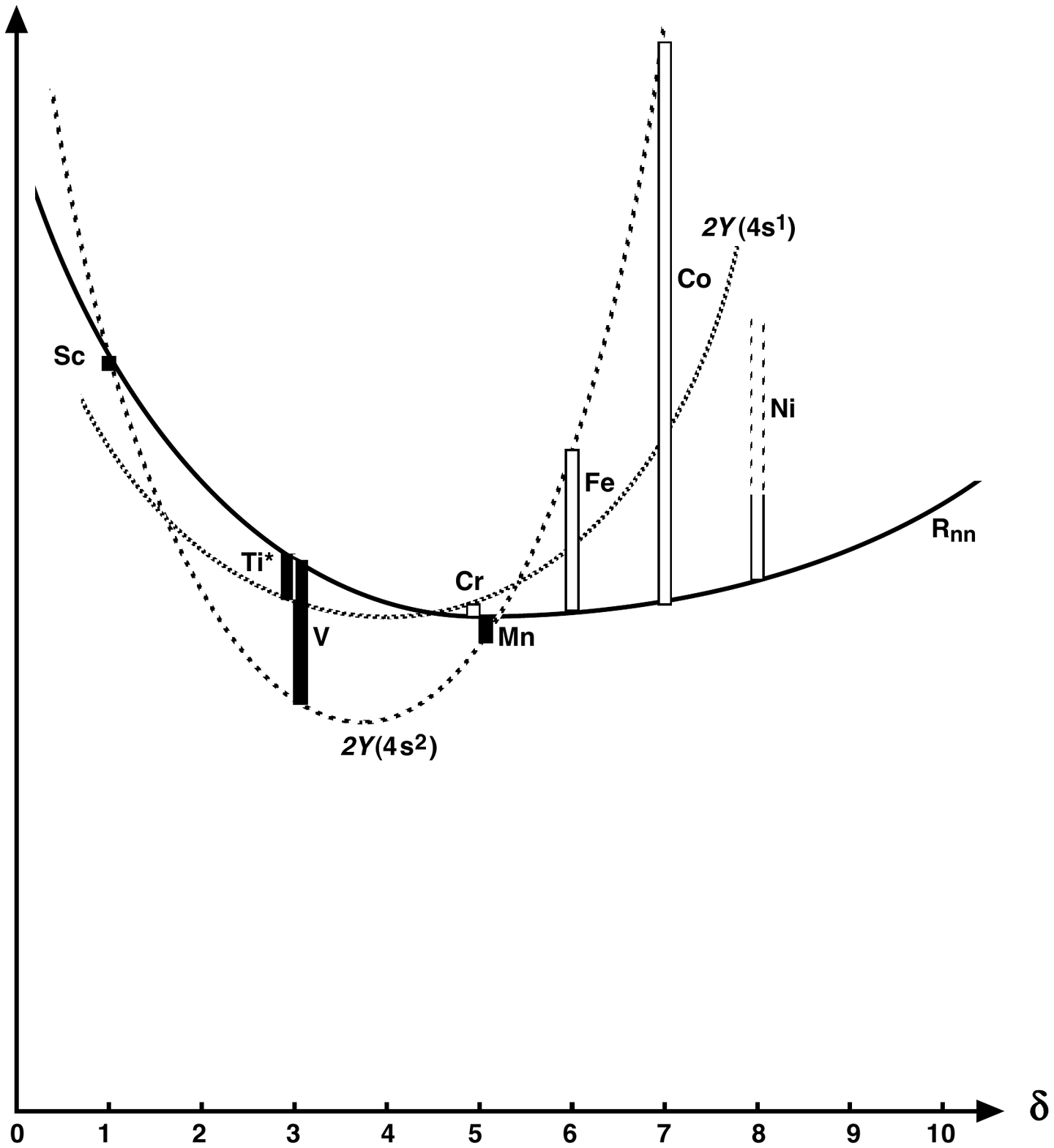}

\noindent FIG. 16.  Lattice spacing and Coulomb oscillations in the $first$ transition-metal series $vs$. $d$-shell occupation $\delta $.  Curve $R_{nn}$ shows the observed nearest-neighbor distance.  Curves $2Y$ schematically show the lateral oscillation width (dotted for half-filled $s$ shell, dashed for closed $s$ shell).  Black bars represent the hypothetical lateral oscillation gap, $\Delta{y} \propto T_{c}$.  Hollow bars signify lateral oscillation overswing.  The asterisk denotes that the electron configuration of free $Ti$ atoms has been altered by the crystal field.
\pagebreak

\includegraphics[width=5.6in]{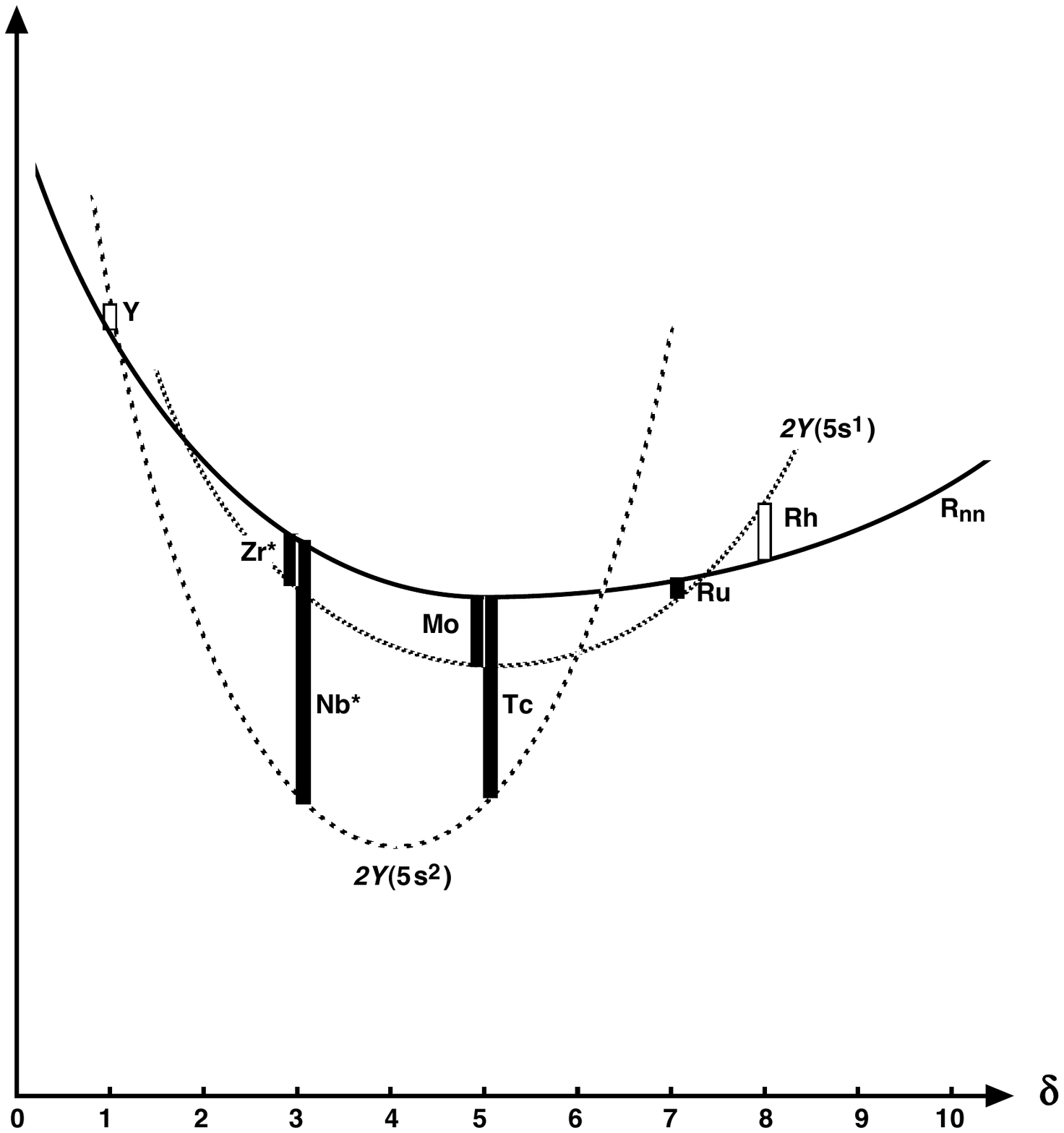}

\noindent FIG. 17.  Lattice spacing and Coulomb oscillations in the $second$ transition-metal series $vs$. $d$-shell occupation $\delta $.  Curve $R_{nn}$ shows the observed nearest-neighbor distance.  Curves $2Y$ schematically show the lateral oscillation width (dotted for half-filled $s$ shell, dashed for closed $s$ shell).  Black bars represent the hypothetical lateral oscillation gap, $\Delta{y} \propto T_{c}$.  Hollow bars signify lateral oscillation overswing.  The asterisk denotes that the electron configuration of free $Zr$ and $Nb$ atoms has been altered by the crystal field.

\pagebreak
\noindent is indicated by the dotted curve $2Y(5s^{1})$.  It gives a small lateral oscillation gap for $Mo$ and $Zr^{*}$ (black bars) and lateral overswing for $Rh$ (hollow bar) in reasonable qualitative agreement with the $T_{c}$ display in Fig. 13.  The free-atom electron configuration of $Nb$,  $5s^{1}4d^{4}$, gives an oscillation gap much too small.  Therefore we assume an alteration to $5s^{2}4d^{3}$ for crystalline niobium, denoted as $Nb^{*}$.  Use of the $2Y(5s^{2})$ curve for atoms with closed $s$ shell gives a very large oscillation gap $\Delta{y}$ for $Nb^{*}$ but lateral overswing (hollow bar) for the element $Y$,  in qualitative agreement with the critical temperatures $T_{c}$ in Fig. 13. The converse situation is encountered with $Zr$, where the free-atom electron configuration, $5s^{2}4d^{2}$, would give a lateral oscillation gap $\Delta{y}$ much too large. However, an alteration to $5s^{1}4d^{3}$ for crystalline $Zr^{*}$ yields a small oscillation gap, in qualitative agreement with a low $T_{c}$  (compare Figs. 17 and 13).  

Likewise, the previous large oscillation gaps in Fig. 15, resulting from $Hf$ and $W$ with free-atom electron configuration, diminish drastically when the altered configuration is used for crystalline $Hf^{*}$ and $W^{*}$, shown corrected in Fig. 18.  In the latter case the half-filled $d$ shell, $\delta = 5$, although insensitive to crystal fields, is still better suited than the free-atom configuration $s^{2}d^{4}$ whose four-fold $d$-shell occupancy jars with hexagonal or cubic symmetry of the crystal field.

In hindsight, of the five cases where we altered the electron configuration in order to obtain agreeable oscillations gaps, four of them share a common feature, namely a redistribution to a \emph{triply} occupied $d$ shell---from the free-atom electron configuration $s^{2}d^{2}$ of $Ti$, $Zr$ and $Hf$ to $s^{1}d^{3}$, and from $s^{1}d^{4}$ of free $Nb$ to $s^{2}d^{3}$.  The crystal field in the hexagonal or, respectively, cubic lattice (see Fig. 14) favors an orientation of $d$ orbitals in the three spatial directions, forcing the accompanying redistribution of outer $s$-shell electrons.\cite{10}

It is the principle of \emph{minimal energy} that causes such electron reconfigurations.  The ensuing low $T_{c}$ of elemental $Ti$, $Zr$ and $Hf$ metals is a consequence, via $\Delta{y}$, of the characteristically large lateral oscillation width $2Y(s^{1})$.  However, the possibility remains that the crystal field in the less symmetric crystalline environment of an alloy may fail to force such electron reconfiguration.  In this case the free-atom configuration of $Ti$, $Zr$ and $Hf$ would invoke the alternate curve $2Y(s^{2})$ and result, via ‚àÜ$\Delta{y}$, in much higher critical temperatures $T_{c}$.  This is indeed the case for certain alloys of $Zr$ and $Hf$ with other transition metals as will be seen below.

\noindent.

\pagebreak

\includegraphics[width=5.6in]{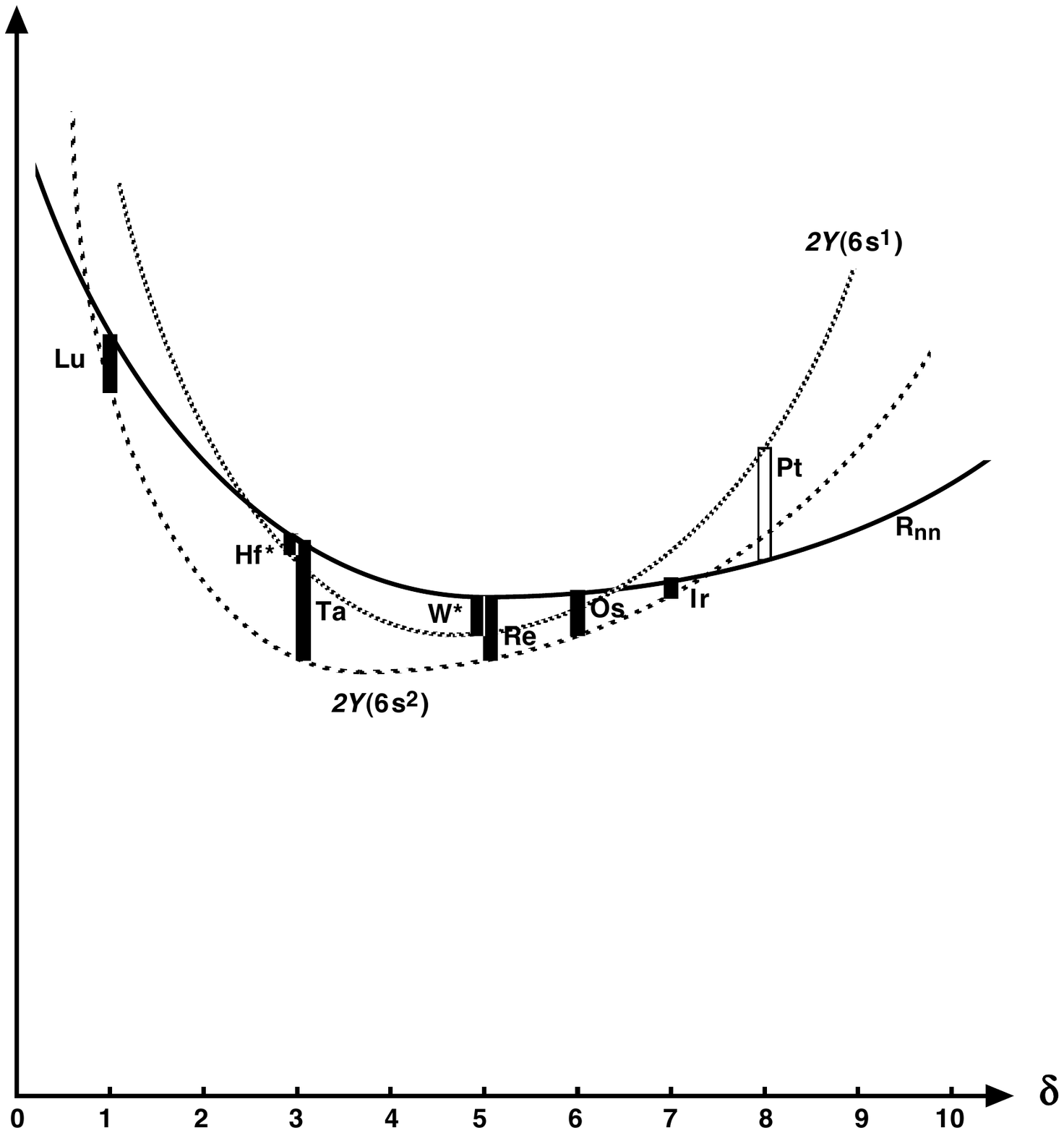}

\noindent FIG. 18.  Lattice spacing and Coulomb oscillations in the $third$ transition-metal series $vs$. $d$-shell occupation $\delta $.  Curve $R_{nn}$ shows the observed nearest-neighbor distance.  Curves $2Y$ schematically show the lateral oscillation width (dotted for half-filled $s$ shell, dashed for closed $s$ shell).  Black bars represent the hypothetical lateral oscillation gap, $\Delta{y} \propto T_{c}$.  Hollow bars signify lateral oscillation overswing.  The asterisk denotes that the electron configuration of free $Hf$ and $W$ atoms has been altered by the crystal field.

\pagebreak
In summary, the main findings in the application of the Coulomb-oscillator model of superconductivity to transition metals are:  (1) A competition of nuclear-based contraction and $d$ shell-based expansion of Coulomb oscillations in free $Tr$ atoms.  (2) Consequent concave curves for the lateral oscillation width, $2Y$, of outer $s$ electrons in $Tr$ crystals---enhanced for closed $s$ shells---in their dependence on $d$-shell occupancy $\delta$.  (3) A moderation of the concave shape for the nearest-neighbor distance, $R_{nn}$.  (4)  Variegated sizes of oscillation gaps resulting from the difference of those curves, $\Delta{y} = R_{nn} - 2Y$, in qualitative agreement with critical temperature, $T_{c} \propto \Delta{y}$.  (5) For $Ti$, $Zr$, $Hf$, $Nb$ and $W$ reasonable agreement with the trend of observed $T_{c}$ is achieved only if the electron configuration is altered from the free-atom situation.  In all cases but one ($W$), the altered configuration involves a triply occupied $d$ shell, $d^{3}$, favored by crystal-lattice symmetry.

Of the four transition metals with the highest critical temperature, $T_{c}(Nb^{*}) > T_{c}(Tc) > T_{c}(V) > T_{c}(Ta)$---landmarks in Fig. 13---three aspects are noteworthy:  (1)  All four of them have an outer \emph{closed} $s$ shell, $s^{2}$.  (2)  The occupancy of their outer $d$ shell straddles the value $\delta = 4$, with half-filled $d$ shell, $d^{5}$, for $Tc$ and triply filled $d$ shell otherwise, $d^{3}$.  (3) Niobium has the highest critical temperature of all elemental metals, $T_{c}(Nb^{*}) = 9.25$ K, which serves as a standard in any comparison of superconducting materials.

Finally, one wonders why there are no transition metals with still higher $T_{c}$.  This could be expected for $Tr$'s with $d$-shell occupancy $\delta = 4$, where the widest gap exists between the curves $R_{nn}$ and $2Y(s^{2})$ (see Figs. 16 - 18).  However, an inspection of all $Tr$'s reveals that no \emph{crystal} atoms exist with four-fold occupied $d$ shell, $d^{4}$, which jars with high crystal symmetry.

\section{INTER-TRANSITION METAL ALLOYS}

The alloys formed by \emph{two} transition metals, denoted as $Tr_{1-x}Tr'_{x}$, raise new aspects, complications and surprises.  Among the surprises are compounds whose critical temperature $T_{c}$ is considerably higher than those of the parent elements $Tr$ and $Tr'$.\cite{11}  Conversely, an alloy formed by two parent superconductors may be a normal conductor.  As a new aspect, neighbor atoms in the alloy lattice can be of different species.  Their difference in atomic number $Z$, and possibly in principal quantum number $n$ of the outer $s$ electrons (which form the Coulomb oscillators), gives rise to an asymmetry of the lateral oscillation as well as to two additional counteracting trends. 

If the alloy partners are from the same $Tr$ series (common $n$ of outer $s$ electrons), then the atom with the larger atomic number $Z$ exerts the larger attraction in the Coulomb oscillation.  As a result, the intersection of axial and lateral oscillation---the previous ``midpoint''---is attracted closer to the larger-$Z$ atom.  The situation becomes more complicated when the alloy partners are from different $Tr$ series (different $n$ of outer $s$ electrons).  However, one special case stands out, namely  when the electron configuration $s^{\sigma}d^{\delta}$ of the partner atoms agrees, $\sigma = \sigma'$, $\delta = \delta'$, while $n \ne  n'$.  An example may illustrate this:  The outer $s$ electrons of a $V$ atom (first $Tr$ series) have $n = 4$ and those of an $Nb^{*}$  atom (second $Tr$ series) have $n = 5$.  If the $(n = 4)$-electron from $V$ would swing across the neighboring $Nb^{*}$ atom, it would encounter an inner $(n = 4)$-electron there.  However, that inner electron is too firmly bound to participate in Coulomb oscillations.  By Pauli's exclusion principle,  the $s$ electron from the $V$ atom, intruding into the $Nb^{*}$ core, must rearrange its wave function.  The result is, generally, a repulsion of the lower-$n$ $s$ electron from an atom with higher-$n$ outer $s$ electrons, possibly to the extent of lateral overswing.  This scenario holds for all $TrTr'$ alloys where the parent atoms have the same $d$-shell occupancy $\delta$ but different principal quantum numbers of the outer $s$ shell, $n \ne n'$---same column but different row of transition metals in the Periodic Table---as can be seen by the corresponding lack of superconductivity in Figs. 19 - 24.  

Otherwise, the balance of the counteracting nuclear-based attraction and the non-orthogonality-based $n$-$n'$ repulsion introduces an asymmetry of the lateral oscillation with respect to the nuclear axis.  The asymmetric path can lead to a higher lateral energy barrier $\Delta{\epsilon}$---effectively a larger lateral oscillation gap $\Delta{y}$---which increases the critical temperature $T_{c} \propto \Delta{\epsilon} \propto \Delta{y}$.  However, if the $n$-$n'$ repulsion is too strong, then the reconstructed wave function extends the lower-$n$ $s$ electron to lateral overswing, which destroys superconductivity.

The crystal structure of a $TrTr'$ alloy results from minimization of binding energy.  While the assortment of lattice types among the \emph{elemental} $Tr's$ is astounding (see Fig. 14), the variety of lattices of sheer endless alloy combinations is truly mindboggling.  It is the subtle balance of various counteracting effects that prevents mono-causal explanations.  In essence a prediction of alloy structure is still beyond the accuracy of present first-principles calculations, due to uncertainties from the approximations involved.  Particularly vexing is the delicate balance of outer $s$ and $d$ shell occupancy, as already witnessed in the variation 
\pagebreak

\includegraphics[width=8.0in]{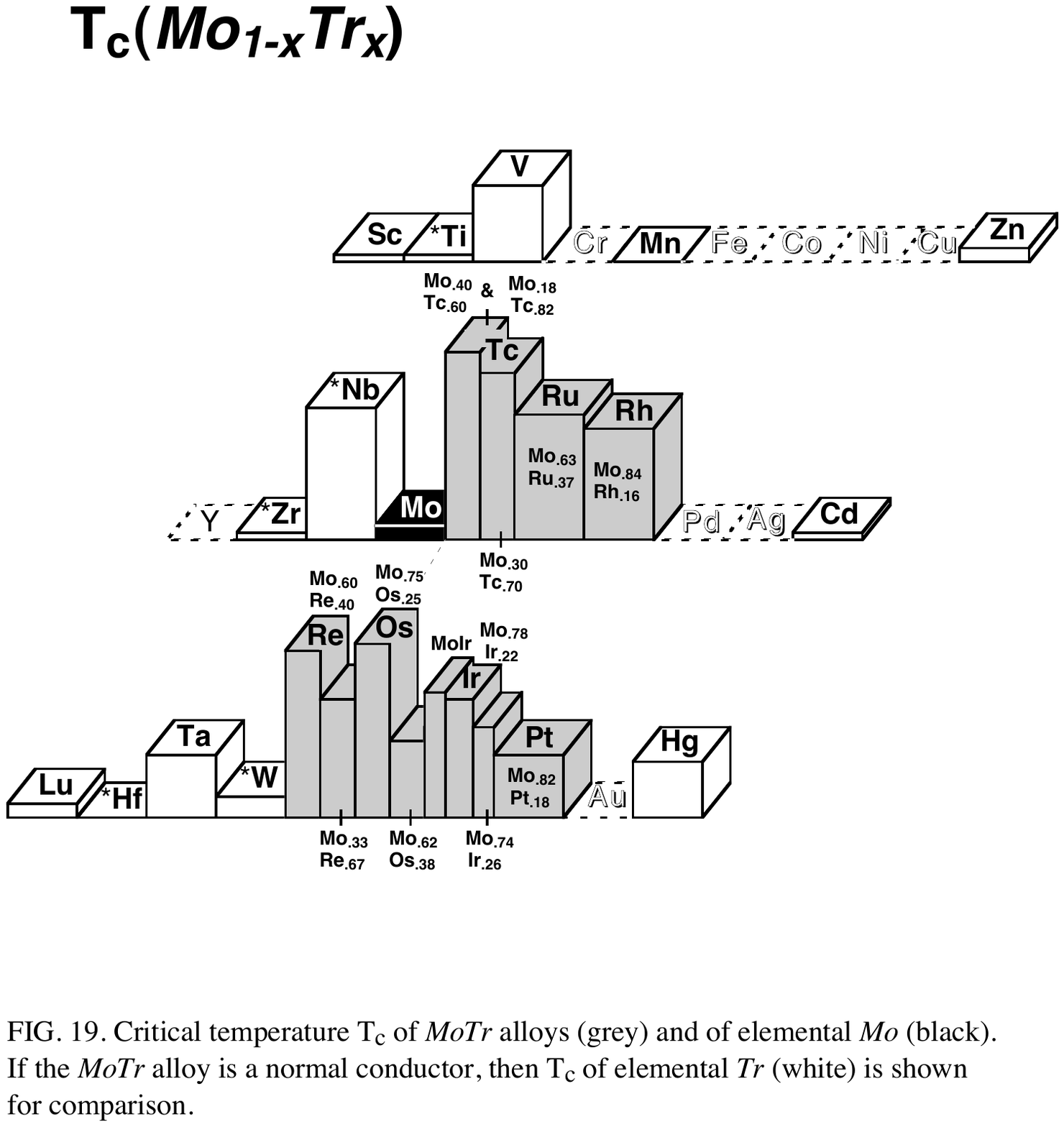}

\pagebreak
\noindent of free-atom electron configuration (see Fig. 13) and the reconfiguration by crystal field, encountered in the elemental $Tr's$, but now subject to even stronger influences in the alloy lattices. 

Despite such difficulties, let's explore some trends in the critical temperature $T_{c}$ of $TrTr'$ alloys.  Since the outer $s$ shell occupancy has a large influence on the width of axial Coulomb oscillation (and proportionally, on \emph{lateral} oscillation)---extended for lone electrons, $s^{1}$, and contracted for closed-shell electrons, $s^{2}$---we start with cases where the electron configuration of the common alloy parent is well established.  Such is the case for $Mo$ whose outer electron configuration $s^{1}d^{5}$ is stabilized by the half-filled $d$ shell.  As Fig. 19 shows, there are only superconducting $MoTr$ alloys to the right (and lower right) of $Mo$ when displayed in the fashion of the Periodic Table.  Generally, when Coulomb oscillators are formed by both lone electrons $s^{1}$ from one partner atom and closed-shell electrons $s^{2}$ from the other, then the extended $s^{1}$ oscillator is determinant with respect to lateral overswing.  Notice in Fig. 19 that almost all the superconducting $MoTr$ alloys are rich in $Mo$ concentration, which typically adds weight to alloy properties by the majority atoms. The dominant influence on the high $T_{c}$ is the nuclear-based contraction of the (extended) $s^{1}$ oscillations with increasing atomic number $Z$ of the alloy partners to the right.  The same rationale explains the lack of superconductivity in $MoTr$ alloys to the left, when the $s^{1}$ electron is subject to nuclear-based extension (with diminishing $Z$ to the left).  The lack of superconductivity (lateral overswing) in $MoTr$ with alloy partners from the first $Tr$ series could be a consequence of the smaller size of those partner atoms (shorter neighbor distance $R_{nn}$).

Essentially the same explanations hold for the $WTr$ alloys displayed in Fig. 20, although to a lesser degree.  As a result of larger atomic number, $Z(W) > Z(Mo)$, and related inner shell screening, the contraction of the $s^{1}$ oscillation is less than in the previous $MoTr$ alloys, leading to correspondingly lower $T_{c}$ and a lack of superconductivity in the $WRh$ and $WPt$ alloys.

Another case where the electron configuration of the common alloy partner is incontrovertible is the element $Tc$ with $s^{2}d^{5}$.  Here we find only superconducting alloys with $Tr$ partners to the left (and lower left) of $Tc$ (as well as $TcSc$ in the top row) as displayed  in Fig. 21. Notice that all superconducting $TcTr$ alloys are rich in $Tc$ concentration.  We assume that all partner atoms have a closed $s$ shell, including the free-atom state of $Zr$ and $Hf$ and the reconfigured state of $Nb^{*}$.  The  Coulomb oscillators formed by such $s^{2}$ electrons 

\pagebreak
\includegraphics[width=8.0in]{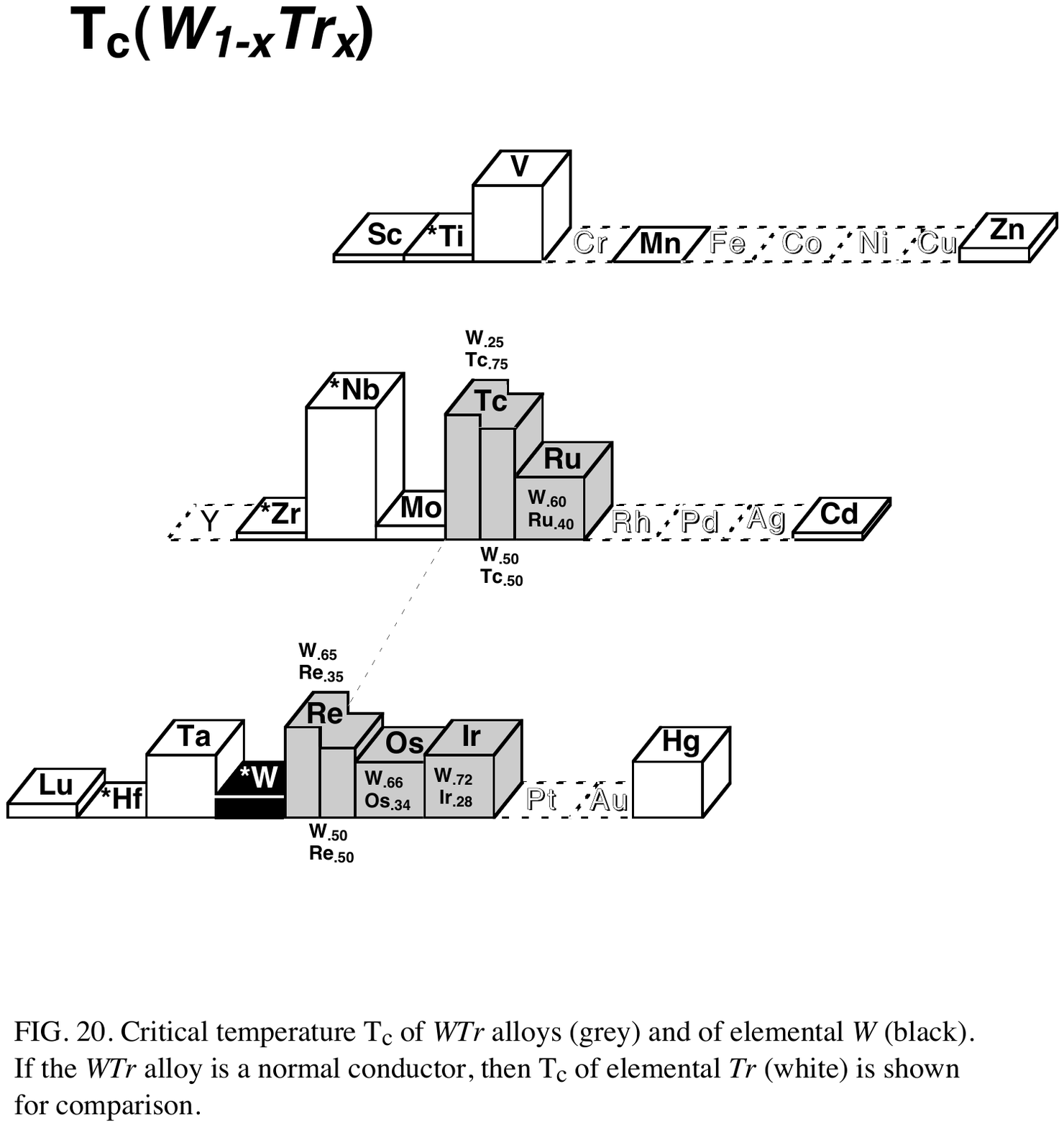}

\includegraphics[width=8.0in]{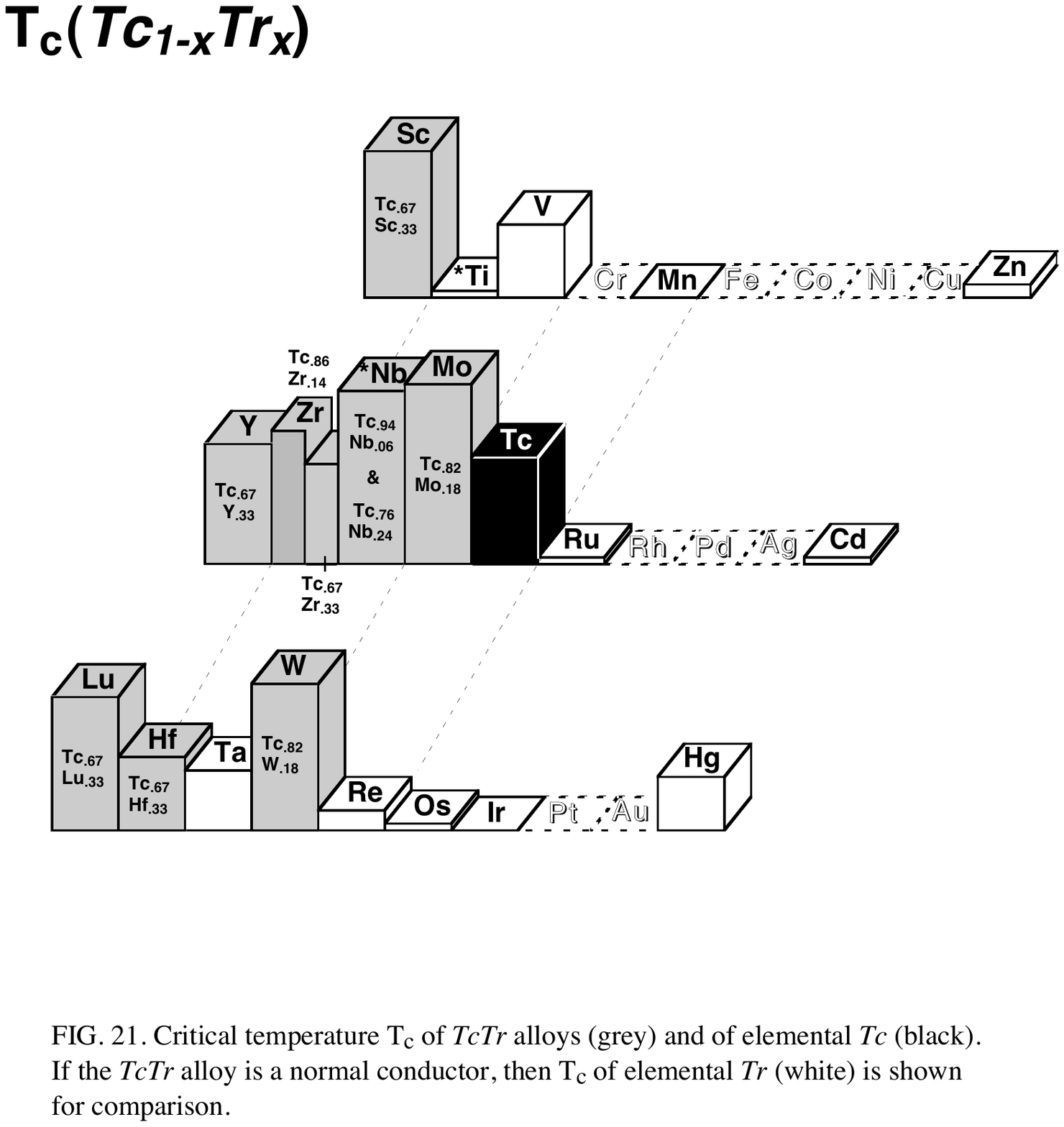}

\includegraphics[width=8.0in]{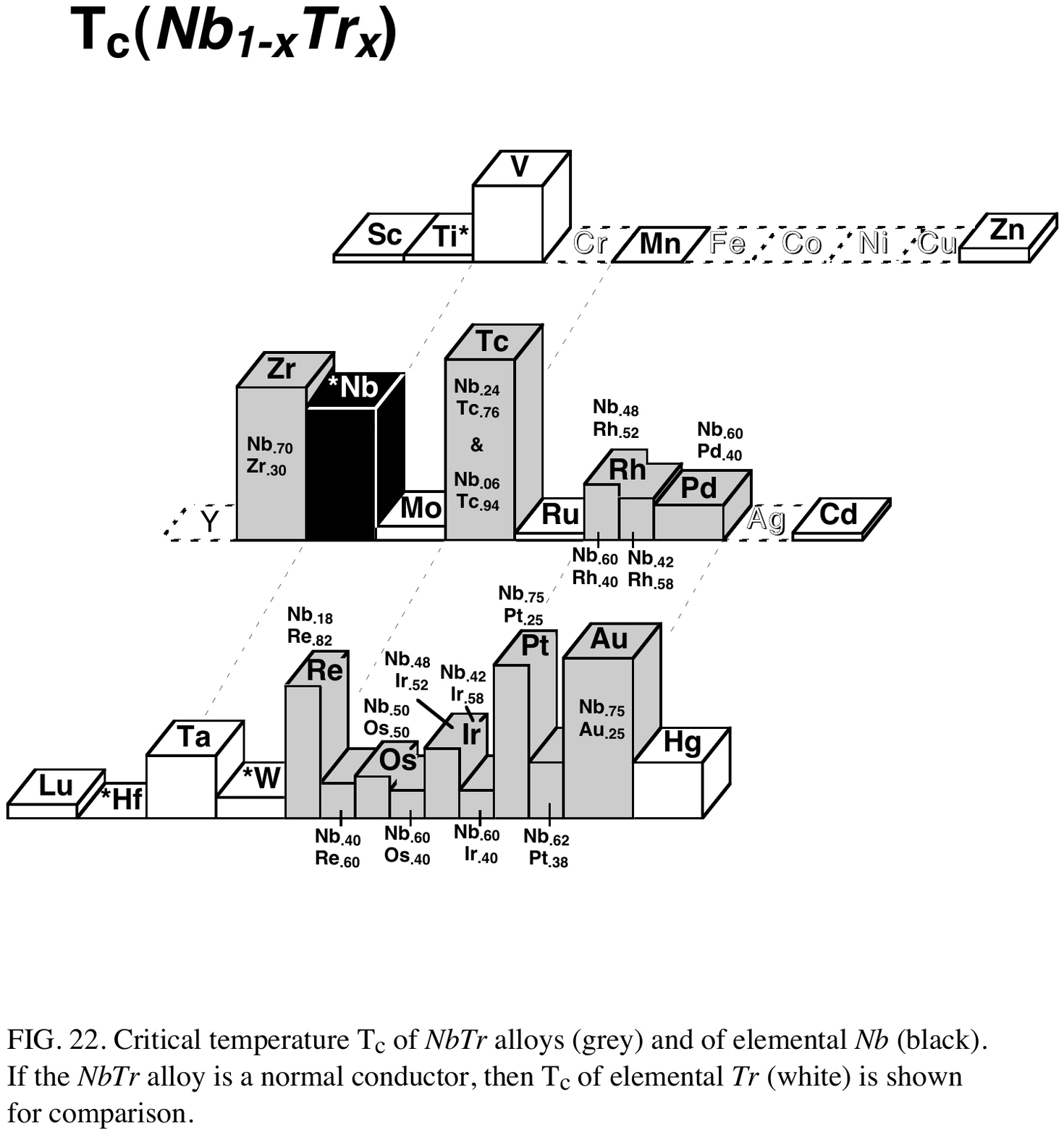}

\pagebreak

\noindent are contracted and therefore additionally dominated by the asymmetric location of the lateral oscillation, huddling the parent $Tc$ atom.  The asymmetric lateral oscillation encounters a higher energy barrier $\Delta{\epsilon}$---effectively a wider lateral oscillation gap $\Delta{y}$---and thus a higher critical temperature $T_{c} \propto \Delta{\epsilon} \propto \Delta{y}$.  Increasing with the difference of atomic number, $Z(Tc) - Z(Tr)$, the asymmetry factor seems strong enough to form superconducting alloys of $Tc$ with $Y$, $Lu$ and $Sc$---elements that are barely superconducting by themselves and which seldom form superconducting alloys.  Turning to the $TcTr$ alloys on the right, the asymmetric attraction of the contracted $s^{2}$ oscillators to the partner atom is counteracted by repulsion
from $d$-shell screening, resulting in lateral overswing and lack of superconductivity.

Dominance of the asymmetry factor also explains qualitatively the critical temperature $T_{c}$ of $Nb^{*}$ alloys shown in Fig. 22.  Because of the position of $Nb^{*}$ relative to $Tc$, the asymmetry of the lateral oscillation is weaker to the left (leaving only superconducting $Nb^{*}Zr$) and stronger to the right where it seems to overcome repulsion from $d$-shell screening.  The lack of superconductivity in $Nb^{*}Mo$, $Nb^{*}Ru$ and $Nb^{*}W^{*}$ (with reconfigured $s^{1}d^{5}$ in $W^{*}$) is due to the extended $s^{1}$ oscillator from the respective partner atom.  Remarkable are the $Nb^{*}Tr$ alloys on the far right and lower far right.  The $Pd$ atom has an $s^{0}d^{10}$ electron configuration, stabilized by premature closing of the $d$ shell.  That means that only $s^{2}$ electrons from $Nb^{*}$ oscillate in the $Nb^{*}Pd$ alloy where the $Pd$ atoms effectively widen the $Nb^{*}$-$Nb^{*}$ neighbor distance $R_{nn}$, compared to elemental $Nb^{*}$.  A reconfiguration to $s^{0}d^{10}$ may also occur in $Pt$ which would explain the high $T_{c}$ of the $Nb^{*}Pt^{*}$ alloy.  The largest surprise in Fig. 22 is the high $T_{c}$ of the $Nb^{*}Au$ alloy.  An explanation for it seems to be beyond the present considerations of asymmetry or electron configuration.  Instead it very likely is caused by the squeeze effect, to be explained in the next chapter.

As can be expected from the closely related electron configuration of $Nb^{*}$ and $V$, some features of the $Nb^{*}Tr$ alloys are recognizable in the $VTr$ alloys, shown in Fig.  23.  Those are the high $T_{c}$ of $VZr$ and $VHf$ to the lower left, and of $VRe$ and $VOs$, as well as of $VPt$ and $VAu$, at the bottom.  Remarkable exceptions are the lack of superconductivity of $VTc$ and the high $T_{c}$ of $VRu$.  The author has no explanation for those exceptions.  

%\alloyed{alloyed} not sure what this was but changed to alloyed...

For the $ZrTr$ alloys in Fig. 24 we assume free-atom electron configuration of \emph{alloyed} $Zr$, $s^{2}d^{2}$.  The lack of superconductivity in $ZrMo$ and $ZrW$ and the low $T_{c}$ of $ZrRu$ originates with the outer $s^{1}$ shell of these partner atoms.  The high $T_{c}$ of $ZrRh$ and $ZrPt$ possibly results from reconfiguration to $s^{0}d^{9}$ in crystalline $Rh^{*}$ atoms and to $s^{0}d^{10}$ in crystalline $Pt^{*}$ 
\pagebreak

\includegraphics[width=8.0in]{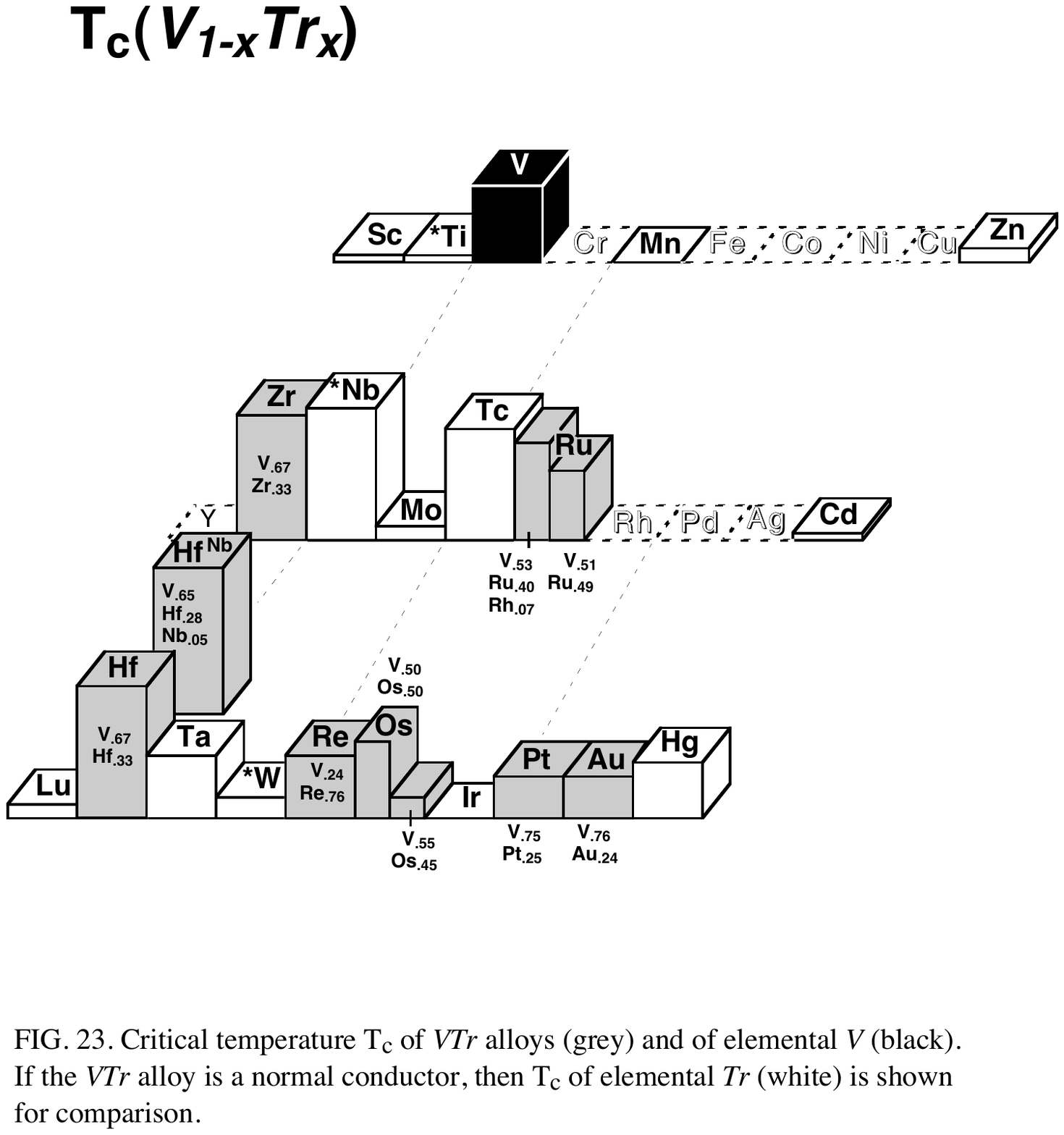}

\includegraphics[width=8.0in]{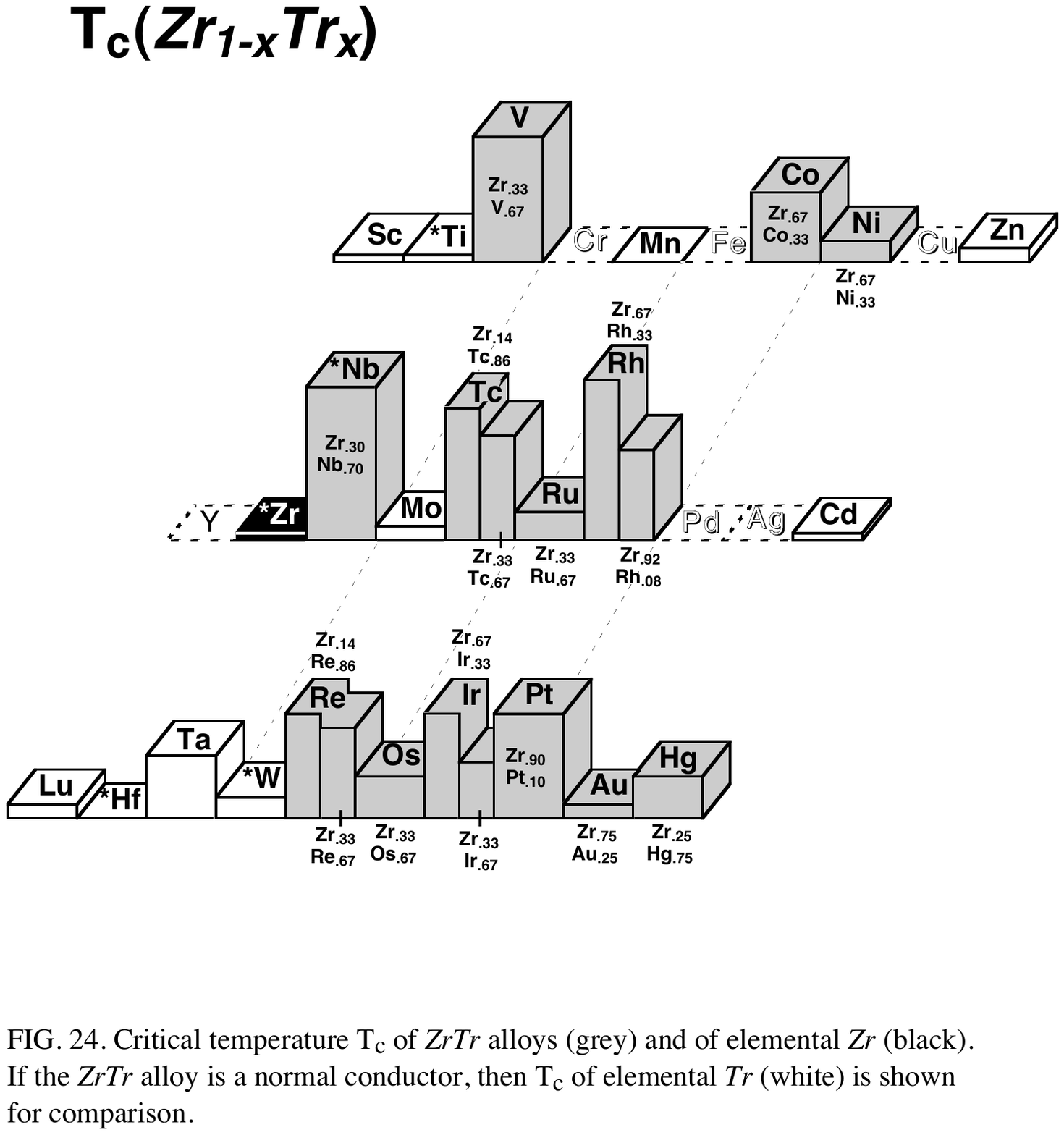}

\pagebreak

\noindent atoms.  Lacking $s$ electrons, those atoms don't participate in Coulomb oscillations but merely widen the $Zr$-$Zr$ distance, compared with elemental $Zr$ metal, causing wider lateral oscillations gaps for the $s$ electrons of the $Zr$ atoms.  

A surprise is the superconductivity of $Zr_{2}Co$ and $Zr_{2}Ni$, shown in the rop right region of Fig. 24.  Here we possibly see the indication of a ``participating-squeeze effect''---one where not only the squeezed atoms furnish the outer $s$ electrons of the superconducting Coulomb oscillators but, in contrast to regular squeeze, the \emph{squeezing} atoms participate, with their outer $s$ electrons, in Coulomb oscillations, too.  As we will see below, such cases can occur in $TrTr'$ alloys where the alloy partners $Tr$ and $Tr'$ possess a large difference in \emph{atomic} (not crystalline!) hardness, indicated by far horizontal separation in the Periodic Table. The $Zr_{2}Co$ and $Zr_{2}Ni$ alloys form tetragonal crystals (equivalent $a$ and $b$ axes) where each small but hard $Co$ or $Ni$ atom---$r(Co)$ = $r(Ni)$ = 1.25 \AA---sits inside a cage of eight large but soft $Zr$ atoms, $r(Zr)$ = 1.61 \AA.  It seems that the alloys' critical temperature $T_{c}$ is caused by a mild squeeze of the $Co$ and $Ni$ atoms, respectively, by their $Zr$ cage.  An inspection of the crystals' $a$ and $c$ axes shows that the cage, although equal in volume, is narrower and taller in shape for $Zr_{2}Co$ than for $Zr_{2}Ni$ ($\Delta a$ = 0.12 \AA, $\Delta c$ = - 0.25 \AA), indicating more mutual squeeze in the former case.  Whereas in elemental $Co$ and $Ni$ metal the atoms' outer $s$ electrons laterally overswing, the lateral oscillation width $2Y$ is reduced by the squeeze of the $Co$ atoms---and less of the $Ni$ atoms---resulting in $T_{c}(Zr_{2}Co)$ $>$ $T_{c}(Zr_{2}Ni)$ $>$ 0.

Reduction of lateral $s^{\sigma}$ oscillation width by the participating-squeeze effect, acting on the $(n+1)s^{\sigma}nd^{10}$ electron configuration of element group IB ($Cu$, $Ag$, $Au$, $\sigma = 1$) and IIB ($Zn$, $Cd$, $Hg$, $\sigma = 2$), in concert with lateral extension ($\sigma = 1$) or contraction ($\sigma = 2$) from screening effects, could account for the respective absence or presence of superconductivity in those metals---in contrast to groups IA (alkali metals) and IIA (alkali-earth metals) which \emph{both} lack superconductivity, apart from $Be$ with $T_{c} = 0.03$ K (see Fig. 13).

\section{COMPOUNDS OF TRANSITION METALS WITH OTHER ELEMENTS}

Much higher critical temperature $T_{c}$ is found for compounds of transition metals, particularly $V$ and $Nb$, with non-transition elements, labeled $Nt$, particularly of groups III, IV, and V.  Sometimes $T_{c}$ of the compound is more than twice that of the parent $Tr$ (see Figs. 25 and 26).  The dominant cause for the higher $T_{c}$ is the squeeze effect, which we already 
\pagebreak

\includegraphics[width=8.0in]{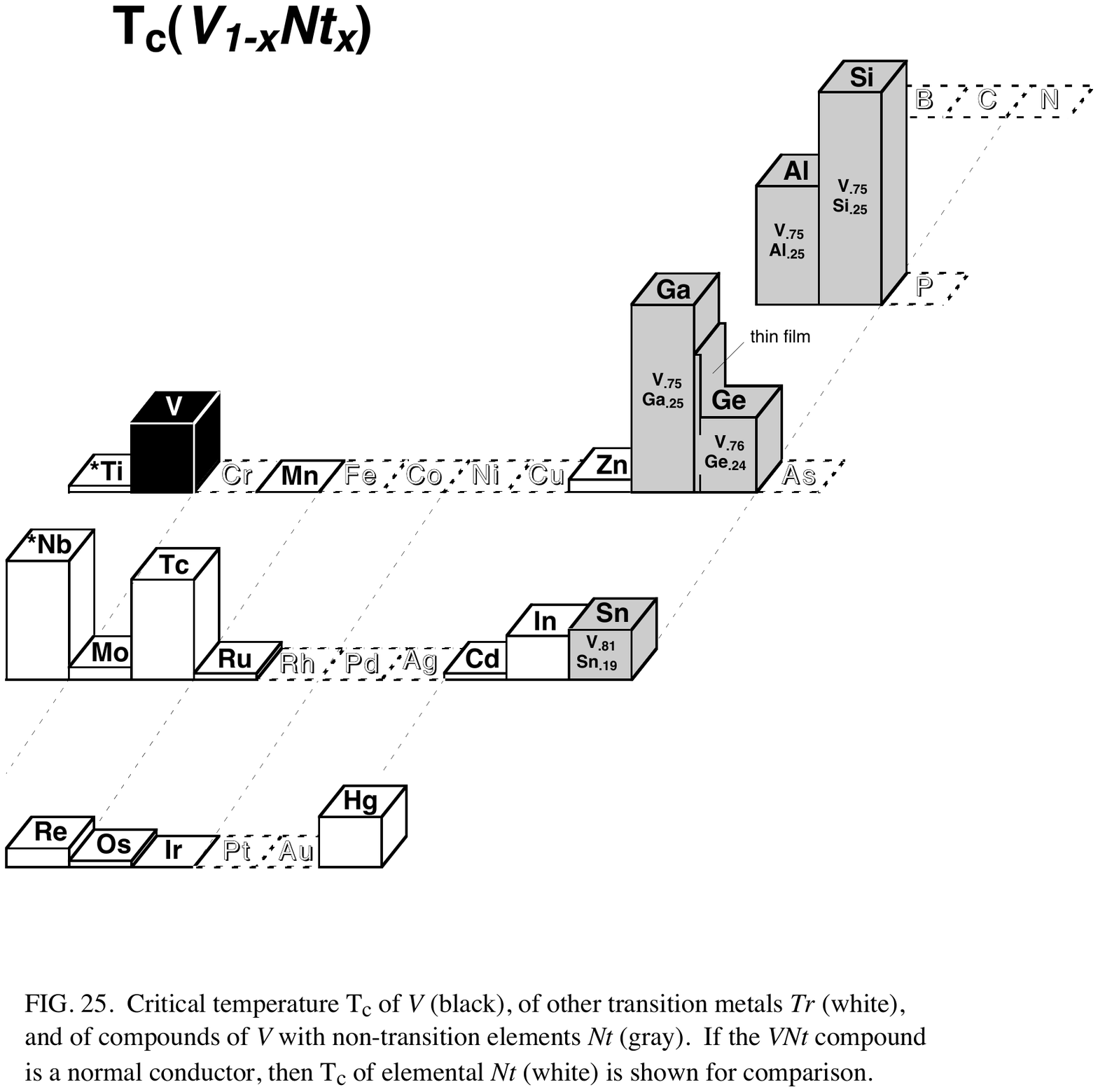}

\includegraphics[width=8.0in]{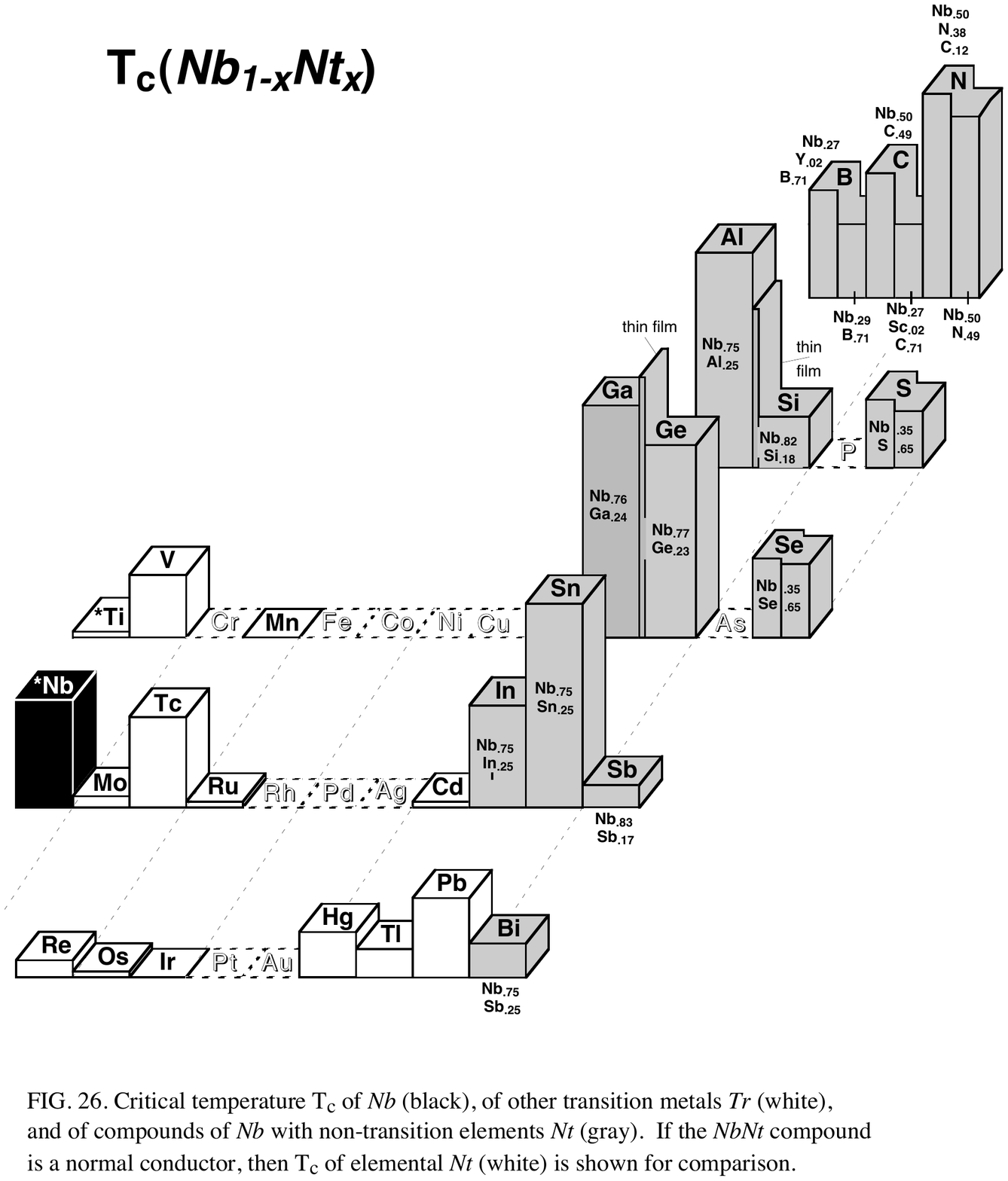}

\includegraphics[width=6.6in]{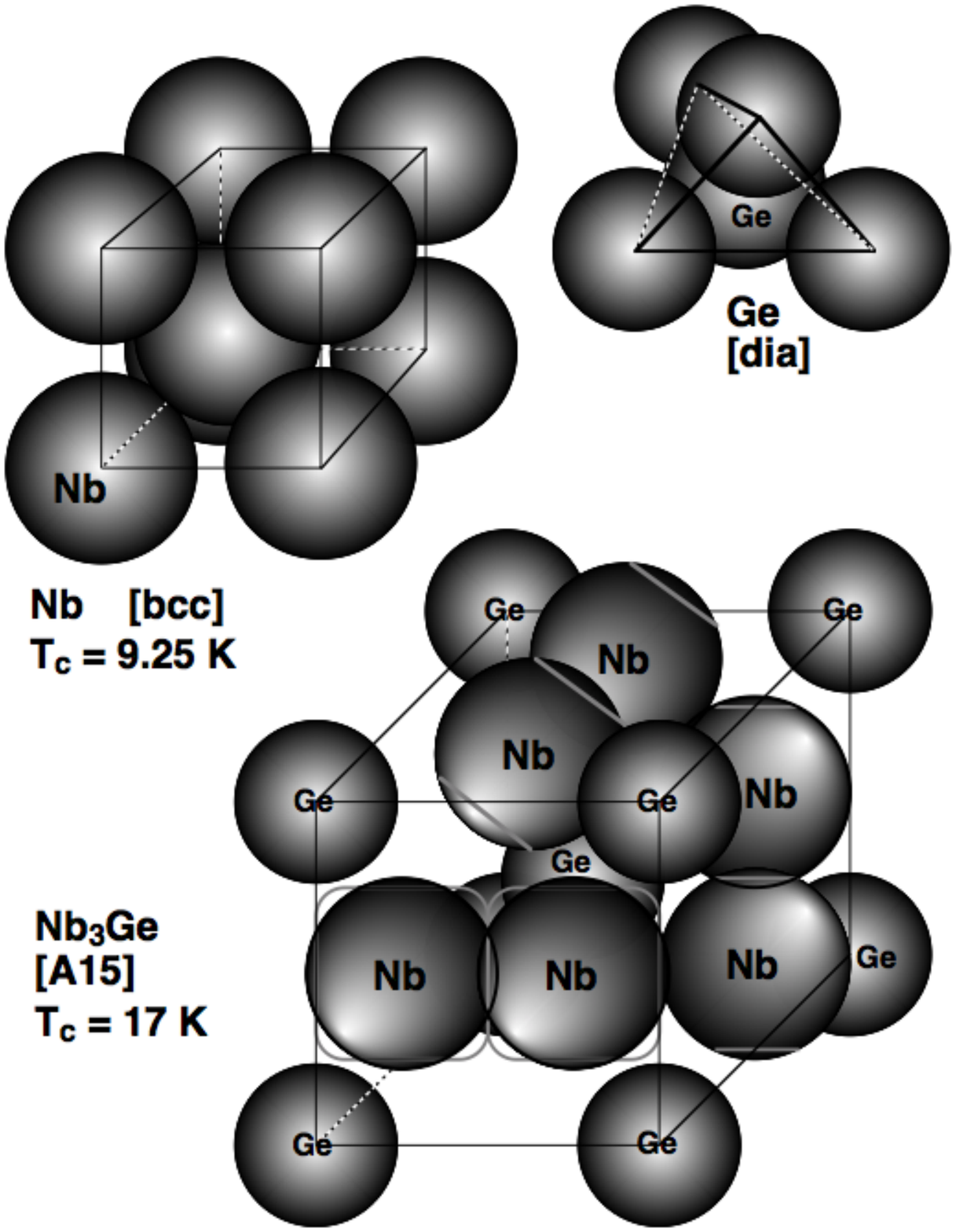}

\noindent FIG. 27. Lattice of elemental $Nb$, elemental $Ge$, and $Nb_{3}Ge$, with hard-sphere display of atoms, all on the same scale.  The deformation (``squeeze'') of the $Nb$ atoms is indicated.

\pagebreak

\noindent encountered with the alkali-doped fullerenes.  A typical example is illustrated in Fig. 27 for the compound $Nb_{3}Ge$.  The atomic size and the crystal structure of both parent elements and of the compound are shown to scale. A hard-sphere display is chosen to represent the atomic size of elemental $Nb$ in its $bcc$ lattice and of elemental $Ge$ in its diamond lattice (where the nearest-neighbor nuclei span tetrahedrons).  The structure of the $Nb_{3}Ge$ compound, known as $A15$, consists of a $bcc$ host lattice of $Ge$ atoms (at the eight corners and the center of the cell) and mutually orthogonal pairs of $Nb$ atoms with their nuclei located on the faces of the cell.  A close inspection of Fig. 27 reveals that the hard-sphere $Nb$ atoms would overlap in the $Nb_{3}Ge$ compound.  In order to fit those $Nb$ atoms into the $A15$ lattice, they have to be deformed (``squeezed'') with flattened $Nb$-$Nb$ interfaces as indicated at the three $Nb$ pairs in Fig. 27.  Note that the hard-sphere display for the elemental crystals is an approximation, here chosen to visualize the squeeze of the $Nb$ atoms in the compound.\cite{12}

The $Nb$ pairs across the cell faces (and neighbor-cell faces) constitute sets of atomic strings (``superwires'') through whose nuclei axial Coulomb oscillations occur.  The concomitant lateral oscillations are in the plane of the flattened $Nb$-$Nb$ interfaces.  Superconductivity persists if the lateral oscillation, say of the $Nb$ pair at the front face, swings over to, say, the orthogonal pair at the top of the cell (where it continues in axial Coulomb oscillation).  Conversely, superconductivity ceases when lateral overswing to a \emph{parallel} string of $Nb$ pairs occurs, say from the hip of the front $Nb$ pair in Fig. 27 to the hip of the parallel $Nb$ pair in the front cell face above (not shown in the figure).  By the squeeze effect, lateral oscillation is reduced in favor of axial Coulomb oscillation.  In the case of  $Nb_{3}Ge$ such squeezing is exerted on the front $Nb$ pair by the four $Ge$ atoms at the corners of the cell face while the  $Ge$ atoms in the center of this cell and of the next cell in front (not shown) prevent the $Nb$ pair from leaving the face plane.  

The $Nt$ atoms from groups III - V  are typically hard atoms due to the nuclear-based shrinking of atomic size across the non-transition columns of the Periodic Table, mentioned above.  The $Tr$ atoms are, by comparison, soft as a result of $d$-shell screening.\cite{13}  This makes $Nt$ atoms from groups III - V typically good ``squeezers'' and $Tr$ atoms good ``squeezees'' with consequent high $T_{c}$'s of the $TrNt$ compounds.  A confirmation that the squeeze effect is responsible for the higher $T_{c}$ of many $TmNt$ compounds is the frequent finding that still higher $T_{c}$ is achieved in thin film (see Figs. 25 and 26) or under pressure, both of which cause more atomic squeeze.
\pagebreak

\includegraphics[width=8.0in]{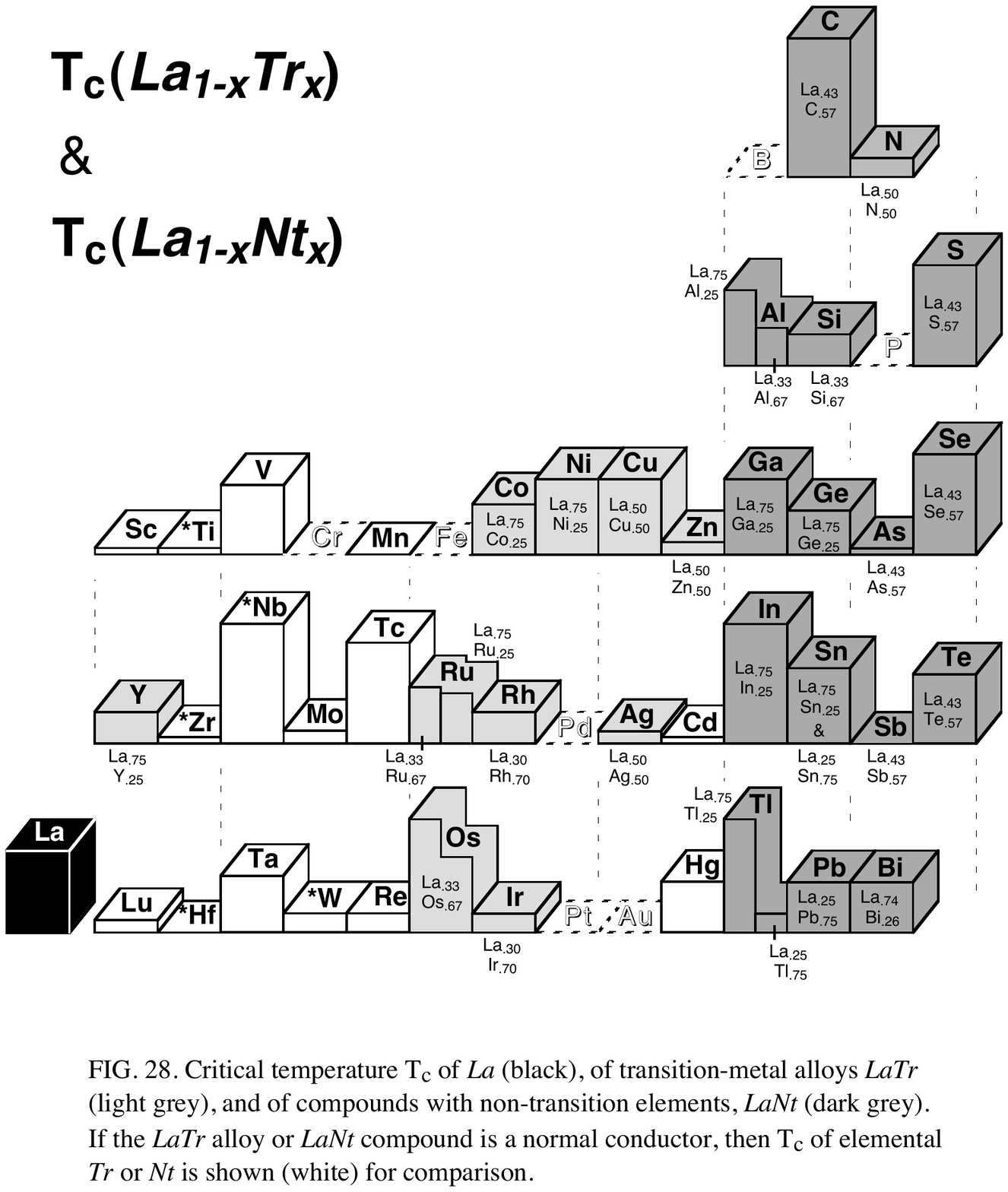}

\includegraphics[width=8.0in]{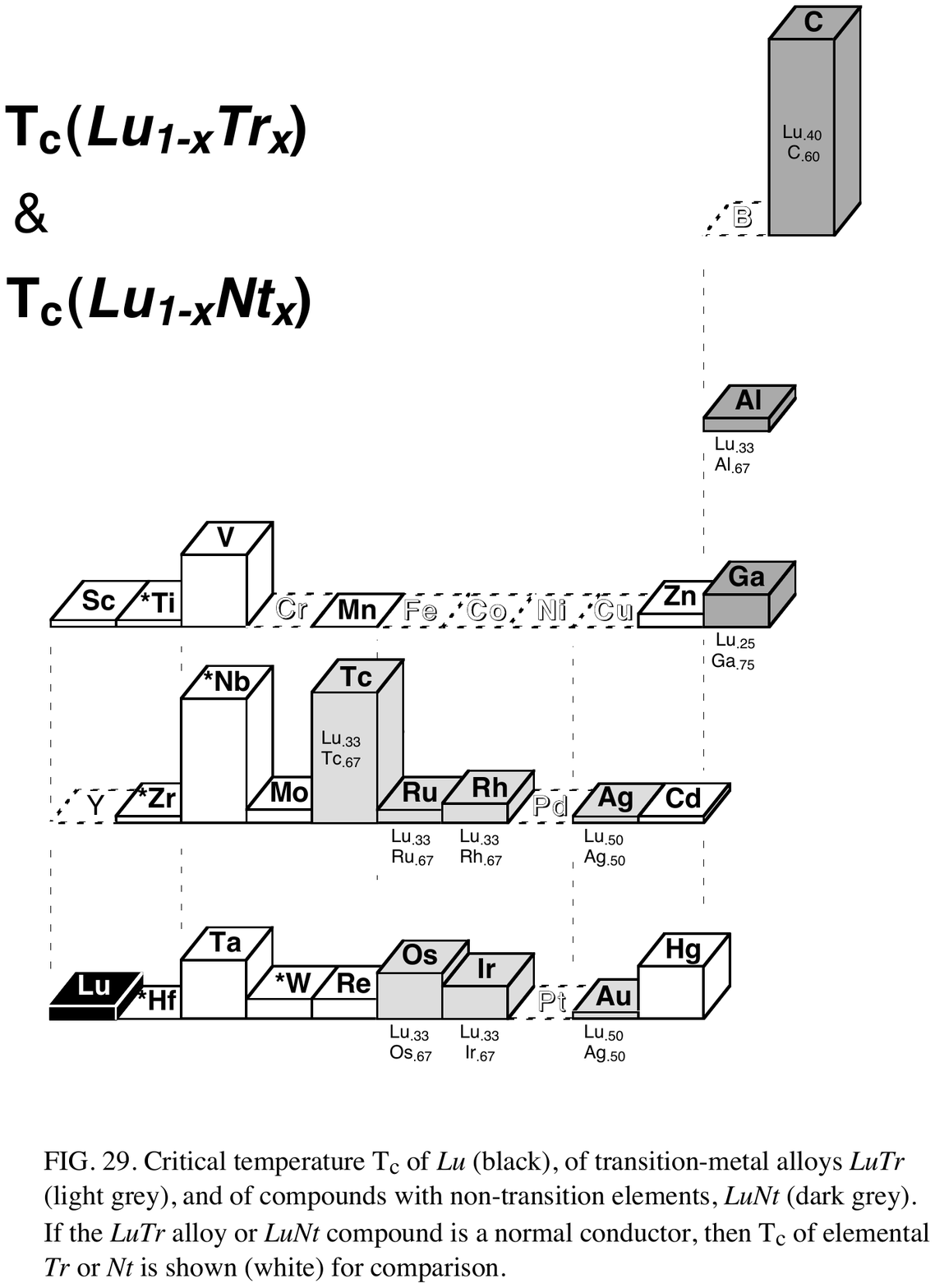}

\pagebreak

The surprisingly high $T_{c} = 11.5$ K of $Nb_{0.75}Au_{0.25}$, encountered earlier in Fig. 22, is very likely a result of the squeeze effect (be it regular or participating).  This $NbAu$ 
compound has the same crystal structure ($A15$) as $Nb_{0.75}Ge_{0.25}$ and is visualized by Fig. 27 if one replaces the small, hard $Ge$ atoms by larger, but less hard $Au$ atoms.  The larger size of $Au$ than $Ge$ affects the lattice constant, $a(Nb_{0.75}Au_{0.25}) = 5.20$ {\AA} $>$ $a(Nb_{0.75}Ge_{0.25}) = 5.16$ \AA.  The lesser hardness of the $Au$ atom than the $Ge$ atom causes less squeeze of the $Nb$ atoms resulting in a lower $T_{c}(Nb_{0.75}Au_{0.25}) = 11.5$ K $<$ $T_{c}(Nb_{0.75}Ge_{0.25}) = 17$ K.\cite{14}

The pattern of $T_{c}$ in alloys of $La$ with transition metals, $LaTr$, and with non-transition elements, $LaNt$, in Fig. 28 sheds more light on the ``participating squeeze effect,'' encountered in passing above.  Both $La$ and $Lu$ could be regarded as the first member of the third transition period, due to their common $6s^{2}5d^{1}$ electron configuration but with $4f^{0}$ and, respectively, $4f^{14}$ subshells beneath.  The latter subshell is present in all other third-period transition metals, which makes $Lu$ a natural member, as mentioned before.  Chemically similar, the two atom species differ in size, $r(La)$ = 1.88 {\AA} $>$ $r(Lu)$ = 1.76 {\AA} (lanthanide contraction) and in atomic hardness, rendering the $La$ atom rather soft.

The high critical temperatures $T_{c}$ of $LaNt$ compounds in the right part of Fig. 28 are the result of the regular squeeze effect, where very hard $Nt$ atoms squeeze the soft $La$ atoms, thereby reducing the lateral oscillation width $2Y$ of the outer $s$ electrons of $La$ atoms.  (The squeezing $Nt$ atoms don't provide $s$ electrons for Coulomb oscillations---hence the term ``regular'' squeeze effect.)  In contrast, the $T_{c}$ of the $LaTr$ alloys in the center part of Fig. 28 can be regarded as a manifestation of the ``participating-squeeze effect,'' where the softer $La$ atoms and the less softer $Tr$ atoms mutually squeeze each other according to their atomic hardness while \emph{both} participate with their outer $s$ electrons in superconducting Coulomb oscillations.

Note that the corresponding $La$ and $Tr$ partners are horizontally far apart in the Periodic Table with consequent differences in atomic hardness.  The pattern of $T_{c}(LaTr)$ in Fig. 28
suggests that the underlying mechanism (here, a participating-squeeze effect) is different than the mechanisms (asymmetry factor, etc.) that gives rise to the clusters of $T_{c}(TrTr')$ in Figs. 19 - 24 for alloy partners that are close neighbors in the Periodic Table.

In passing, another peculiarity of $La$ should be pointed out:  This is the superconductivity of $La_{0.75}Y_{0.25}$---an alloy where both partner atoms have electron configurations $(n+1)s^{\sigma}nd^{\delta}$ with equal subshell occupancy, $\sigma = 2$ and $\delta = 1$, but different principal quantum numbers 

\includegraphics[width=6.4in]{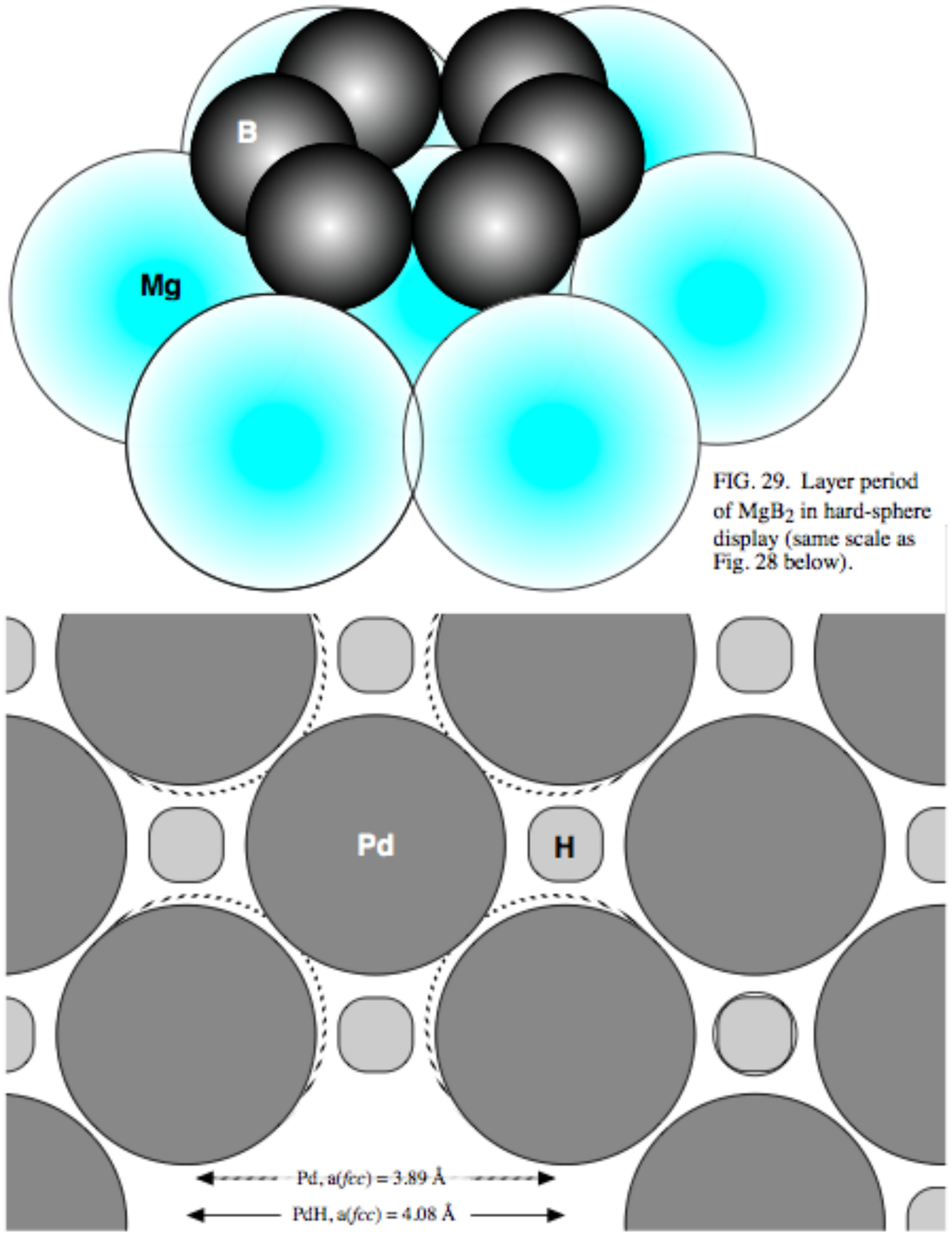}

\noindent FIG. 30. Squeeze of $H$ atoms in the $NaCl$ lattice of $PdH$.  The hard-sphere display of nearest neighbors in \emph{pure} $Pd$ metal is depicted by dashed lines about the marked $Pd$ atom.  The small circle represents a free $H$ atom.

\pagebreak

\noindent $n$ = 5 and 4, respectively.  No such case occurs in the alloys discussed above, Figs. 19 - 24.

The superconductivity of $Lu$ alloys in Fig. 29 is a weak echo of the $La$ alloys in Fig. 28 apart from two outstanding cases with high ${T_c}$, namely $LuTc_{3}$ and $Lu_{2}C_{3}$.  Discussed  earlier, the high $T_{c}$ of $LuTc_{3}$ is attributed to asymmetry of the $Lu$-$Tc$ Coulomb oscillator (see Fig. 21).  Otherwise, the transition temperature of the inter-transition metal alloys $LuTr$ is interpreted again in terms of the participating-squeeze effect, but on a more moderate scale than for the $LaTr$ alloys due to a less softer $Lu$ atom than $La$ and less horizontal $Lu$-$Tr$ distance than $La$-$Tr$ in the Periodic Table.

\section{UNCONVENTIONAL SUPERCONDUCTORS}

\emph{ \underline {Palladium hydride.}} A good example of the squeeze effect is at hand with $PdH$.  This compound crystallizes in the $NaCl$ structure, resulting from the residence of $H$ atoms at the interstitial sites of the $fcc$ lattice of elemental $Pd$.  The $Pd$ atom has an electron configuration $4d^{10}5s^{0}$.  Accordingly it does not participate in Coulomb oscillation.  Instead the P$d$ atoms, although of the typically medium softness of $Tr$ species, squeeze the much softer $H$ atoms.  This is shown in Fig. 30 by a comparison of lattice constants, $a(PdH)$ $>$ $a(Pd)$, together with atomic sizes and shapes. The resulting critical temperature is $T_{c}(PdH) = 9.62$ K.  The $H$-$H$ separation in $PdH$ is $1.44$ \AA $ \, = 2.73$ $a. u.$---a nuclear distance that permits Coulomb oscillation through both $H$ nuclei 
in the $H_{2}^{+}$ molecule ion (see Figs. 3 and 4).  By the squeeze effect, the atomic deformation of $H$ atoms in $PdH$ both enhances the axial Coulomb oscillation and reduces the lateral oscillation of each $H$ atom's electron.\cite{11,15}

\emph{ \underline {Alkaline-earth borides.}} The squeeze effect is also responsible for high $T_{c}$'s of various compounds composed of non-transitional elements.  An impressive example is provided by the alkaline-earth borides, such as $MgB_{2}$ with $T_{c} = 39$ K---a layered compound where the hard $B$ atoms deform the softer $Mg$ atoms, shown to scale in Fig. 31.\cite{11}

\emph{ \underline{Rare-earth nickel borocarbides}}, $RNi_{2}B_{2}C$, present another case of the squeeze effect.  By chemists' convention, the group of rare-earth elements comprises the elements $Sc$ and $Y$ together with the lanthanides, $R$ = \{ $Sc$, $Y$, $Ln$ \}, where $Ln$ = $La$, $Ce$, ..., $Yb$, $Lu$.  The crystal structure of $RNi_{2}B_{2}C$ is tetragonal (equivalent $a$ and $b$ axes) and is diagrammatically shown in Fig. 33.  The crystal consists of atomic layers in the $ab$ plane, stacked in the $c$ direction.  Note that in the diagram the right column is a periodic repetition of the left 
\pagebreak

\includegraphics[width=7in]{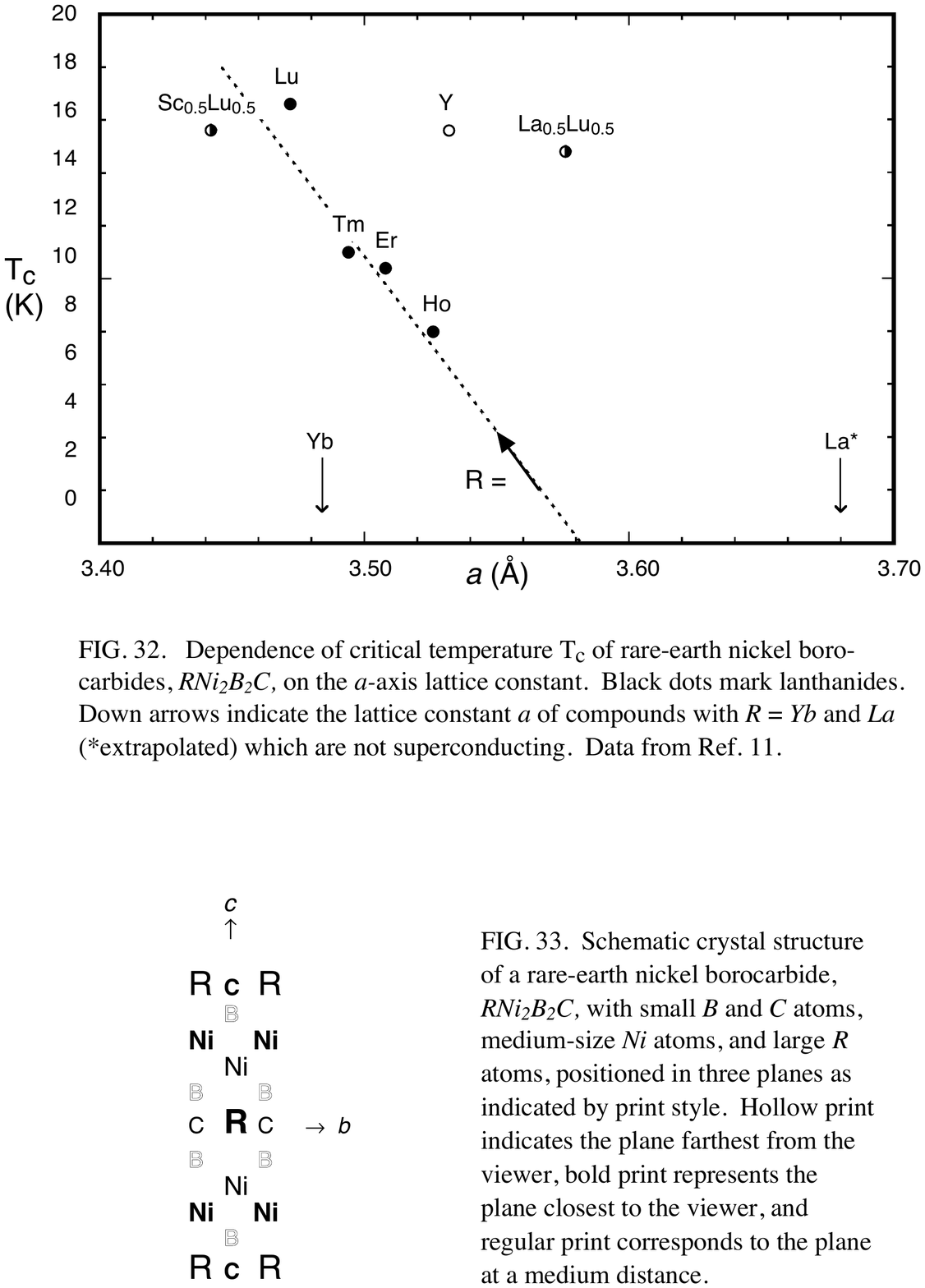}

\pagebreak

\noindent column.  Likewise the top row is a periodic repetition of the bottom row. The atoms in the first quadrant of the $bc$ coordinate system, except the top row, comprise the chemical unit, $RNi_{2}B_{2}C$.  The print style indicates atomic planes perpendicular to the $a$ direction (toward the viewer), with hollow print being farthest, bold print closest, and regular print in between.

The compounds' $Ni$ atoms, whose outer $s$ electrons constitute the Coulomb oscillators, are squeezed---directly according to the hardness of the $R$ atoms and also indirectly through the influence of the $R$ atoms on the compound's lattice spacing.  As can be inferred from Fig. 33, the size of the rare-earth atoms, $r(R)$, determines, cumulatively with the size of the $C$ atoms, the $b$-axis lattice constant of the crystal ($b$ = $a$).  This change of lattice constant furnishes a convenient demonstration of the squeeze effect, particularly for the lanthanide compounds since the size of lanthanide atoms, $r(Ln)$, decreases monotonously with increasing atomic number $Z$ (lanthanide contraction).  The \emph{indirect} squeeze of the compounds' $Ni$ atoms results in an increase of critical temperature $T_{c}(LnNi_{2}B_{2}C)$ with shrinking lanthanide atoms, $Ln$, shown in Fig. 32.

For an explanation, consider the rare-earth atom $R$ at the origin of the $bc$ coordinate system in Fig. 33.  This atom is bracketed by rigid, boomerang-shaped atom triples $BCB$, formed by strong covalent bond between the small $B$ and $C$ atoms and curved in the $a$-direction.  When, with increasing atomic number by increments of $\Delta Z = 1$, the $R=Ln$ atom shrinks, the flanking $BCB$ boomerangs move closer to $R$, causing a contraction of the $a$-axis lattice constant by $\Delta a \approx - 0.01$ \AA. It is observed that the $c$-axis lattice constant then expands, in a volume-preserving fashion, by $\Delta c \approx -2 \Delta a$.  This makes the triangle of the $B$ atoms in the second and fifth row of the diagram narrower and taller, thereby squeezing the enclosed $Ni$ atom in the forth row.  (The periodic repetition of the hollow-print $B$ atoms, located closer to the viewer than the bold-print plane, likewise squeezes the $Ni$ atoms in the third row.)  The indirect squeeze by $\Delta a$ on the $Ni$ atoms reduces the lateral oscillation width of their outer $s$ electrons, thereby increasing $T_{c}(LnNi_{2}B_{2}C)$ in the sequence $Ln=Ho$, $Er$, $Tm$, $Lu$ (see Fig. 32). 

Included in that critical temperature is a constant contribution on account of the \emph{direct} squeeze from the large-size central $R = Ln$ atom in Fig. 33 on the nearby medium-size $Ni$ atom in the diagram's forth row due to similar hardness of the mentioned late $Ln$ atoms.  The $Yb$ atom, however, is noticebly softer than its $Tm$ and $Lu$ neighbors in the Periodic Table.\cite{16}  The ensuing lack of direct squeeze from the central $Yb$ atom on the close (next-nearest) neighbor $Ni$ atom explains the lack of superconductivity of the $YbNi_{2}B_{2}C$ compound.

According to the linear trend in Fig. 32, the \emph{mixed} lanthanide compound $La_{0.5}Lu_{0.5}Ni_{2}B_{2}C$ would be expected to have a critical temperature $T_{c} \approx 0$.  However, the lattice distortion arising from the large size difference of the atoms at the terminal ends of the $Ln$ series, $r(La)$ = 1.88 \AA \, and $r(Lu)$ = 1.76 \AA, skews the $B$ triangles of narrow base and tall height and of wide base and lesser height such that their effective squeeze on the enclosed $Ni$ atom is comparable to the one in the $LuNi_{2}B_{2}C$ compound.  This results in  $T_{c}(La_{0.5}Lu_{0.5}Ni_{2}B_{2}C) \approx T_{c}(LuNi_{2}B_{2}C)$ despite the different lattice spacing of these compounds.

The mixed compound $Sc_{0.5}Lu_{0.5}Ni_{2}B_{2}C$ deviates only slightly---like the $Lu$ compound---from the linear trend in Fig. 32.  But the $Y$ compound, $YNi_{2}B_{2}C$, having the same $a$-axis and $c$-axis lattice constants as the $Ho$ compound, starkly deviates from that trend.  Recall that the linear relationship holds for indirect squeeze of $Ni$ atoms at practically constant direct squeeze on account the $Ln$ atoms' hardness.  A general tendency in the Periodic Table, atoms are the softer, the further down a given column.  This holds not only for the sequence $Sc$-$Y$-$La$, but extends, to a lesser degree, into the lanthanides, $Sc$-$Y$-$Ln$.  Despite the same lattice spacing, $a(YNi_{2}B_{2}C) = a(HoNi_{2}B_{2}C)$, the larger hardness of the $Y$ atom shows up in the nearest-neighbor distance of the elemental metals, $R_{nn}(Y) = 3.65$ \AA \, $> 3.58$ \AA \, $= R_{nn}(Ho)$.  It is the stronger \emph{direct} squeeze form the harder central $Y$ atom in Fig. 33 on the close (next-nearest) $Ni$ neighbor, compared to the lesser squeeze from the softer $Ho$ atom, that causes the high $T_{c}$ of the $Y$ compound in Fig. 32.  Overall, the critical temperatures of the rare-earth nickel borocarbides are not spectacular, 7 $K$ $<$ $T_{c}$ $<$ 17 $K$, nor are the changes, $ \Delta T_{c}/ \Delta Z \approx 1$ $K$.  But the systematic trend of the lanthanide compounds is noteworthy and serves as a subtle example of the squeeze effect.

\emph{ \underline{Ferropnictides}} are superconducting compounds containing iron, $Fe$.  Prominent examples are the rare-earth compounds $RFePnO_{1-y}$ and $RFePnO_{1-x}F_{x}$ with $T_{c}$ up to 55 K where $Pn$ = $P$, $As$; and the alkaline-earth compounds $Ae_{1-x}A_{x}Fe_{2}Pn_{2}$ with $T_{c}$ up to 40 K where $Ae$ = $Ca$, $Sr$, $Ba$; $A$ = $Li$, $Na$, $K$.  Their crystal structure comprises layers of $Fe$, sandwiched between layers of $Pn$.  Those $Pn$-$Fe$-$Pn$ planar sheets are separated by layers of rare-earth oxides, $RO$, or alkaline-earth atoms, respectively, when oxygen sufficient ($y=0$) or undoped ($x=0$). Otherwise the separating layers have $O$ atoms missing, or include substitutional 
\pagebreak

\includegraphics[width=7in]{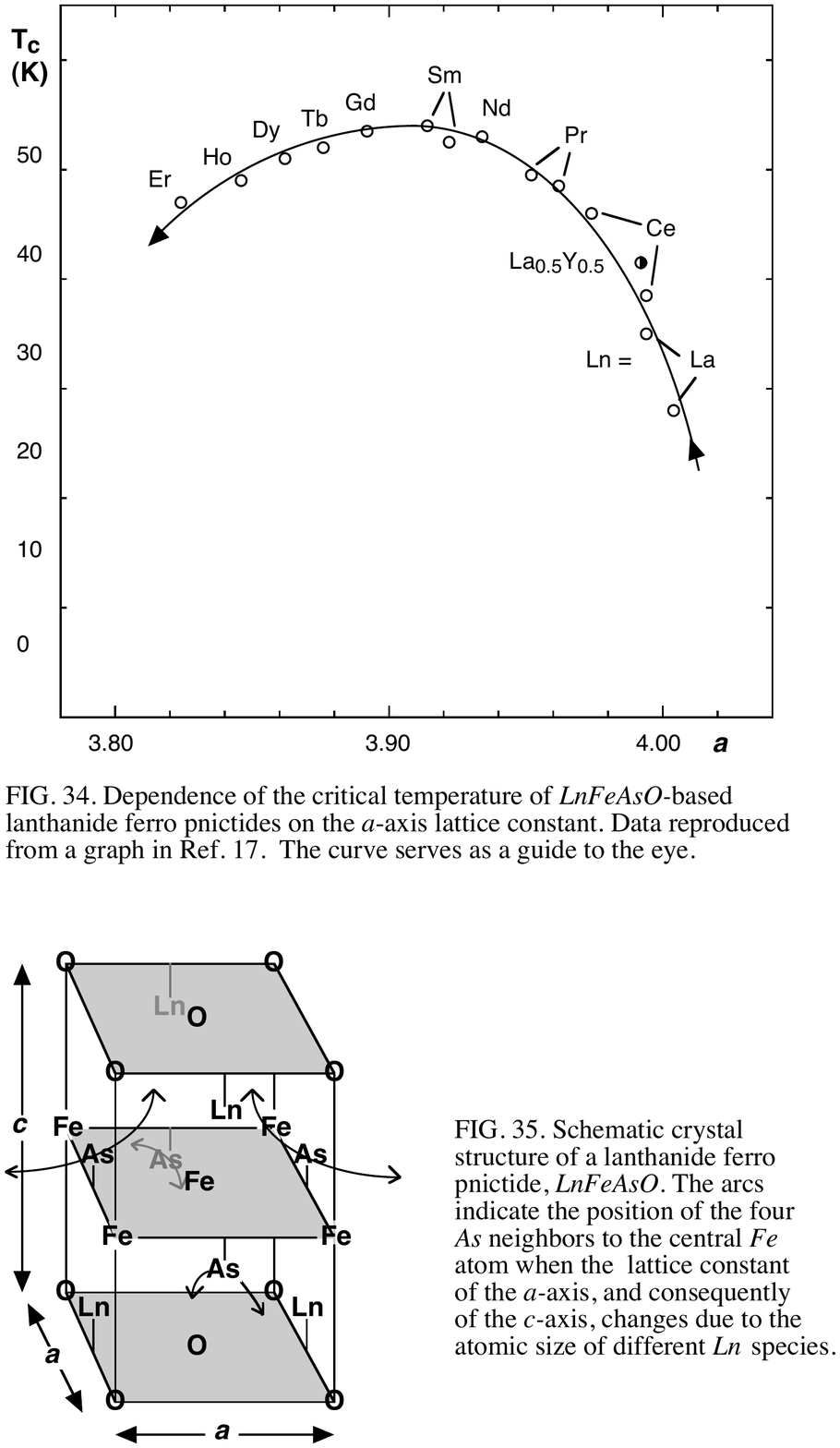}

\pagebreak

\noindent alkali atoms, $A$, or fluorine atoms, $F$.  In most cases the oxyen-sufficient or undoped compounds are normal (semi-) conductors.  Onset of superconductivity can be achieved with external pressure (squeeze effect) or stoichiometrically.  The latter method employs oxygen deficiency or doping with monovalent fluorine, $F$, or alkali atoms, $A$, in lieu of divalent $O$ or respectively $Ae$, typically in the range from about $x=0.05$ to $0.15$.  The internal electric field from those ionized lattice defects affects indirectly the $4s$ electrons of the $Fe$ atoms through its direct influence on the energy levels of the $Fe$ $3d$ orbitals and their associated magnetic properties.  Superconductivity is carried in the $Fe$ plane by Coulomb oscillators from the $4s$ electrons of the $Fe$ atoms.  The high critical temperatures $T_{c}$ are achieved through squeeze of the $Fe$ atoms by the much harder $Pn$ atoms. Corroborating the squeeze effect, many oxygen-deficient or doped ferropnictides show still higher $T_{c}$ under external pressure.

A new variant of the squeeze effect is encountered in the lanthanide ferroarsenides, 
$LnFeAsO_{1-x}F_{x}$.  Their crystal structure is tetragonal as shown diagrammatically in Fig. 35.  Alternate planes of $O$ atoms and $Fe$ atoms are stacked along the $c$-axis with those atoms residing at the plane corners and plane center of the unit cell.  Arsenic atoms, $As$, hover alternately above and below the edge centers of the $Fe$ unit plane.  Likewise, (ionized)  $Ln$ atoms, hover above and below the edge centers of each $O$ unit plane.  The lanthanide ions $Ln^{2+}$ determine, cumulatively with the $O^{2-}$ ions, the $a$-axis lattice constant and, cumulatively with the $Fe$ atoms, the $c$-axis lattice constant.  The change of those lattice parameters with successive $Ln$ replacement is $\Delta c \approx  + 2\Delta a$, that is, \emph{no} cell-volume conservation holds as in the $LnNi_{2}B_{2}C$ compounds discussed above.  However, it is observed that with increasing atomic number $Z$---and consequent shrinking of the $a$-axis spacing---the \emph{height}, $h$, of the $As$ atoms above the $Fe$ plane (see Fig. 35), and likewise of the $As$ atoms beneath, \emph{increases}.\cite{18}  Thus with \emph{decreasing} lattice constant $a$, all four neighboring $As$ atoms to the central $Fe$ atom in Fig. 35 shift inward and simultaneously upward concerning the elevated left and right $As$ neighbor, but down concerning the lowered back and front $As$ neighbor.  Those shifts of the $As$ neighbors to the central $Fe$ atom are indicated in Fig. 35 by arcs.  

As seen from the central $Fe$ atom, the four As neighbors span a tetrahedron.  The $As$-$Fe$-$As$ bond angle $\alpha$ is determined by the horizontal spacing of $As$ neighbors, $a$, in concert with an essentially constant $Fe$-$As$ bond length $\ell  = 2.40 \pm 0.02$ \AA \, for all $Ln$ ferroarsenides.  As a result the vertical distance $h$ of the $As$ atoms from the $Fe$ plane varies accordingly.  For the cases displayed in Fig. 34 the bond angle ranges from $\alpha  = {114^ \circ }$ for the $La$ compound to $\alpha  = {108^ \circ }$ for the $Er$ compound, straddling the angle for a \emph{regular} tetrahedron, $\hat{\alpha} \simeq {109^ \circ }$, which is close to the value of $\alpha  = {110.6^ \circ }$ found for the $Sm$ compound.\cite{19, 20}

It is widely agreed that the hump-shaped curve in Fig. 34 results from deviation of the shape of the $FeAs_{4}$ tetrahedron from regularity,\cite{18, 19, 20} but the associated mechanism of superconductivity is still unresolved. Viewed mechanically, the forces from the four $As$ neighbors on the central $Fe$ atom compress it by their normal component (along regular tetrahedron axes) and shear it by their tangential component.  By the squeeze effect only the \emph{compression} of the $Fe$ atom reduces the lateral oscillation width of its $4s$ electron oscillation.  This would qualitatively explain the convex curve in Fig. 34 for the dependence of $T_{c}$ on the $a$-lattice spacing.  In contrast to the linear increase of $T_{c}$ of $LnNi_{2}B_{2}C$ compounds with shrinking $a$-lattice spacing in Fig. 32, caused by increasing squeeze of $Ni$ atoms, the tetrahedral bonding in the $LnFeAsO_{1-x}F_{x}$ compounds causes the squeeze of $Fe$ atoms to pass through a maximum according to deviation from tetrahedron regularity.

\emph{ \underline{Organic charge-transfer salts}} possess superconductivity in some cases.  One example is bis(ethylenedithio)tetrathiafulvalene, abbreviated as $\beta$-$(ET)_{2}I_{3}$, with $T_{c}= 8$ K.\cite{11}  The compound crystallizes in layers of $I$ atoms, sandwiching the organic molecules in between.  Each such molecules consists---besides a majority of $C$ and $S$ atoms---of four $H$ atoms at each terminal end (see Fig. 36).  The compound ionizes internally by the transfer of \emph{one} electron from a quadruple of terminal $H$ atoms to an opposite iodine triple, rendering the latter as $I_{3}^{-}$.  One of the two pairs of terminal $H$ atoms, deprived of one electron, can be regarded as a \emph{bound} hydrogen-molecule ion, $H_{2}^{+}$, whose lone $s$ electron performs Coulomb oscillations through the corresponding $H$ nuclei.  If the Coulomb-oscillator model of superconductivity is valid, then this scenario would be the mechanism of superconductivity in $\beta$-$(ET)_{2}I_{3}$.  The axial separation of neighbor $H$ nuclei in $\beta$-$(ET)_{2}I_{3}$ is about $3$ \AA.  This is just below the separation---3 {\AA}  $ \cong  5.7 \, a. u.$---where the electron becomes untrapped from one $H$ nucleus to perform Coulomb oscillations through both $H$ nuclei of a \emph{free} $H_{2}^{+}$ as well as concomitant lateral oscillations (see Figs. 3 and 4).  It is likely that the squeeze of $H$ atoms in $\beta$-$(ET)_{2}I_{3}$ by neighboring $C$ and $I$ atoms reduces the lateral oscillation and correspondingly enhances the axial Coulomb oscillation through adjacent $H$ nuclei.

\noindent .

\noindent .

\noindent .

\pagebreak

\includegraphics[width=7in]{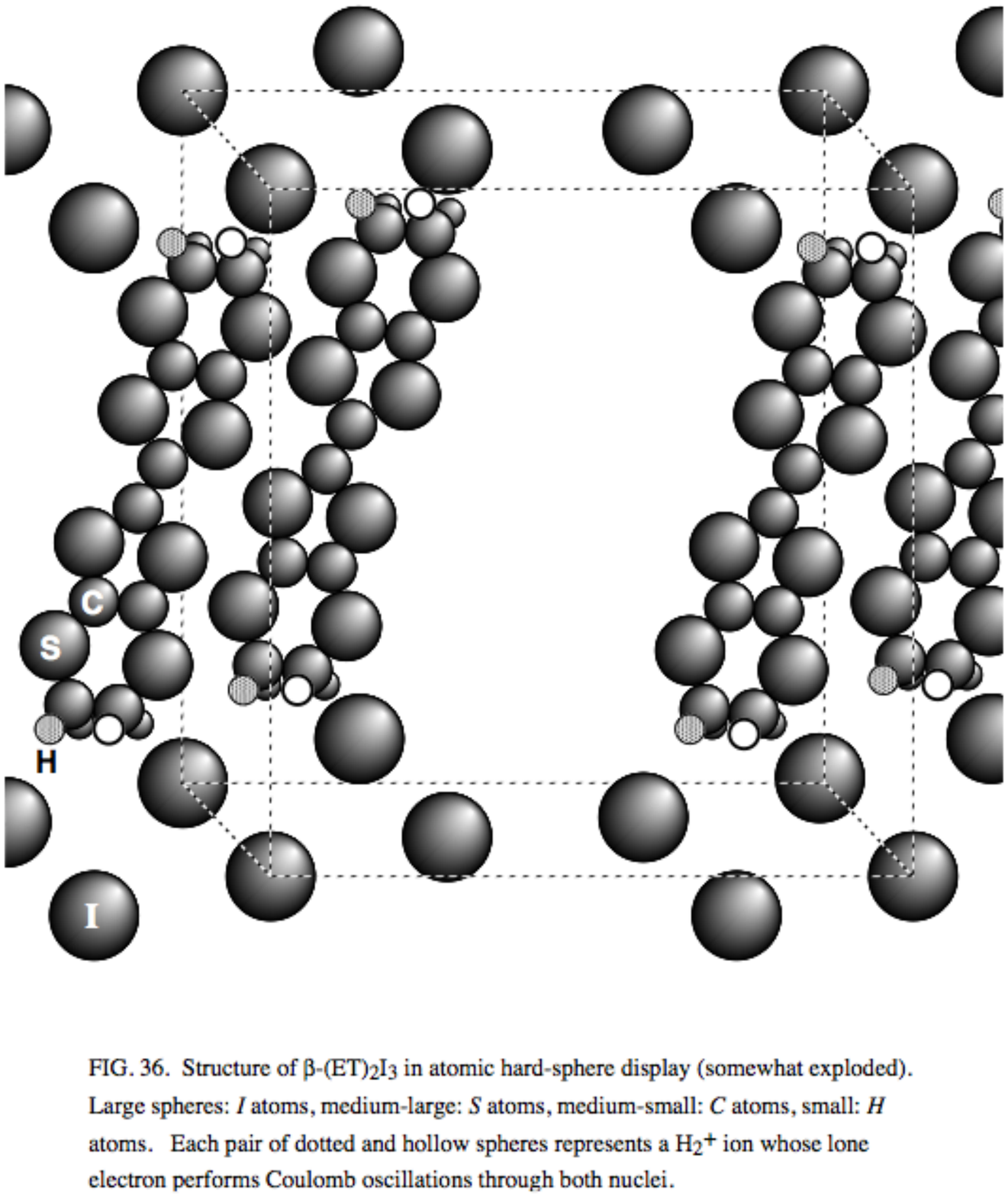}

\pagebreak

\emph{ \underline{Oxide superconductors}}---most of them cuprates---were discovered in the mid-1980's and are still widely referred to as ``high-$T_{c}$'' superconductors.  Their layered crystal structure can be derived from the perovskite lattice.  Table III gives a selection of such compounds with their critical temperature $T_{c}$ under various conditions, along with an indication of their relation to perovskites, and an acknowledgment of the landmark discoveries.  The oxides under consideration are ionic crystals where the electrostatic part of the binding---the Madelung energy---plays a dominant part.

The once popular compound $YBa_{2}Cu_{3}O_{7}$, included in Table III, will be used to exemplify the underlying physics of oxide superconductors.  Figure 37 shows its crystal structure in a realistic display of ionic radii.  For the radius of the oxygen ion, $r(O^{2-})$---not well-defined, in part because of the instability of free $O^{2-}$---the value from the alkaline-earth oxides was chosen, $r(O^{2-}) \equiv  R_{nn}(AeO) - r(Ae^{2+}) \cong 1.44 $ \AA.  The use of that $O^{2-}$ radius illustrates the squeeze of the $O^{2-}$ ions as they are much more confined in the copper oxides than in the alkaline-earth oxides.  The $YBa_{2}Cu_{3}O_{7}$ lattice is vertically exploded in the right part of Fig. 37 to better show its layers.  It is believed that superconductivity occurs in the $CuO_{2}$ layers, called $conducting$ layers.  Between the conducting layers is a separating layer, here of $Y$ ions, but frequently of $Bi$, $Tl$ or $Hg$ ions (see Table III).  The $BaO$ layer is called a \emph{bridging} layer and the $CuO$ layer is called an \emph{additional} layer.  Usually it suffices to combine those layers by their function.  Then the combination of conducting and separating layers, here
$Cu{O_2}YCu{O_2}$, is called the \emph{conducting slab} and the combination of bridging and 
additional layers, here $BaOCuOBaO$, is called the \emph{binding slab}.\cite{21} 

%$BaO$‚Äî$CuO$‚Äî$BaO$  $CuO_{2}$‚Äî$Y$‚Äî$CuO_{2}$

By the Coulomb-oscillator model, superconductivity is carried by $s$ electrons.  Stoichiometric charge balancing of the compounds in Table III reveals that the cations are stripped off of the $s$ electrons that they possessed as neutral atoms.  This leaves only the $O^{2-}$ ions as candidates to furnish the Coulomb oscillators for superconductivity.  But their $2s^{2}$ electrons are too deeply buried inside the $2p^{6}$ shell to form Coulomb oscillators between neighboring oxygen nuclei.  However, as Fig. 37 indicates, the $O^{2-}$ ions in the conducting layer are squeezed by their neighbor $Cu^{3+}$ ions.  One way to reduce short-range repulsion is provided by an electron reconfiguration in $O^{2-}$ from $2s^{2}2p^{6}$ to $2s^{2}2p^{4}3s^{2}$, that is, by depopulating the two $p$ orbitals that are aligned along the $O^{2-}$-$Cu^{3+}$ axis in favor of two excited $s$ orbitals.  It is those $3s^{2}$ electrons of extremely squeezed $O^{2-}$ ions that form the Coulomb oscillators in the oxide superconductors.  By the squeeze effect, these $s$ electrons dramatically decrease 
\pagebreak

\includegraphics[width=7.5in]{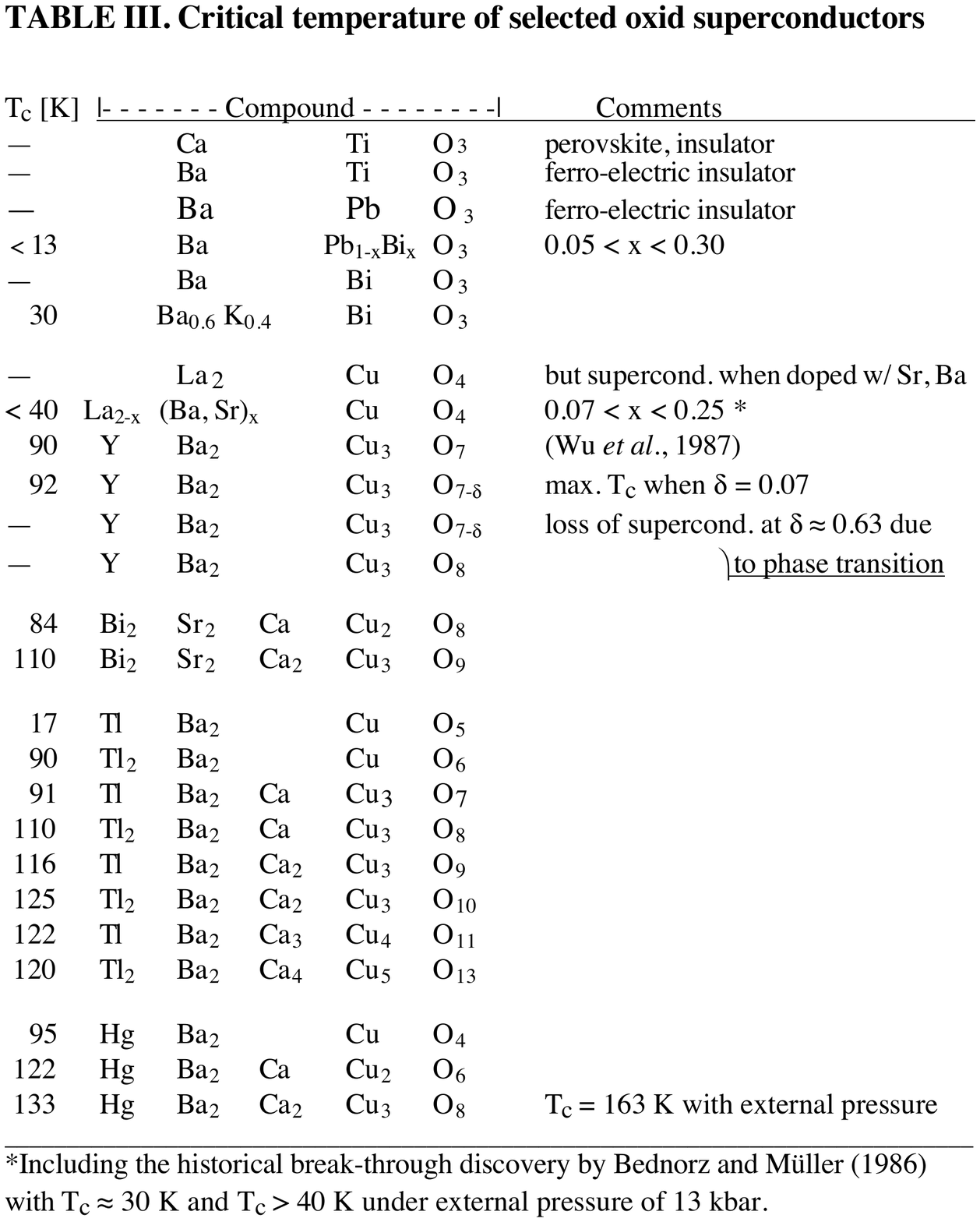}

\includegraphics[width=6in]{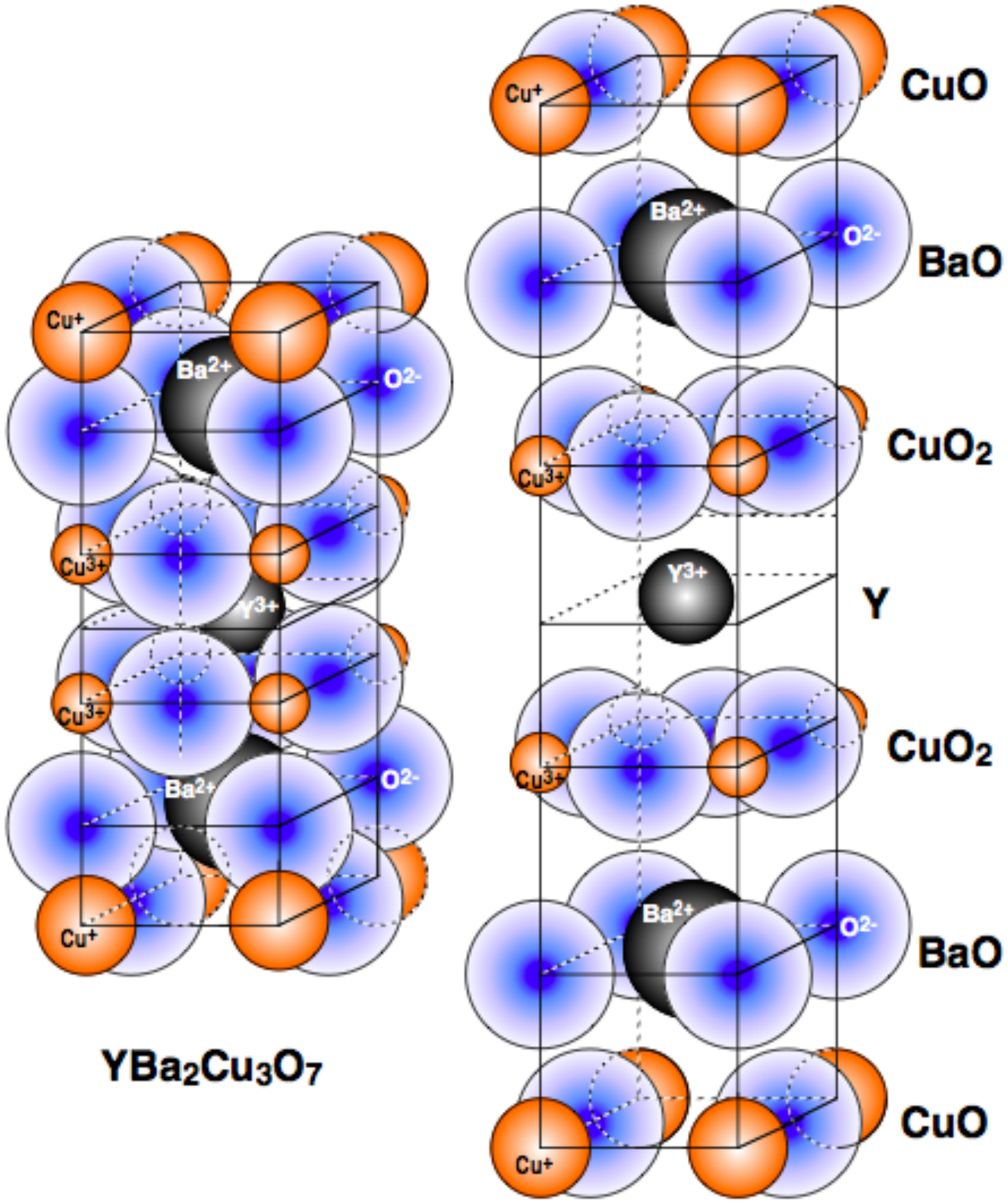}

\noindent FIG. 37. Crystal structure of $YBa_{2}Cu_{3}O_{7}$ in hard-sphere ion display (left), and vertically exploded to show the layers (right).

\pagebreak

\noindent their lateral oscillation to the benefit of enhanced axial Coulomb oscillation.  This, in a nutshell, is the reason for the high $T_{c}$ of the oxide superconductors.

An indication that the squeeze effect is at work with the oxide superconductors is the finding of still higher $T_{c}$ when the sample is under external pressure (see Table III).  Another way to increase $T_{c}$ is by stoichiometric oxygen deprivation.  For example a lack of oxygen in the  compound under consideration, $YBa_{2}Cu_{3}O_{7-\delta}$, gives the highest $T_{c}$  ($= 92$ K) when $\delta = 0.07$.  It is known from alkali halides---the simplest ionic crystals---that lack of anions (or excess of cations) introduces anion vacancies.  Somewhat counterintuitively, the cation neighbors of an anion vacancy are displaced \emph{away} from the vacancy, not toward it, as one would expect by space availability.  The reason is the dominance of the electrostatic (Madelung) energy over the short-range forces in ionic crystals.  In the case of oxygen deficient $YBa_{2}Cu_{3}O_{7-\delta}$, the $Cu^{3+}$ neighbors are displaced away from $O^{2-}$ vacancies and thereby squeeze the remaining $O^{2-}$ ions even more.

One can regard an oxide superconductor as a ``vise'' for Coulomb oscillators.  The $O^{2-}$ ions that furnish the Coulomb oscillators are in the conduction slab.  The force on the jaws of the vise is provided by the binding slab of metal oxides.  Research in oxide superconductivity has shown that increased sophistication (and complication) of compound composition increases $T_{c}$ by ``strengthening'' the compounds' binding slab (see Table III).  But it came as a disappointment that $T_{c}$ seems to level out at about 130 K.  Once the squeeze effect is expressible in terms of quantum-mechanical molecular orbitals, it may be possible to determine its limit with respect to high $T_{c}$.  Technologically, a severe short-coming of the oxide superconductors is their brittleness (lack of mallebility) due to strong ionic contributions in the crystal binding.  Thus, it is hard to avoid the impression that the pursuit of oxide superconductors has proved a \emph{cul-de-sac}, instead of the road to room-temperature superconductivity.  Other means are necessary to achieve this goal.

Nevertheless, some lessons can be learned from the discussed high-$T_{c}$ superconductors: For once, they all have layered crystal lattices.  Why do such layers favor superconductivity?  Two reasons stand out:  (1) The conducting layers in the $ab$ planes prevent lateral overswing of the superconductivity-enabling $s$ electrons in the \emph{perpendicular} $c$ direction.  (2)  Chemical pressure from the binding slabs on the conducting layers prevents, via the squeeze effect, lateral overswing of those $s$ electrons \emph{within} the $ab$ layers.  Since the crystal structure of, say, ferropnictides and copper oxides is typically tetragonal, the $a$ and $b$ directions in the 

\pagebreak

\includegraphics[width=6.25in]{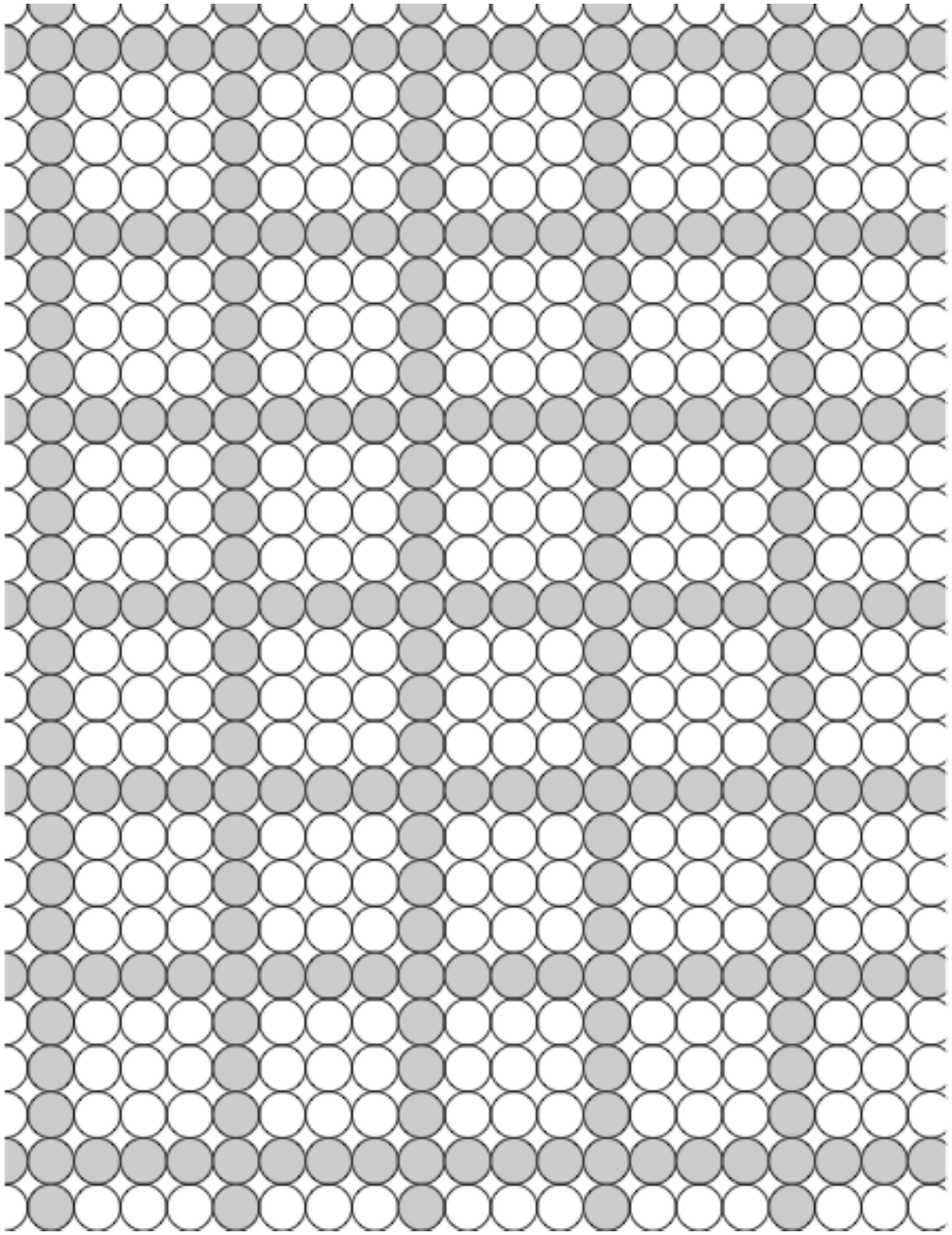}

\noindent FIG. 38. Network of monatomic wires (``superwires'', gray) embedded in a matrix of insulating atoms (white). Perpendicular superwires extent likewise from the intersections in the plane.

\pagebreak

\noindent conducting layers are equivalent.  However, applying the idea of directional hierarchy ($c \gg a$) of the tetragonally layered structures once more \emph{within} the conducting layer ($b > a$), one could search for materials of orthorhombic lattices with \emph{conducting stripes} in the $a$ direction, separated by insulating stripes of large width in the $b$ direction, possibly with mutual attraction to exert squeeze on the intervening conducting stripes.  Conceivably this could lead to higher critical temperatures than achieved with tetragonal materials.  The price, albeit, would be even more complicated crystals of, very likely, high brittleness---something more of academic interest than of practical use.

\section{ROOM-TEMPERATURE SUPERCONDUCTORS}
As the present overview of all major groups of superconductors shows, elevated critical temperature $T_{c}$ is achieved in most cases---certainly for $T_{c} > 20$ K---by the squeeze of Coulomb-oscillator providing, soft atoms by neighboring hard atoms or molecules.  Exceptions are the elemental metals and most alloys of transition metals due to a lack of, or minimal \emph{difference} in atomic hardness, although a mild version, the ``participating squeeze effect,'' can be found in some $TrTr'$ alloys.

The traditional strategy for finding better superconductors has been the employment of elemental metals, metallic alloys, or systematically synthesized compounds, typically grown from the melt, and more recently with high-pressure techniques.  It can be regarded as the search for better ``bulk superconductors.'' Since nature optimizes \emph{binding energy}, not superconductivity, this approach frequently misses the target.  Instead of melting mixtures of chemicals in order to crystallize in the structure of maximum binding energy, one should \emph{fabricate} superconductors---if necessary atom by atom---according to the requirements of the mechanism of superconduction.  If the Coulomb-oscillator model of superconductivity is valid, then the task is to \emph{avoid lateral overswing of outer} s \emph{electrons}.  Instead of suppressing lateral oscillation by atomic squeeze, as in all present superconductors of elevated $T_{c}$, one could employ \emph{``superwires''}, that is, single monatomic metallic filaments, supported by enveloping insulating sheaths.  Better still would be a sufficiently widely spaced network of monatomic filaments---a superwire lattice---embedded in an insulating matrix as indicated in Fig. 38.  Such a lattice of superwires would be an emulation of the network of the superconducting $Nb$-$Nb$ chains in $Nb_{3}Ge$ (see Fig. 26), although with wider spacing.  If the 

\pagebreak

\includegraphics[width=6.25in]{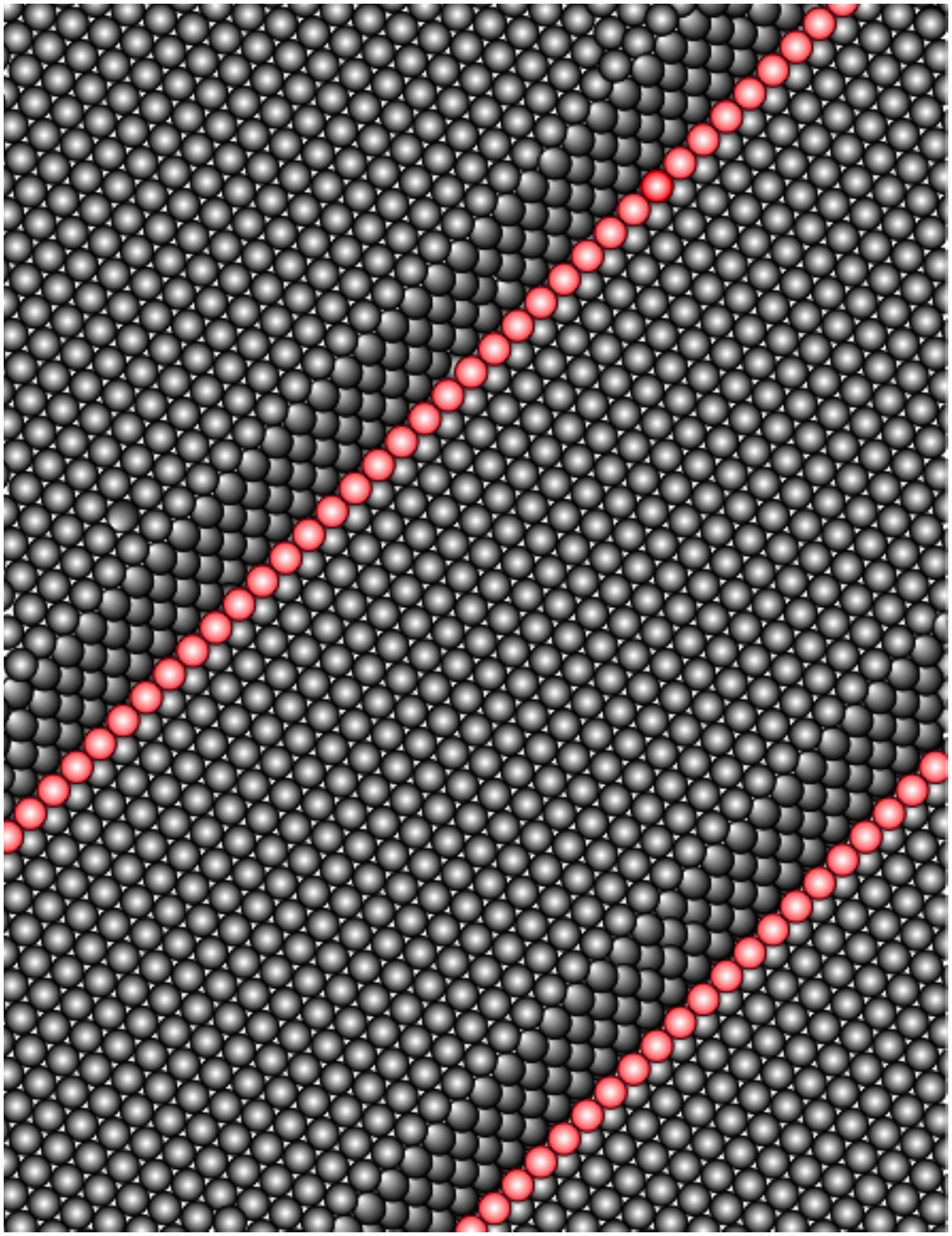}

\noindent FIG. 39. Monatomic filaments of metal (red/gray) along terrace corners of an insulating substrate (dark).

\pagebreak

\noindent separation of parallel embedded superwires is sufficiently large---something like three or four atomic neighbor distances, as in Fig. 38---then the width of the outer $s$ electrons' lateral oscillation doesn't matter and it doesn't need to be suppressed by atomic squeeze.

A crucial test would be the conductivity of a \emph{single superwire}, obtained, possibly, by atomic deposition of some metal on the terraced surface of an insulating substrate, as schematically indicated in Fig. 39.  Activated by heat treatment, the deposited atoms (grey) would aggregate in monatomic filaments along terrace corners of the substrate (dark atoms).  In that configuration each deposited atom benefits cohesively from adhering to \emph{two} substrate neighbor atoms.  Short monatomic wires would be added, say with an atomic-force manipulator, extending to macroscopic leads for standard four-probe measurement of electric resistance.  If the signal from one superwire is too weak, then a parallel array of several superwires---two are shown in Fig. 39---should be employed.

It may well be that this test has already been done \emph{in essence}, although unwittingly, accidentally, and with little control.  As recently reported, room-temperature superconductivity has been found in hydrogen-doped granular graphite\cite{22} and is supported by granular superconductivity in highly oriented pyrolytic graphite.\cite{23,24}  (Graphite powder was exposed to a hydrogen plasma or soaked with water, filtered and heated to dry.  Superconductivity was detected by measurement of magnetization $M$, not of resistivity $\rho $.  Pressing the treated powder reduced the superconduction signal or suppressed it completely.)

Similar to the alignment of deposited atoms along terrace corners (Fig. 39), the doped hydrogen atoms would aggregate along the corners that are formed where the graphite grains (size: several tens of micrometers) are in contact, as illustrated in Fig. 40.  The resulting monatomic hydrogen loops are typically unconnected to each other.  That's why only magnetic measurements can be performed, not the standard four-probe resistance measurement.  When the powder sample is pressed, then the grain-contact areas are reduced and hydrogen is driven out.  This scenario would be corroborated if hydrogen gas were detected when the powder is pressed inside a chamber.

\noindent .

\noindent .

\noindent .

\noindent .

\noindent .

\pagebreak

\includegraphics[width=6.25in]{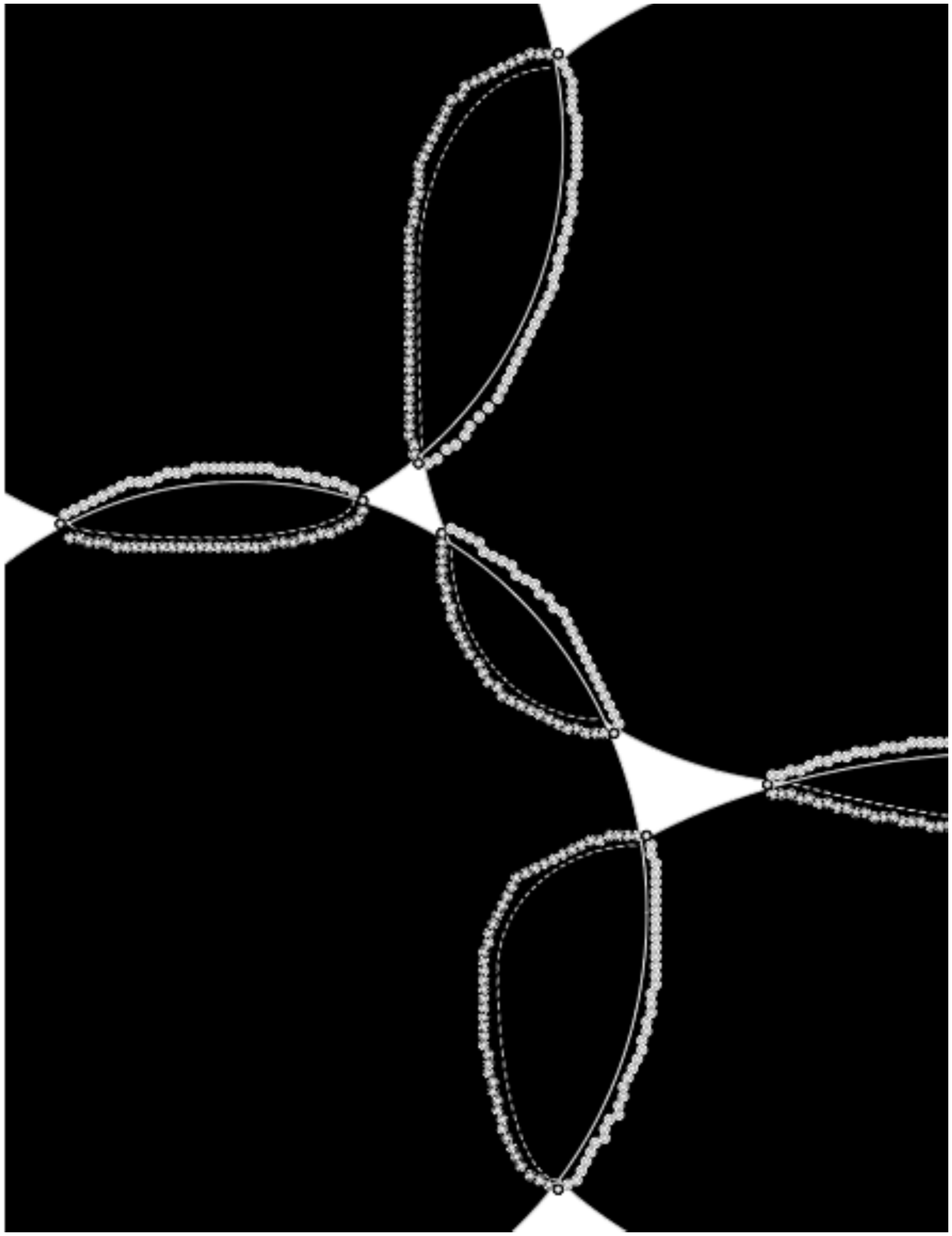}

\noindent FIG. 40. Monatomic hydrogen filament loops along the contact edges of graphite grains (schematic).

\pagebreak

\centerline{ \textbf{ACKNOWLEDGMENTS}}

I thank Ernst Mohler, Duane Siemens, and Kurt Estel for listening to my thoughts on superconductivity and giving me feedback.  I also thank Preston Jones for help with LaTeX.


\begin{thebibliography}{24}
\bibitem{5} The $R_{nn}$ value of $K$ metal is not fully warranted by atomic size $r(K)$ but is affected by relative weak van der Waals attraction, compared to other $A$ metals.  An orbital-moments analysis of the underlying ion cores can be found in Ref. 6)
\bibitem{10} No reconfiguration to a $d^{3}$ shell is possible for $s^{2}d^{4}$ of a free $W$ atom, hence the change to $s^{1}d^{5}$ for crystalline $W^{*}$.
\bibitem{12} More realistically, the parent atoms are already deformed in their elemental crystal lattice---rounded octahedrons in metallic $Nb$ and rounded tetrahedrons in semiconducting $Ge$.  However, the change of predeformed atoms in their elemental parent lattices to the new deformations in a compound lattice would be much harder to visualize and harder to recognize.
\bibitem{15} When alloyed with group IB elements---$Cu,Ag,Au$---still higher critical temperatures are achieved, $T_{c}(Pd_{0.55}Cu_{0.45}H_{0.70}) = 16.6$ K, $T_{c}(Pd_{0.70}Ag_{0.30}H_{0.80}) = 15.6$ K, and $T_{c}(Pd_{0.84}Au_{0.16}H_{0.90}) = 13.6$ K.  It is unlikely that the higher $T_{c}$'s are caused by the $s$ electrons of the noble metals, because of their far separation (low concentration).  Instead, the distorted $fcc$ lattice of the alloy---still more distorted by $H$ vacancies---exerts more squeeze on the present $H$ atoms.
\bibitem{16} The softness of the $Yb$ atom, resulting from its closed subshell, $4f^{14}$, gives rise to the looser $fcc$ lattice of elemental $Yb$ metal, in contrast to the hexagonal lattice of its neighbor elements, $Tm$ and $Lu$ metal.
\end{thebibliography}
\end{document}